\documentclass[a4paper,fleqn]{cas-dc}

\usepackage[numbers]{natbib}
\usepackage{cepcphysics}
\usepackage{subfigure}
\usepackage{makecell}
\usepackage{graphicx}
\usepackage{bm}

\def\tsc#1{\csdef{#1}{\textsc{\lowercase{#1}}\xspace}}
\tsc{WGM}
\tsc{QE}

\begin{document}
\let\WriteBookmarks\relax
\def\floatpagepagefraction{1}
\def\textpagefraction{.001}

\shorttitle{Expected $\Higgs \to \gamma\gamma$ branching ratio at the CEPC}    

\shortauthors{Fangyi Guo et.al}  

\title [mode = title]{The expected measurement precision of the branching ratio of the Higgs decaying to the di-photon at the CEPC}  



%

\author[1,2]{Fangyi Guo}[ type=editor, auid=000, orcid=0000-0002-3864-9257]
\ead{guofangyi@ihep.ac.cn}

\author[1,2] {Yaquan Fang}
\author[1,2] {Gang Li}
\author[1,2,3] {Xinchou Lou}






\affiliation[1]{organization={Institute of High Energy Physics (IHEP), Chinese Academy of Science},
            city={Beijing},
            postcode={100049},
            country={China}}
\affiliation[2]{ organization={University of Chinese Academy of Science},
            city={Beijing},
            postcode={100049},
            country={China}}
\affiliation[3]{ organization={University of Texas at Dallas},
            city={Richards},
            postcode={75080-3021},
            state={TX},
            country={USA}}









\begin{abstract}
This paper presents the prospects of measuring $\sigma(\ee\to\ZH)\times\Br(\Higgs \to \gamgam)$ in 3 \Z decay channels 
$\Z \to \qqbar/ \mumu/ \nnbar$ using the baseline detector with $\sqrt{s} = 240$ GeV at the Circular Electron Positron Collider (CEPC). 
The simulated Monte Carlo events are generated and scaled to an integrated luminosity of 5.6 \iab to mimic the data. 
Extrapolated results to 20 \iab are also shown.
The expected statistical precision of this measurement after combining 3 channels of Z boson decay is 7.7\%. 
With some preliminary estimation on the systematical uncertainties, the total precision is 7.9\%. 
The performance of CEPC electro-magnetic calorimeter (ECAL) is studied by smearing the photon energy resolution in simulated events in $\ee \to \ZH \to \qqyy$ channel.
In present ECAL design, the stochastic term in resolution plays the dominant role in the precision of Higgs measurements in $\Higgs \to \gamgam$ channel. 
The impact of the resolution on the measured precision of $\sigma(\ZH)\times\Br(\ZH \to \qqyy)$ as well as the optimization of ECAL constant term and stochastic term are studied for the further detector design.
\end{abstract}




\maketitle

\section{Introduction}
\label{sec:Intro}

In 2012, the ATLAS and CMS collaboration announced the discovery of Higgs Boson at the Large Hadron Collider (LHC)~\cite{Higgs_ATLAS, Higgs_CMS}. 
In the following years the precise measurements of Higgs properties become one of the main goals in particle physics, hoping to answer the remaining basic questions in nature and find new physics. 
For this purpose, hadron colliders like the LHC may not be the best choice due to large amount of background processes and corresponding lower ratio between the signals and backgrounds. 
Instead, a lepton collider can provide cleaner experiment environment and well-known initial states, which is crucial for high precision studies to find the hints of new physics. 
Thus several future lepton collider experiments are proposed, including the International Linear Collider (ILC)~\cite{ILC_TDR}, 
the Circular Electron Positron Collider (CEPC)~\cite{CEPC_CDR_V2}, the Future Circular Collider $\ee$ (FCC-ee)~\cite{FCC-ee}, 
and the Compact Linear Collider (CLIC)~\cite{CLIC}.

The CEPC is designed to be a circular lepton collider hosted in a tunnel with a circumference of 100 km and operate at a center of mass energy $\sqrt{s} = 240$ GeV as a Higgs factory. 
After 10 years running period, the CEPC will collect 5.6 \iab data, corresponding to more than 1 million Higgs boson. 
With this clean and large Higgs sample, the precision for the measurements of the Higgs properties is expected to be enhanced with one order of magnitude comparing with the LHC~\cite{An:2018dwb}. 

The Higgs boson interacts with photon through the top quark loop and massive boson loop. 
This mechanism gives low $\Higgs\to \gamgam$ branching ratio in the Standard Model (SM) but also makes it a good channel to test the new physics beyond the SM. 
Besides, high energy photons from the Higgs boson decay can be identified and measured well experimentally. 
So this channel also serves as a good benchmark for the performance of the electromagnetic calorimeter (ECAL) study. 
Current measurement of the inclusive Higgs boson signal strength in the diphoton channel in LHC is $1.04^{+0.10}_{-0.09}$ in ATLAS~\cite{arxiv.2207.00348} 
and $1.03^{+0.11}_{-0.09}$ in CMS~\cite{CMS-PAS-HIG-19-015} with the $pp$ collision data collected by ATLAS and CMS from 2015 to 2018. 
The results are consistent with the SM prediction with the present precision. 
In the HL-LHC period the ATLAS expects to collect 3 \iab data. 
The projected precision of the $\Higgs\to \gamma\gamma$ measurement is 6\% to 4\% depending on different considerations of the systematic uncertainties S1 or S2 in~\cite{arxiv.1902.00134}.
Combined with CMS, a precision of 2.5\% can be reached in the optimistic systematic scenario S2. 

A previous analysis studied the expected Higgs precision in various Higgs decay channels~\cite{An:2018dwb} including $\Higgs\to \gamgam$. 
A precision of 6.8\% is expected for the measurement of $\sigma(\ZH) \times \Br(\Higgs\to \gamgam)$ with CEPC-v4 conceptual detector.
However, this result is based on the fast simulation of Monte Carlo samples and cut-based analysis method. 
In recent study~\cite{CEPC_Accelerator_Snowmass} the CEPC accelerator study group updated on the radiation power, resulting in an increase of the instantaneous luminosity of 66\%. 
Based on this update, a new nominal data-taking scenario is proposed. 
It aims at ten years of data taking at $\sqrt{s}$ = 240 GeV with two interaction points (IPs), 
accumulating an integrated luminosity of 20 \iab Higgs data~\cite{CEPCPhysicsStudyGroup:2022uwl}. 
And one new conceptual detector design is also on-going. 
A homogeneous ECAL is considered to replace the previous silicon-tungsten sampling calorimeter~\cite{CEPCPhysicsStudyGroup:2022uwl, instruments6030040, Liu_2020_CrystalCalo}. 
So it is worth to re-study the $\Higgs\to \gamgam$ process with the latest benchmark, 
investigating the impact from the larger statistics and the new detector.

This paper is organized as following. 
Sec.~\ref{sec:det_MC} briefly introduces the CEPC detector and the simulated Monte-Carlo samples used in this analysis. 
Sec.~\ref{sec:selection} presents the object reconstructions and event selections. 
Sec.~\ref{sec:MVAanalysis} describes the developed MVA method for this work. 
Sec.~\ref{sec:models} studies the signal and background models. 
The results are summarized in Sec.~\ref{sec:results}. 
In Sec.~\ref{sec:EcalRes} we investigate how these results can be influenced by the CEPC ECAL resolution, that can provide guidelines for the detector optimization. 
The conclusions are drawn in Sec.~\ref{sec:conclusion}.

\section{CEPC detector and Monte-Carlo simulation}
\label{sec:det_MC}

The CEPC detector is designed to match the physics goals that all final states can be identified and reconstructed with high resolution. 
The baseline detector concept utilizes the particle flow approach (PFA) idea~\cite{PandoraPFA_2009}, 
with a precise vertex detector, a Time Projection Chamber (TPC), a silicon tracker, a high granularity Silicon-Tungsten sampling ECAL and a GRPC-based high granularity hadronic calorimeter (HCAL). 
All the system is imbedded in 3 Tesla magnetic field. The outermost of the detector is a muon chamber. 
The details can be found in Ref.~\cite{CEPC_CDR_V2}.

\begin{figure}[htbp]
\centering
  \subfigure[]{\includegraphics[width=0.30\linewidth]{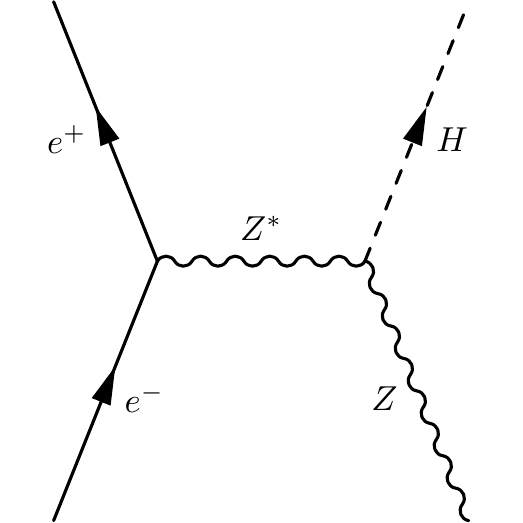}}
  \subfigure[]{\includegraphics[width=0.30\linewidth]{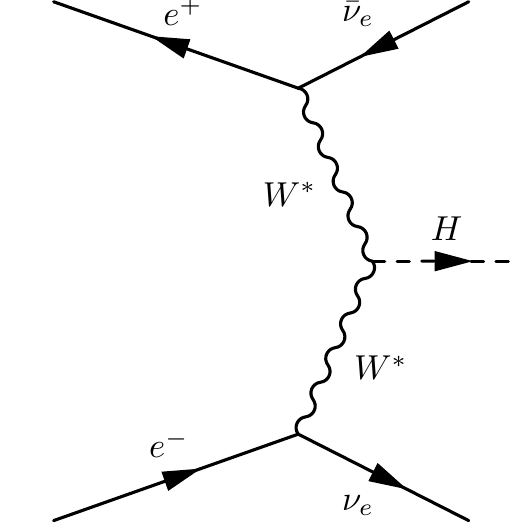}}
  \subfigure[]{\includegraphics[width=0.30\linewidth]{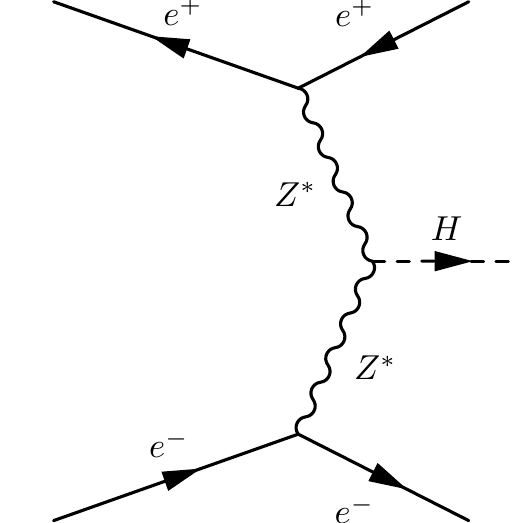}}
  \caption{Feynman diagrams of the Higgs boson production processes at the CEPC: (a) $\ee\to \ZH$, (b) $\ee\to \nnbar \Higgs$ and (c) $\ee\to \ee \Higgs$.}
  \label{fig:FeynmanDiagram}
\end{figure}

The Higgs production mechanisms at the CEPC are Higgs-strahlung $\ee\to \ZH$, \W/\Z fusion $\ee\to \nnbar \Higgs$ and $\ee\to \ee \Higgs$ as illustrated in Figure~\ref{fig:FeynmanDiagram}. 
In this analysis Higgs production via \ZH process decaying to diphoton final state $e^{+}e^{-} \to \ZH \to f\bar{f}\gamma\gamma$ at $\sqrt{s}=240$ GeV is considered as the dominant signal. 
It is further divided into 3 sub-channels, depending on \Z decaying to $q\bar{q}$, $\mu^{+}\mu^{-}$ and $\nu\bar{\nu}$. 
$\Z\to e^{+}e^{-}$ channel is abandoned due to the known extremely large Bhabha background, and $\Z\to \tau^{+}\tau^{-}$ channel is dropped as well because of the complexity of $\tau$ identification. 
\W/\Z fusion process is counted in $\ZH, \Z\to\nnbar$ sub-channel. 
The only considered background process is the 2-fermion background $\ee\to \ffbar$ in CEPC 
with at least two photons from initial and final state radiations.
The Higgs resonant background, 4-fermion processes and possible reducible background in the experiment are expected to be negligible. 
These SM physics processes are generated with Whizard~\cite{Kilian:2007gr} at leading order (LO) interfaced with Pythia 6~\cite{PYTHIA6} for parton showering and hadronization with parameters based on the Large Electron Positron Collider (LEP) \cite{LEP} data. 
Initial state radiation (ISR) and final state radiation (FSR) effects are taken into account.
The total energy spread caused by beamstrahlung and synchrotron radiation is studied by Monte-Carlo simulation and determined to be 0.1629\% at CEPC~\cite{CEPC_CDR_V1}. 
Table~\ref{tab:MC_xs_stat} lists the cross sections of physics processes and MC sample statistics used in the analysis. 
Event yields are normalized to 5.6 \iab. Detailed configurations can be found in Ref.~\cite{Mo:2015mza}.

\begin{table}[htbp]
\centering
\begin{tabular}{|l|l|l|}
\hline
Process                   & $\sigma$        & statistics     \\ \hline
\multicolumn{3}{|c|}{\qqyy sub-channel}                     \\ \hline
$\ee\to \ZH \to \qqyy$    &  0.31 \fb       & 100 k          \\ \hline
$\ee \to \qqbar$          &  54.1 \pb       & ~20 M          \\ \hline
\multicolumn{3}{|c|}{\mmyy sub-channel}                     \\ \hline
$\ee\to \ZH \to \mmyy$    &  0.15 \fb       & 100 k          \\ \hline
$\ee \to \mumu$           &  5.3 \pb        & ~20 M          \\ \hline
\multicolumn{3}{|c|}{\nnyy sub-channel}                     \\ \hline
\makecell[l]{$\ee\to \ZH \to \nnyy$ \\ $\ee\to \nnbar \Higgs \to \nnyy$} 
                          &  0.11 \fb       & 100 k          \\ \hline
$\ee \to \nnbar$          &  54.1 \pb       & ~20 M          \\ \hline
\end{tabular}
\caption{Cross sections and the simulated MC sample statistics. In \qqyy and \mmyy channels \ZH is the only considered process, and in \nnyy channel both \ZH $\Z\to inv.$ and \W/\Z fusion processes are considered. }
\label{tab:MC_xs_stat}
\end{table}

The simulation of detector configuration and response is handled by MokkaPlus~\cite{Mokka}, a GEANT4~\cite{GEANT4} based framework. 
The full detector simulation is performed for signal process only. 
The background processes are simulated by smearing the truth particles with the parameterized detector resolution and efficiency to save the computing resource.

\section{Object reconstruction and event selection}
\label{sec:selection}

The CEPC follows the PFA scheme in the event reconstruction, with a dedicated tookit ARBOR~\cite{ARBOR_2014,Ruan:2018yrh}. 
The tracks are firstly reconstructed with the hits in the tracking detector by Clupatra module~\cite{Clupatra}. 
Then ARBOR collects the tracks from Clupatra and hits in calorimeter, and composes the Particle Flow Objects (PFOs) by its clustering and matching modules. 
These PFOs are identified as charged particles, photons, neutral hadrons and unassociated fragments. 
With this approach the photon is identified in ARBOR with the shower shape variables obtained from the high granularity calorimeter, without any matched track. 
Converted photons are not considered yet, which counts 5-10\% in central region and 25\% in forward region~\cite{CEPC_CDR_V2}. 
The lepton (\epm, \mupm) is defined by a track-matched particle. 
A likelihood-based algorithm, LICH~\cite{Yu_2017_LICH}, is implemented in ARBOR to separate electrons, muons and hadrons. 
Jets are formed from the particles reconstructed by ARBOR with the Durham clustering algorithm \cite{CATANI1991432_eekt} after excluding the interested particles. 
The jet energy is calibrated by the MC simulation currently, while is foreseen to be re-calibrated with physics events like $\W\to \qqbar$ and/or $\Z\to \qqbar$ in CEPC. 
No flavor tagging approach is used in this analysis for simplicity. 

The event selections are applied to improve the signal significance and background modeling.
In three sub-channels individual strategies are considered depending on the topology of the physics process. 
In $\ZH\to \nnbar\gamgam$ channel 2 photons are required inclusively in the final state. 
In $\ZH\to \mumu\gamgam$ channel the 2 leading photons and 2 muons are selected exclusively, requiring a veto of other particles, 
the missing energy $E_{missing}$ and missing mass $M_{missing}$ less than 10 GeV and the invariant mass of the muon pair close to \Z boson mass.

In $\ZH\to q\bar{q}\gamma\gamma$ channel, 2 leading photons are firstly selected, and other particles are reconstructed into 2 jets with Durham algorithm. 
Some dedicated cuts are applied on the kinematic variables of these final state objects as listed in Table~\ref{tab:cutflow_qqyy}, \ref{tab:cutflow_mmyy}, \ref{tab:cutflow_nnyy}, along with the final efficiency and expected event yields. 

\begin{table}[htbp]
\centering
\scriptsize
\begin{tabular}{l|c|c}
\hline \hline
Selections                         & Higgs signal   & \qqyy background   \\ \hline
Exclusive 2 jets and 2 photons     & 85.56\%        & 69.57\%     \\
$E_{\gamma 1} >$ 25 GeV            & 100.00\%       & 2.35 \%     \\
$E_{\gamma 2} \in [35,95]$ GeV      & 98.37\%        & 35.33\%     \\
$\cos\theta_{\gamgam}>$ -0.95      & 95.20\%        & 68.01\%     \\
$\cos\theta_{jj} >$-0.95           & 90.86\%        & 85.54\%     \\
$pT_{\gamma 1} > $20 GeV           & 93.42\%        & 56.94\%     \\
$pT_{\gamma 2} > $30 GeV           & 93.25\%        & 54.54\%     \\
$\myy \in [110,140]$ GeV           & 97.50\%        & 21.14\%     \\
$E_{\gamgam} >$ 120 GeV            & 99.47\%        & 98.41\%     \\
$min|\cos\theta_{\gamma j}| <$0.9  & 71.67\%        & 48.05\%     \\  \hline
Total eff                          & 44.08\%        & 0.01\%      \\  \hline
Yields in 5.6 \iab                 & 766.64         & 26849.38    \\
\hline 
\end{tabular}
\caption{Selection criteria and the corresponding efficiencies in \qqyy channel. $\gamma 1 (\gamma 2)$ is defined as the photon with lower (higher) energy. $\cos\theta_{\gamgam} (\cos\theta_{jj})$ is the polar angle of di-photon (di-jet) system. $min|\cos\theta_{\gamma j}|$ is the minimum $\cos\theta$ of the photon-jet pairs. }
\label{tab:cutflow_qqyy}
\end{table}

\begin{table}[]
\centering
\scriptsize
\begin{tabular}{l|c|c}
\hline \hline
Selections                                 & Higgs signal    & \mmyy background        \\ \hline
Exclusive 2 muons and 2 photons            & 70.18\%         & 5.18\%      \\
$E_{\gamma} >$35 GeV                       & 99.21\%         & 8.39\%      \\
$|\cos\theta_{\gamma}|<$0.9                & 83.79\%         & 38.14\%     \\
$pT_{\gamma 1} \in [10, 70]$ GeV           & 99.84\%         & 86.30\%     \\
$pT_{\gamma 2} \in [30, 100]$ GeV          & 99.96\%         & 95.59\%     \\
$\myy \in [110,140]$ GeV                   & 98.08\%         & 37.62\%     \\
$M_{\gamgam}^{recoil}\in [85,105]$ GeV     & 80.12\%         & 21.29\%     \\
$E_{\gamgam} \in [125,145]$ GeV            & 99.88\%         & 95.86\%     \\  \hline
Total eff                                  & 45.69\%         & 0.01\%      \\  \hline
Yields in 5.6 \iab                         & 39.32           & 2662.77   \\ 
\hline 
\end{tabular}
\caption{Selection criteria and the corresponding efficiencies in \mmyy channel. $\gamma 1 (\gamma 2)$ is defined as the photon with lower (higher) energy. $M_{\gamgam}^{recoil}$ is the recoil mass of di-photon system in CEPC $\sqrt{s}=240$ GeV: $(M_{\gamgam}^{recoil})^2 = (\sqrt{s}-E_{\gamgam})^2 - p^2_{\gamgam} = s-2E_{\gamgam}\sqrt{s}+m_{\gamgam}^2$. }
\label{tab:cutflow_mmyy}
\end{table}

\begin{table}[]
\centering
\begin{tabular}{l|c|c}
\hline \hline
Selections                       & Higgs signal   & \nnyy background       \\ \hline
Inclusive 2 photons              & 85.51\%        & 0.34\%     \\
$E_{\gamma}>$ 30 GeV             & 99.81\%        & 20.13\%    \\
$|\cos\theta_{\gamma}|<$ 0.8      & 70.48\%        & 11.56\%    \\
$pT_{\gamma}>$ 20 GeV            & 99.97\%        & 99.26\%    \\
$M_{missing}>$ 60 GeV            & 98.17\%        & 99.71\%    \\
$\myy \in [110,140]$ GeV         & 97.51\%        & 22.86\%    \\
$E_{\gamgam} \in [120,150]$ GeV  & 99.16\%        & 99.58\%    \\ \hline
Total eff                        & 57.08\%        & 0.002\%     \\ \hline
Yields in 5.6 \iab               & 335.89         & 3640.20  \\ \hline
\end{tabular}
\caption{Selection criteria and the corresponding efficiencies in \nnyy channel. $M_{missing}$ is the missing mass calculated from the total visible objects. }
\label{tab:cutflow_nnyy}
\end{table}

\section{MVA-based analysis}
\label{sec:MVAanalysis}

The Multi-Variate Analysis (MVA) method is employed to further suppress the background. 
It uses the machine learning (ML) techniques to combine the separation power from several variables into a unique variable. 
In this analysis we choose the Gradient Boosted Decision Tree (BDTG) method with TMVA toolkit~\cite{TMVA}. 
In each sub-channel the \ZH and 2 fermion processes are considered as the signal and background for the BDTG. 
All events from MC are separated into 2 sets for the 2-fold validation~\cite{k-fold-validation} to avoid the risk of overtraining. 
Following principles are considered while constructing the input variables for BDTG: 

\begin{itemize}
	\item The basic information is the Lorentz vector of the final state particles. These include the momentum (P), transverse momentum ($p_{T}$), energy ($E$), polar angle ($\cos\theta$), recoil mass for photons, fermions, systems, and the $\Delta P$, $\Delta E$, $\Delta \Phi$, $\Delta \cos\theta$, $\Delta R$ for 2 objects or systems, and the missing mass $M_{missing}$. 
  \item Use the separation $\left\langle S^2 \right \rangle$ defined in Eq.~\ref{eq:SeparationPower} to quantify the discrimination power between signal and background of a given variable, where $y$ represents the discriminating variable, and $\hat{y}_{s}(y)$ and $\hat{y}_{b}(y)$ are the corresponding probability distribution function of the variable for signal and background samples.

\begin{equation}
\left\langle S^2 \right \rangle = \frac{1}{2}\int \frac{(\hat{y}_{s}(y)-\hat{y}_{b}(y))^2}{\hat{y}_{s}(y)+\hat{y}_{b}(y)}dy.
\label{eq:SeparationPower}
\end{equation}
	\item To ensure the application of 2D model described in Sec.~\ref{sec:models}, which requires an assumption of independence between the BDTG response and \myy, the constructed variable should have low linear correlation with \myy: $|\text{Corr}_{v-\myy}|<30\%$. 
	\item To reduce the redundance for the training, the linear correlation between any two variables should be small: $|\text{Corr}_{v1-v2}|<40\%$. The one with lower separation power is removed. 
\end{itemize}

Table~\ref{tab:MVAvars_qqyy}-\ref{tab:MVAvars_nnyy} lists the selected variables along with their definition and $\langle S^{2} \rangle$ for BDTG. 

\begin{table}[hbt]
\centering
\scriptsize
\begin{tabular}{|l|l|l|}
\hline
Variable                      & Definition                                                                                                    & Separation \\ \hline \hline
$pT_{\gamma 1}$               & Transverse momentum of the sub-leading photon                                                                 & 0.209      \\ \hline
$\cos\theta _{\gamma 2}$      & Polar angle of the leading photon                                                                             & 0.197      \\ \hline
$\Delta\Phi_{\gamgam}$        & Azimuthal angle between two photons                                                                           & 0.147      \\ \hline
$\min\Delta R_{\gamma, j}$    & \makecell[l]{Minimum $\Delta R$ between one of the two photons \\ and one of the jets}                        & 0.054      \\ \hline
$E_{j1}$                      & Energy of the sub-leading jet                                                                                 & 0.041      \\ \hline
$\Delta\Phi_{\gamgam, jj}$    & \makecell[l]{Azimuthal angle between the diphoton and \\ dijet system}                                                         & 0.033      \\ \hline
$pT_{j2}$                     & Transverse momentum of the leading jet                                                                        & 0.032      \\ \hline
$\cos\theta_{j1}$             & Polar angle of the sub-leading jet                                                                            & 0.032      \\ \hline
$\cos\theta_{\gamgam, jj}$    & \makecell[l]{Polar angle difference between diphoton and \\ dijet system $\cos(\theta_{\gamgam}-\theta_{jj})$}                  & 0.024      \\ \hline
$\cos\theta_{\gamma 1, j1}$   & \makecell[l]{Polar angle difference between sub-leading \\ photon and sub-leading jet $\cos(\theta_{\gamma 1}-\theta_{j1})$}    & 0.023      \\ \hline
\end{tabular}
\caption{Input variables for BDTG in \qqyy channel.}
\label{tab:MVAvars_qqyy}
\end{table}

\begin{table}[hbt]
\centering
\scriptsize
\begin{tabular}{|l|l|l|}
\hline
Variable                         & Definition                                                                & Separation \\ \hline \hline
$\min\Delta R_{\gamma, \mu}$     & \makecell[l]{Minimum $\Delta R$ between one of the two photons \\ and one of the muons}    & 0.335      \\ \hline
$E_{\mu\mu}$                     & Energy of the di-muon system                                              & 0.259      \\ \hline
$\cos\theta_{\gamma 1, \mu1}$    & \makecell[l]{Polar angle difference between sub-leading \\ photon and sub-leading muon}    & 0.189      \\ \hline
$E_{\gamma 2}$                   & Leading photon energy                                                     & 0.160      \\ \hline
$\Delta\Phi_{\gamgam}$           & Azimuthal angle between two photons                                       & 0.090      \\ \hline
$\cos\theta_{\gamma 2}$          & Polar angle of the leading photon                                         & 0.072      \\ \hline
$\Delta\Phi_{\gamgam, \mu\mu}$   & \makecell[l]{Azimuthal angle between diphoton and dimuon \\ system}                        & 0.034      \\ \hline
$\cos\theta_{\mu 1}$             & Polar angle of the sub-leading muon                                       & 0.014      \\ \hline
\end{tabular}
\caption{Input variables for BDTG in \mmyy channel.}
\label{tab:MVAvars_mmyy}
\end{table}

\begin{table}[hbt]
\centering
\scriptsize
\begin{tabular}{|l|l|l|}
\hline
Variable                  & Definition                                                            & Separation \\ \hline \hline
$pT_{\gamma 1}$           & Transverse momentum of the sub-leading photon                         & 0.089      \\ \hline
$\cos\theta _{\gamma 2}$  & Polar angle of the leading photon                                     & 0.079      \\ \hline
$\Delta\Phi_{\gamgam}$    & Azimuthal angle between two photons                                   & 0.054      \\ \hline
$pTt_{\gamgam}$           & \makecell[l]{Diphoton $p_{T}$ projected perpendicular to the \\ diphoton thrust axis}  & 0.042      \\ \hline
$pT_{\gamma 2}$           & Transverse momentum of the leading photon                             & 0.037      \\ \hline
\end{tabular}
\caption{Input variables for BDTG in \nnyy channel.}
\label{tab:MVAvars_nnyy}
\end{table}

\section{Signal and background models}
\label{sec:models}
\begin{figure}[h]
  \centering
  \subfigure[\qqyy signal]{ \includegraphics[width= 0.9\linewidth]{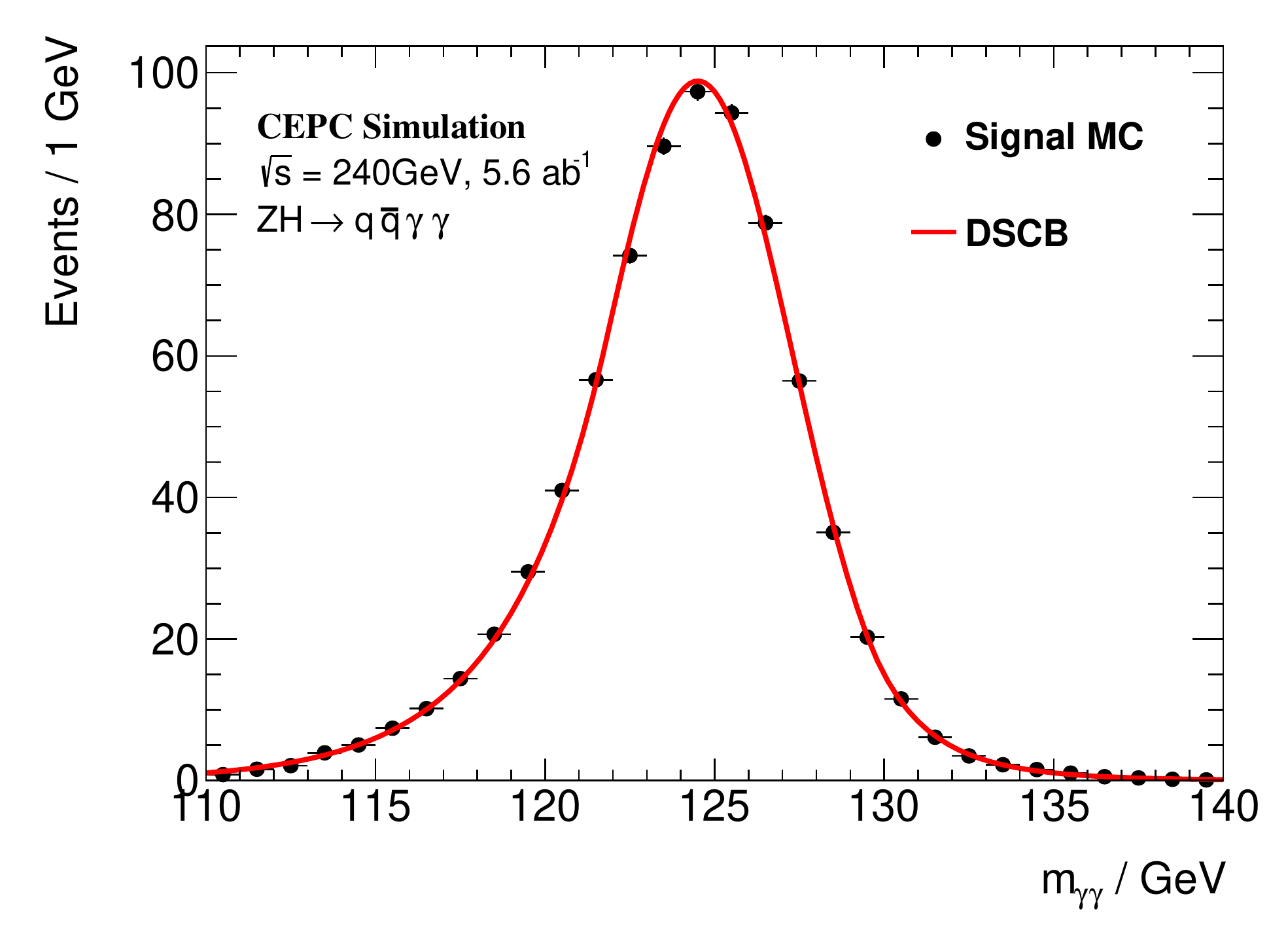} } \\
  \subfigure[\mmyy signal]{ \includegraphics[width= 0.9\linewidth]{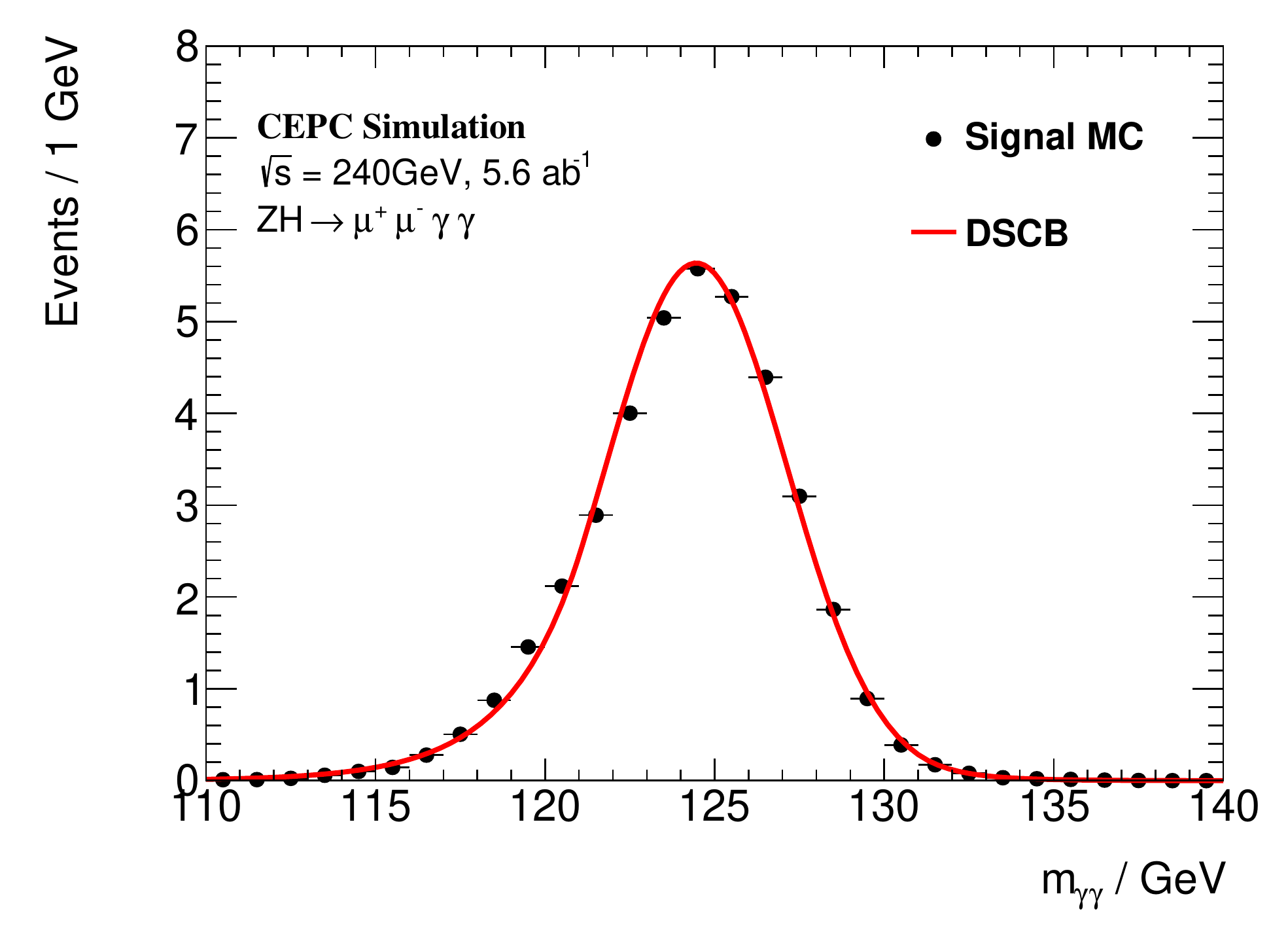} } \\
  \subfigure[\nnyy signal]{ \includegraphics[width= 0.9\linewidth]{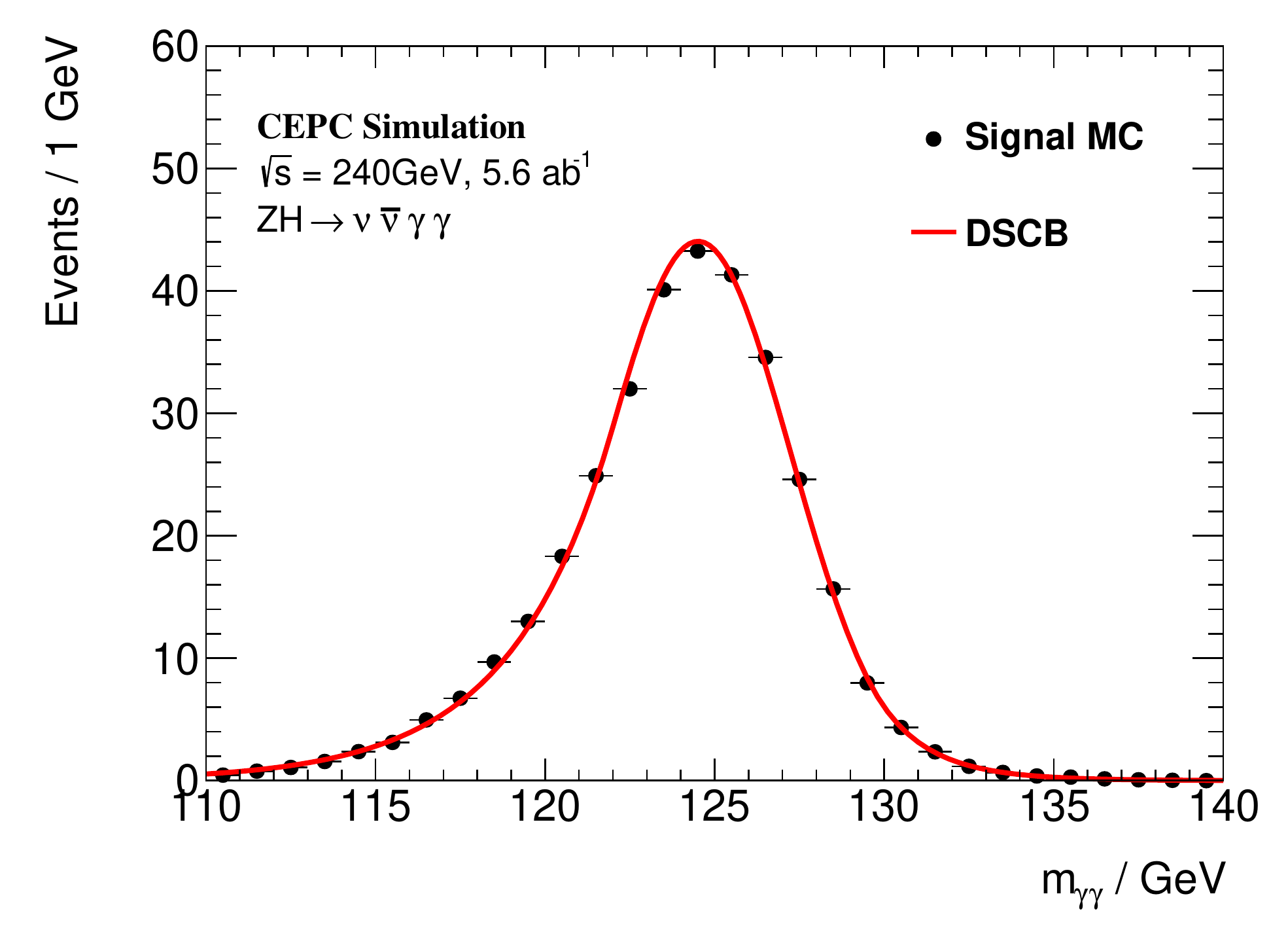} } \\
\caption{The signal MC and the fitted DSCB model in 3 channels. }
\label{fig:signal}
\end{figure}

The Higgs signal is extracted by fitting the \myy and the shape of the BDTG responses. 
The resonant peak above a smooth \myy distribution for the background at around Higgs mass (125 GeV) can be reconstructed through the excellent calorimeter energy resolution in CEPC. 
The signal \myy distribution is fitted with a Double Side Crystal Ball (DSCB) function: 
\begin{align}
\footnotesize
  f(t) = N \times
  \begin{cases}
    e^{-t^{2}/2} & \text{if } -\alpha_{low} \leq t \leq \alpha_{high} \\
    \frac{ e^{-{}^{1}_{2} \alpha_{low}^{2}} } { \left[ \frac{1}{R_{low}} \left(R_{low} - \alpha_{low} - t \right) \right]^{n_{low}} } & \text{if } t < -\alpha_{low} \\
    \frac{ e^{-{}^{1}_{2} \alpha_{high}^{2}} } { \left[ \frac{1}{R_{high}} \left(R_{high} - \alpha_{high} + t \right) \right]^{n_{high}} } & \text{if } t > \alpha_{high} \\
  \end{cases}
  \label{eq:DSCB}
\end{align}
where $N$ is a normalization factor, $t=(\myy - \mu_\text{CB}) / \sigma_\text{CB}$.
Figure~\ref{fig:signal} shows the fitted \myy signal shape in 3 channels. They are well described by the DSCB function. 
The resolution is estimated to be 2.81 / 2.68 / 2.74 GeV in \qqyy / \mmyy / \nnyy channel.  

\begin{figure}[htbp]
  \centering
  \subfigure[\qqyy background]{ \includegraphics[width= 0.9\linewidth]{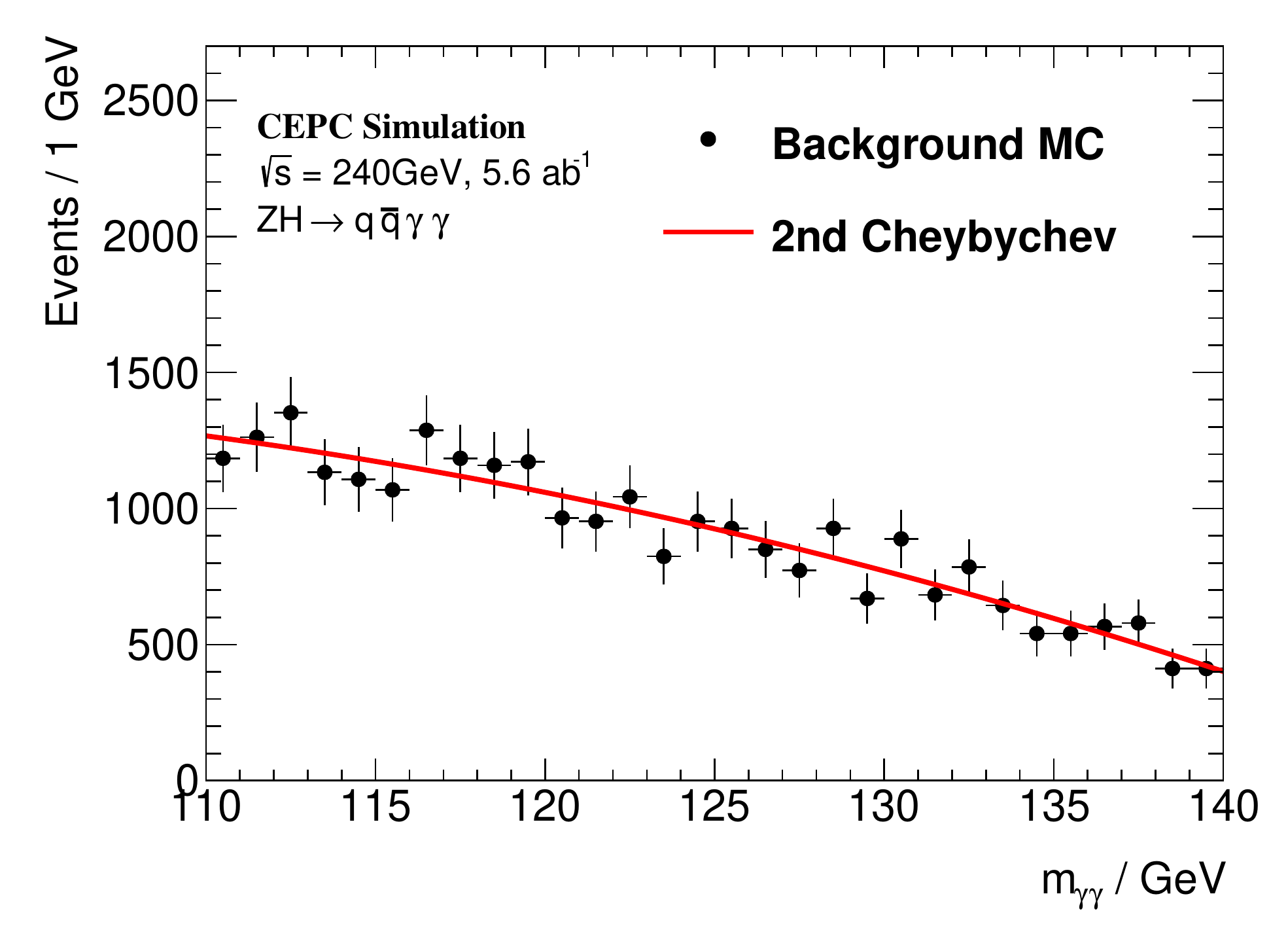} } \\
  \subfigure[\mmyy background]{ \includegraphics[width= 0.9\linewidth]{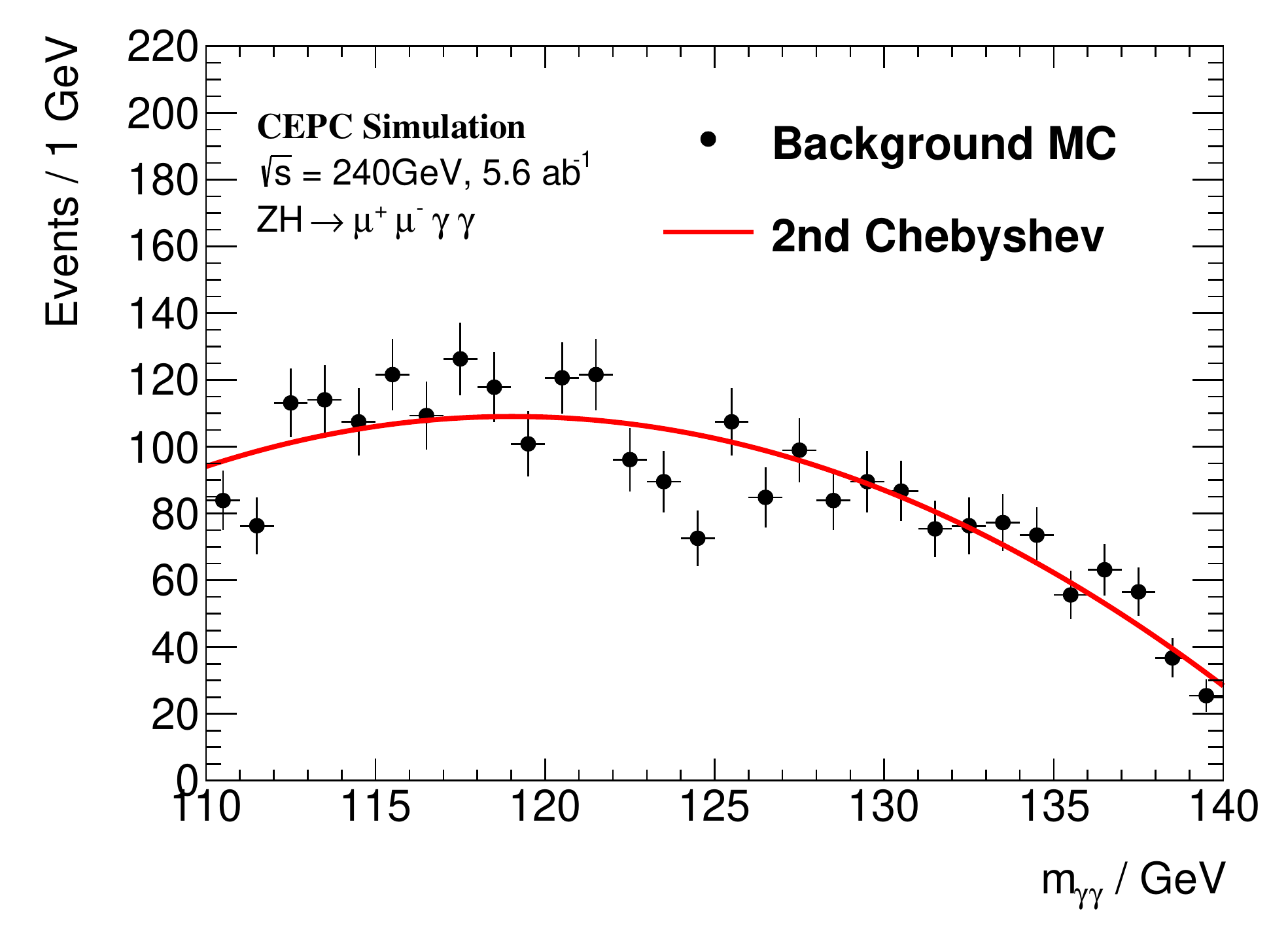} } \\
  \subfigure[\nnyy background]{ \includegraphics[width= 0.9\linewidth]{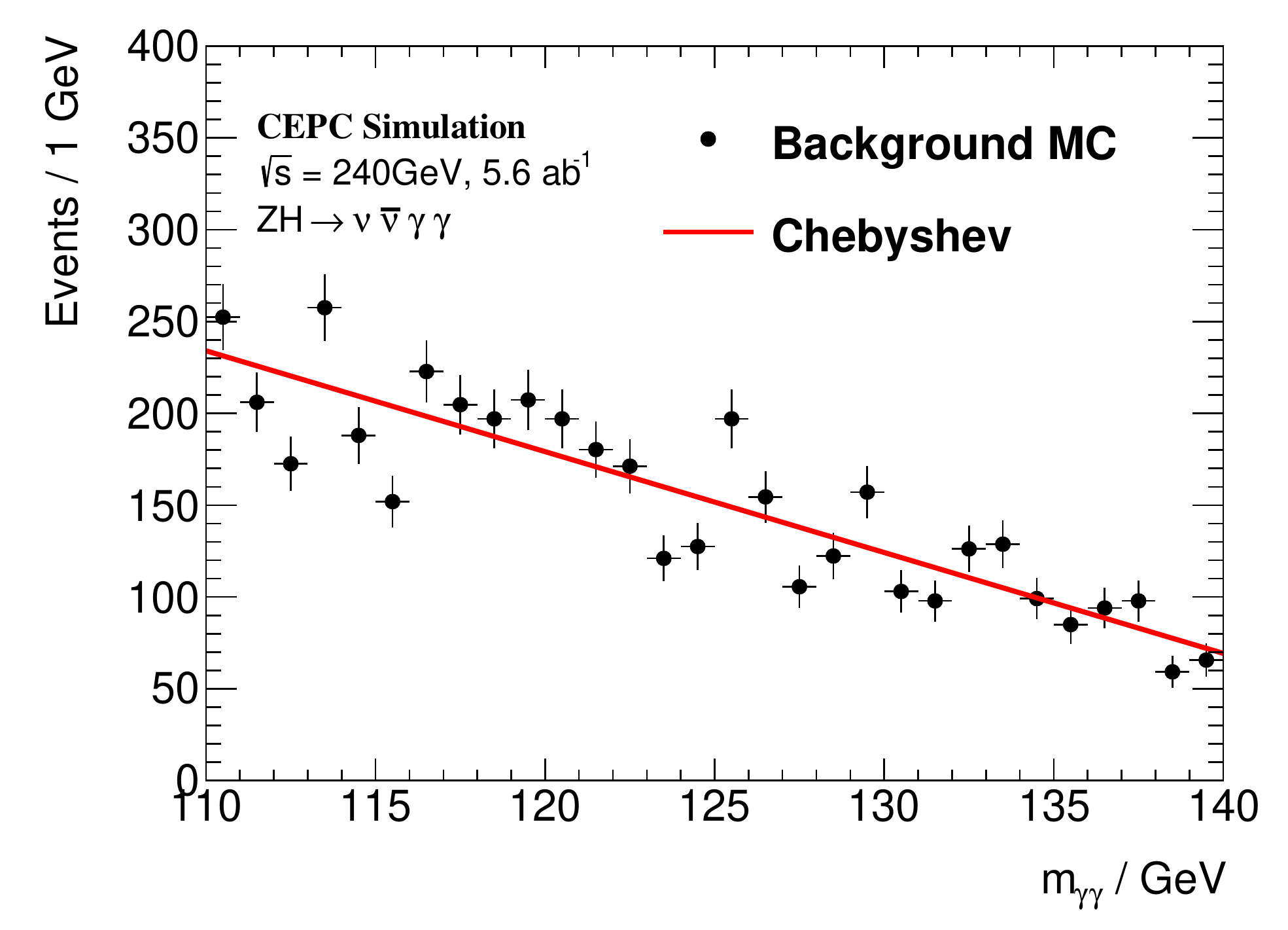} } \\
\caption{The background MC and the fitted \myy models in 3 channels. }
\label{fig:bkgmodel}
\end{figure}

Several smooth functions (Cheybyshev polynomials, exponential families and polynomial families) are tested for the background modeling, 
and the one with the least $\chi^{2}$/Ndof value is finally selected. 
The results are listed in Table~\ref{tab:bkgmodel_chi2} and shown in Figure~\ref{fig:bkgmodel}.

\begin{table}[h]
\centering
\begin{tabular}{|l|l|l|}
\hline
Channel     & Selected function    &  $\chi_{2}$/Ndof   \\ \hline
\qqyy       & 2nd order Chebyshev  &  0.60              \\ \hline
\mmyy       & 2nd order Chebyshev  &  1.79              \\ \hline
\nnyy       & 1st order Chebyshev  &  3.32              \\ \hline
\end{tabular}
\caption{ The decided background model in 3 channels. Tested functions include the exponential, 2nd order exponential polynomial, 1st and 2nd order polynomials, 1st and 2nd order Chebyshev polynomials. }
\label{tab:bkgmodel_chi2}
\end{table}

The histograms from the MC of signal and background are used to build the binned Probability Density Function (PDF), which is used as the model of BDTG distributions. 


The strategies in constructing BDTG ensure the reasonable independence between the BDTG response and \myy. 
Therefore a 2-dimension model from the multiplication of \myy and BDT models is applied to describe the signal and background. 
A high correlation can introduce the mis-modeling of the signal and/or background process. 
The linear correlation coefficients between \myy and BDT are -3.45\%, -11.6\%, 8.33\% for signal in \qqyy, \mmyy and \nnyy chananels. 
The corresponding correlation coefficients for the background are 11.6\%, 28.2\% and 28.4\% respectively.

\section{Systematic uncertainties}
\label{sec:syst}

The systematic uncertainties relevant to this measurement can be caused by several sources. 
However, at this stage most of them are not specifically studied yet for the CEPC. 
So in this paper we only present the methodology of analyzing the systematic uncertainties in CEPC and take the leading terms into account. 
Further quantified analysis awaits updates on the theoretical calculation, more comprehensive detector performance optimization and real data.

Based on the strategy of event modeling in Sec.~\ref{sec:models}, the systematic uncertainties can be grounded into two types: 
uncertainties in the expected signal yields in each channel, and the uncertainties in the modeling of the signal \myy distribution. 
The background yields and \myy model parameters are floated to consider the background mis-modeling effect and the contributions from model-dependent background process cross section calculation. 
The uncertainty of BDT modeling for both signal and background is contained by an envelope, 
which is counted into the signal event yield uncertainty. 

These systematic terms are incorporated into the likelihood model as nuisance parameters. 
For each of these nuisance parameters, a Gaussian or log-normal constraint PDF is included in the likelihood function, 
for the symmetric term like the \myy shape peak position, or non-negative term like the event yield. 
The construction of likelihood with the nuisance parameters are presented in Sec~\ref{sec:results}. 

\subsection{Theoretical uncertainties}
\label{ssec:theo_syst}
In contrast to hadron collider, only few theoretical uncertainty can enter the measurement in lepton collision experiment like CEPC. 
The theoretical calculations are less dependent on higher order QCD radiative correction.
And there is no influence from the Parton Distribution Functions or \alphas.
In this $\sigma\times\Br$ measurement the observed event yields are directly from the fitting, 
so the uncertainties from signal cross section calculation and $\Br(\Higgs\to\gamgam)$ can be eliminated. 
The only remaining uncertainty is the parton shower uncertainty in the \qqyy channel. 
It can be described by the MC sample difference from a set of generators, and is believed negligible. 
For completeness, a 0.5\% theoretical uncertainty is assumed on the signal yield in \qqyy channel.

\subsection{Experimental uncertainties}
\label{ssec:exp_syst}
The experimental systematic uncertainties affecting this measurement can include: integrated luminosity, detector acceptance, trigger efficiency, object reconstruction and identification efficiency, object energy scale and resolution.
In CEPC the luminosity can be monitored by the Lumi-Cal with the high statistics BhaBha process, thus a relative accuracy of 0.1\% is expected to be achieved~\cite{CEPC_CDR_V2}. 
Pile-up effects and underlying events should be negligible.  
A well-described detector geometry in the simulation is able to provide precise model of the detector acceptance and response. 
The possible mis-modeling can be further fixed with some data-driven methods. So the uncertainties should be very small. 
The photon reconstruction, identification and energy calibration rely on the dedicated algorithms and the real data. 
In CEPC CDR these uncertainties are studied to be controlled with sub-percent level. 
Furthermore, the known physics processes can be used as standard candles for the calibration, like $Z\to \ee + \gamma$, $\pi^{0}\to\gamgam$, etc. 
Similarly electrons, muons and jets can in principle be described well. 
In this di-photon channel study, the photon related uncertainties should be dominant. 
So we assume a 1\% uncertainty on the photon efficiency and 0.05\% uncertainties on the photon energy scale (PES) and resolution (PER).
Others terms remains to be added with better understanding about the experiments.    

The signal yield is affected by the luminosity, the photon efficiency and the impact of the photon energy scale and resolution uncertainties on the selection efficiency. 
A set of alternative simulation samples are generated, randomly rejecting 1\% photons, or scaling the energy up/down for 0.05\% or smearing the photon energy with 0.05\%. 
The expected signal yields are counted after all the selections, and a relative variation $\delta n^{i} = \frac{|n_{var}^{i}-n_{nom}^{i}|}{n_{nom}^{i}}$ is used to represent the influence from each term. 
This value from photon efficiency is around 2\%, and from photon energy scale and resolution is at the order of 0.01\%. 
They are considered as symmetric uncertainties on the signal yield. 

The signal \myy distribution is described with the double-side crystal ball function. 
The photon energy scale uncertainty is propagated to the peak position of the signal peak, and the photon energy resolution uncertainty is propagated to the signal width. 
They are estimated by refitting the signal shape in the variation samples and comparing with the nominal one: 
$\delta\mu_{CB}=\frac{\mu_{CB, var}-\mu_{CB, nom}}{\mu_{CB,nom}}$, $\delta\sigma_{CB} = \frac{\sigma_{CB, var}-\sigma_{CB, nom}}{\sigma_{CB,nom}}$. 
The impact from PES to the signal peak is between 0.04\% to 0.10\%, in different channels, and the impact from PER to the signal width is 0.004\% to 0.02\%. 
A 5.9 MeV Higgs mass measurement uncertainty is also considered based on CEPC estimation~\cite{An:2018dwb}.

The influence from these aforementioned uncertainties on the BDT modeling is studied by comparing the BDT distribution bin by bin 
between the nominal and variation MC sample. 
The maximum variation value $\delta n = \frac{|n_{var}-n_{nom}|}{n_{nom}}$ in all BDT bins and systematic terms is applied on the signal yield as the uncertainty from BDT, except the bin with low statistics (bin content less than 5\% of total yield). 
The uncertainty from BDT itself is believed to be included in this envelope value. 
In three channels this term is 0.5\% to 0.7\%.

\section{Results}
\label{sec:results}
The number of expected signal events is extracted by a combined fit in three channels with the unbinned maximum likelihood fit method.
The likelihood function is built with the models in Sec.~\ref{sec:models} and the constraints coming from systematic uncertainties in Sec.~\ref{sec:syst}: 


\begin{equation}
\begin{aligned}
\mathcal{L}(\mu,\bm{\theta};(\myy, \text{BDT})) & = \prod_{c}\text{Pois}(n_c|N_c(\mu, \bm{\theta}))\cdot \\
         & \prod_{i}^{n}f_{c}((\myy, \text{BDT})^{i};\bm{\theta}) \cdot \prod_{j} G(\theta_j),
\end{aligned}
\end{equation}

in which: 
\begin{itemize}
	\item $\mu$ is the signal strength $\mu = \frac{N\ (\ee\to \ZH\to \ffbar\gamgam)} {N_{SM}\ (\ee\to \ZH\to \ffbar\gamgam)}$, which is the parameter of interest (POI) in the fit;
  \item $\bm{\theta}$ are nuisance parameters defined for systematic terms;
  \item $n_c$ is the observed event number in the channel $c$ from the data;
  \item $N_c(\mu, \bm{\theta})=\mu S_{SM, c}(\bm{\theta_{yield}}) + B_c$. $S_{SM, c}(\bm{\theta_{yield}})$ is the expected signal yield in the channel, including the relevant nuisance parameters. $B_c$ is the background yield;
  \item $f_{c}((\myy, \text{BDT})^{i};\bm{\theta})$ is the probability density function built with the signal and background model in Sec.~\ref{sec:models}:

\begin{equation}
\begin{small}
\begin{aligned}
& f_{c}((\myy, \text{BDT})^{i};\bm{\theta}) = \frac{1}{N_c}\times \\
& \left[ \mu S_{SM, c}(\bm{\theta_{yield}})f_{c, sig}((\myy,\text{BDT})^i;\bm{\theta}) + B_{c} f_{c, bkg}((\myy,\text{BDT})^i;\bm{\theta}) \right].
\end{aligned}
\end{small}
\end{equation}

  \item The signal yield $S_{SM,c}$, shape peak $\mu_{CB}$ and width $\sigma_{CB}$ are affected by the systematic uncertainties with a response function, so 

\begin{equation}
\begin{small}
\begin{aligned}
& S_{SM,c}(\bm{\theta_{yield}})=S_{SM,c}\prod_{j}e^{\theta_j \sqrt{\ln(1+\delta_j^2)}}, \\
& \mu_{CB}(\bm{\theta_{peak}}) = \mu_{CB}^{nom}\prod_{j}(1+\delta_j \theta_j), \\
& \sigma_{CB}(\bm{\theta_{width}}) = \sigma_{CB}^{nom}\prod_{j}e^{\theta_j \sqrt{\ln(1+\delta_j^2)}} 
\end{aligned}
\end{small}
\end{equation}

  \item $G(\theta_{j})$ is the unitary Gaussian constraint PDF for the nuisance parameter $j$, with mean 0 and width 1. 
\end{itemize}
In the fitting the signal model parameters are fixed to the value in fitting the signal MC.
The background yields, model parameters and all nuisance parameters are floated as mentioned in Sec.~\ref{sec:syst}.

\begin{figure}[hbtp]
  \centering
  \subfigure[\qqyy \myy model]{ \includegraphics[width= 0.4\linewidth]{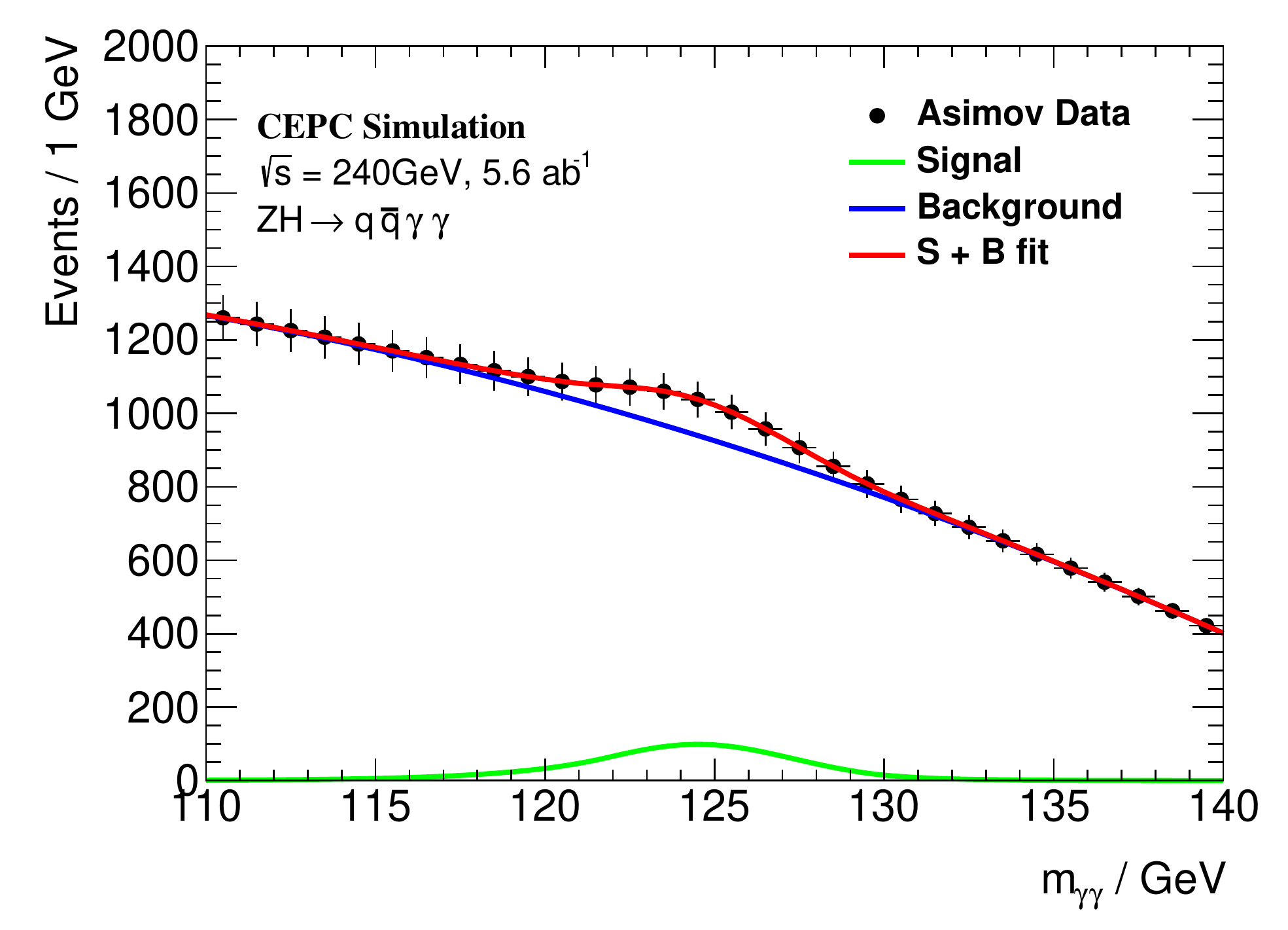} }
  \subfigure[\qqyy BDT model]{ \includegraphics[width= 0.4\linewidth]{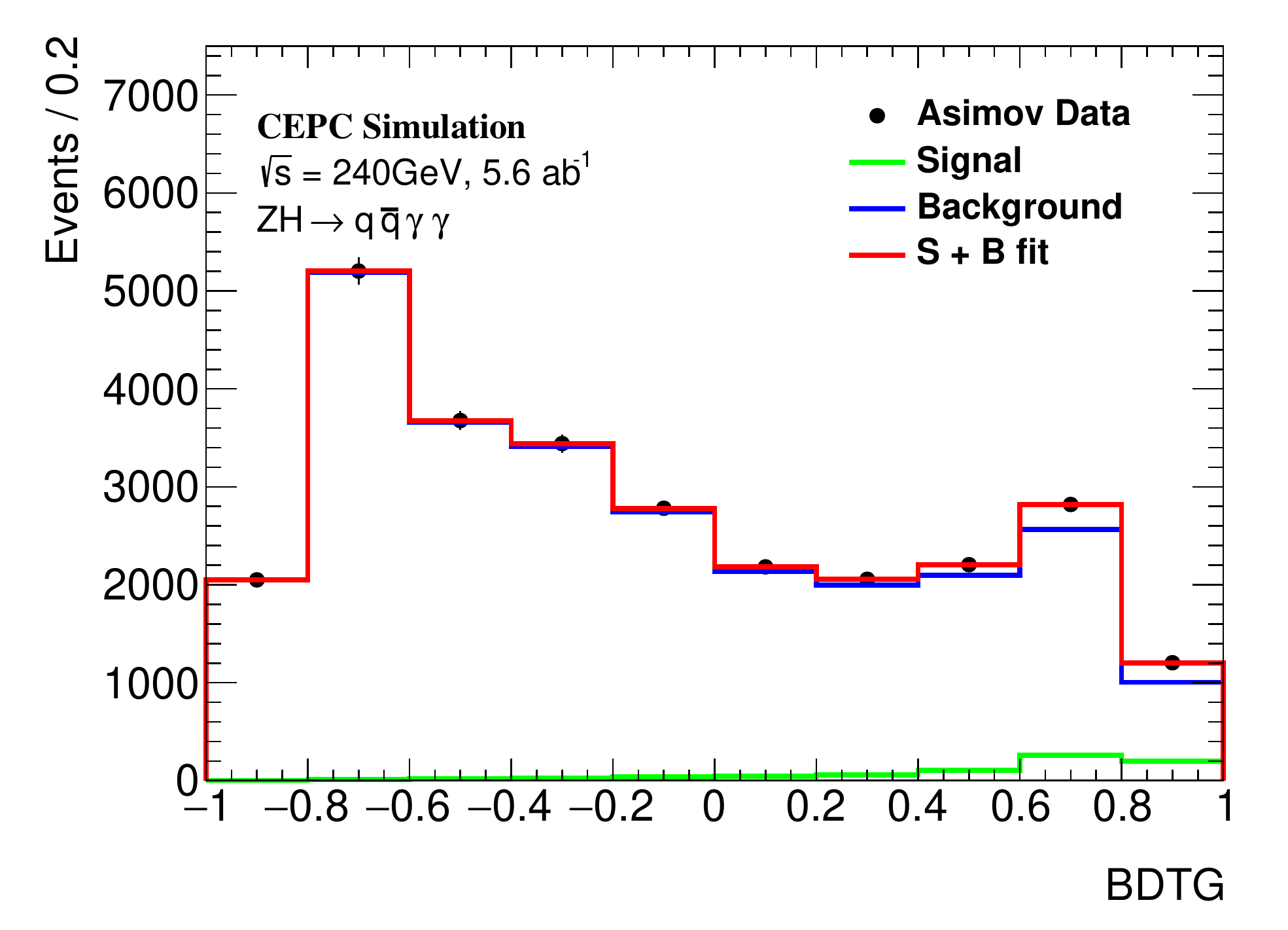} } \\
  \subfigure[\mmyy \myy model]{ \includegraphics[width= 0.4\linewidth]{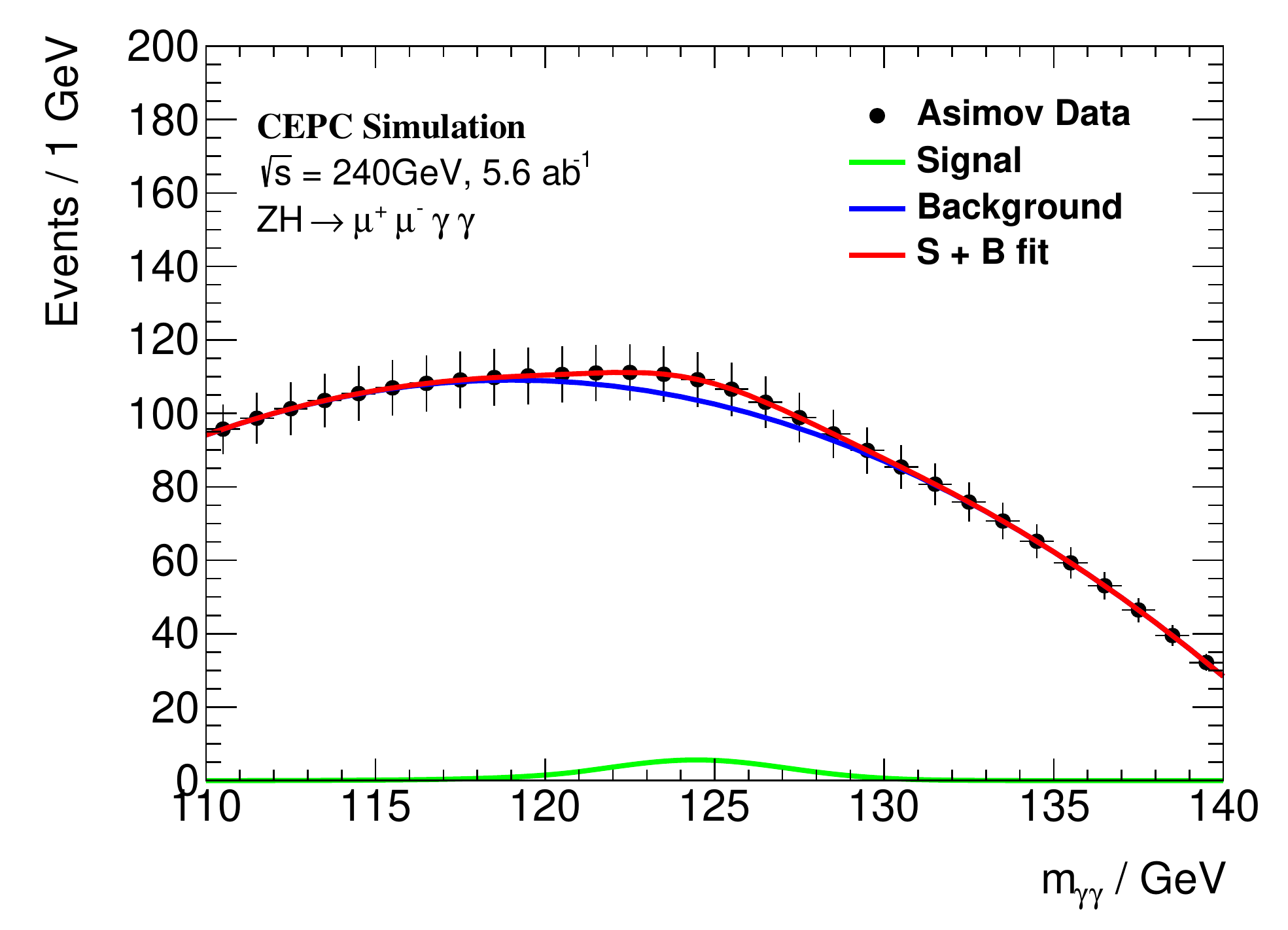} }
  \subfigure[\mmyy BDT model]{ \includegraphics[width= 0.4\linewidth]{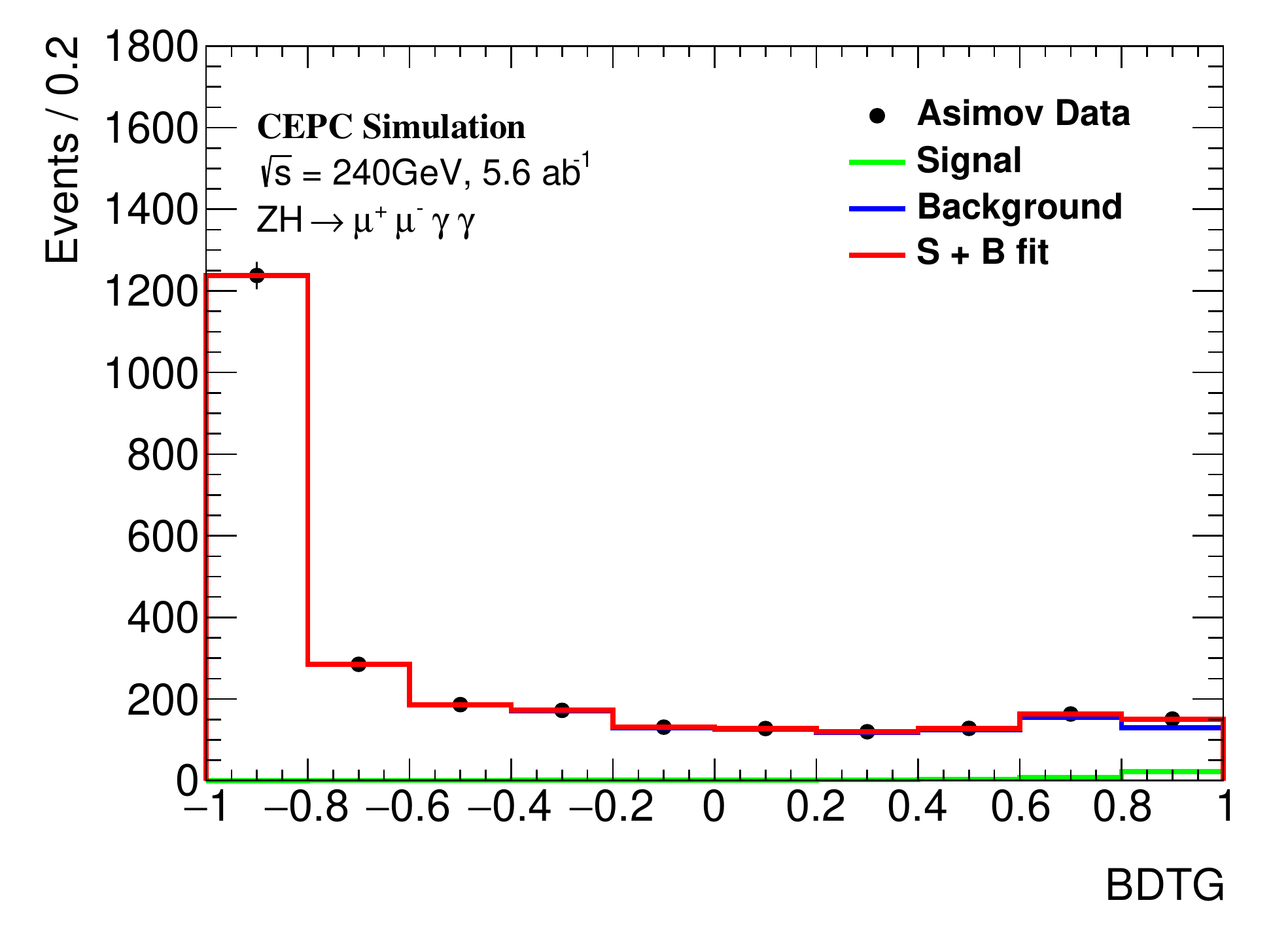} } \\
  \subfigure[\nnyy \myy model]{ \includegraphics[width= 0.4\linewidth]{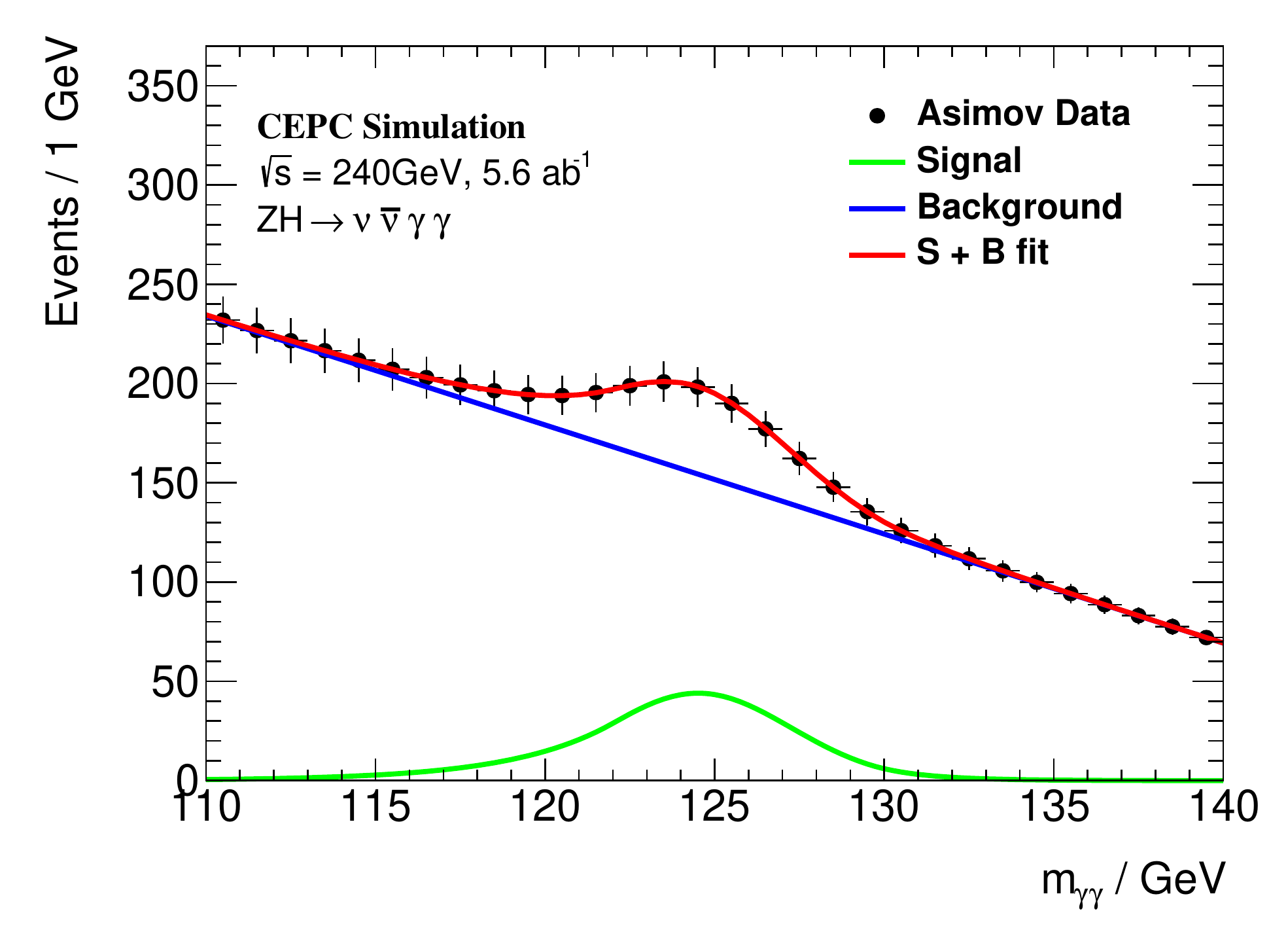} } 
  \subfigure[\nnyy BDT model]{ \includegraphics[width= 0.4\linewidth]{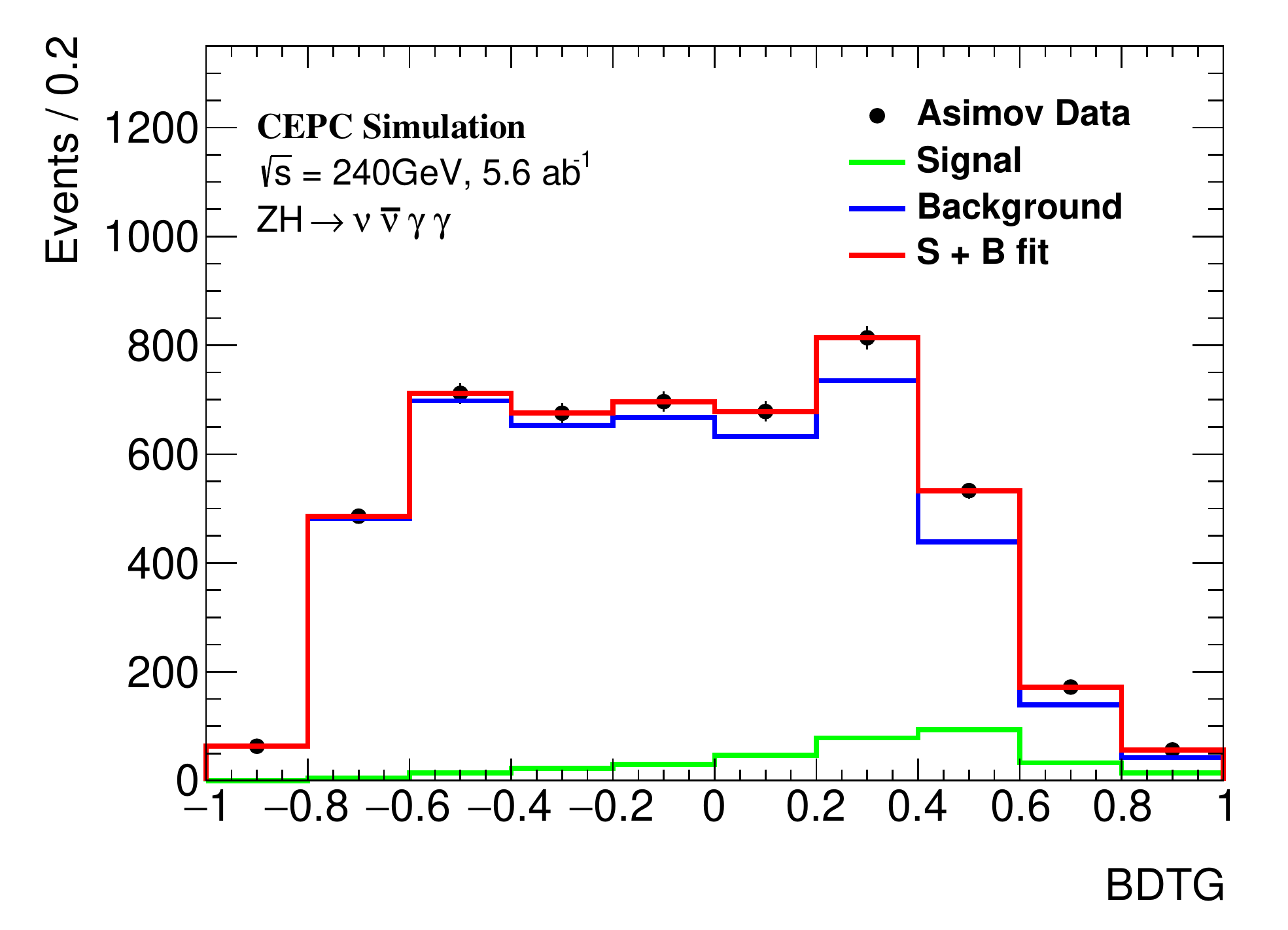} } \\
\caption{The Combined fit to the Asimov data in 3 channels. }
\label{fig:AsiFit}
\end{figure}

In order to mimic the real data and avoid the statistical fluctuations of MC samples, 
a set of Asimov data~\cite{Cowan_2011} are generated from the signal + background models 
and are simultaneously fitted to obtain the expected precision and significance. 
Figure~\ref{fig:AsiFit} shows the \myy and BDTG distributions of the Asimov data and the models in 3 channels. 
A final precision of 7.7\% (stat.)$\pm$ 2.1\% (syst.) for the $\sigma\times\Br$ measurement can be reached in $\Higgs\to\gamgam$ channel in the CEPC with 5.6 \iab data.
With the 20 \iab data of the updated CEPC operation period, the expected precision is 4.0\% (stat.)$\pm$ 2.1\% (syst.).
Table ~\ref{tab:syst} lists the contributions from each systematic terms. 
The contribution from background modeling is decoupled from fixing and floating the background parameters in the fitting, 
and it is included into the statistical precision.
Combined results are summarized in Table~\ref{tab:result}.
With our preliminary assumption this measurement is still statistics dominant in CEPC.

\begin{table}[]
\begin{tabular}{|l|c|c|c|}
\hline
                    &    \qqyy   &   \mmyy   &   \nnyy    \\ \hline
Theo 0.5\%          &   0.005    &     -     &     -      \\ \hline
Lumi 0.1\%          &   0.001    &   0.001   &   0.001    \\ \hline
photon eff 1\%      &   0.019    &   0.020   &   0.020    \\ \hline
PES 0.05\%          &   0.001    &   <0.001  &   0.001    \\ \hline
PER 0.05\%          &   <0.001   &   <0.001  &   <0.001   \\ \hline
mH 5.9 MeV          &   <0.001   &   <0.001  &   <0.001   \\ \hline
BDT                 &   0.006    &   0.006   &   0.007    \\ \hline
Bkg. modeling       &   0.029    &   0.062   &   0.006    \\ \hline 
\end{tabular}
\caption{The decoupled contributions from considered systematic uncertainties of $(\sigma\times\Br) \slash (\sigma\times\Br)_{SM}$ measurement in 3 channels. The 0.5\% theoretical uncertainty is only considered in \qqyy channel.}
\label{tab:syst}
\end{table}

\begin{table}[h]
\centering
\begin{tabular}{|l|cc|cc|}
\hline
\multirow{2}{*}{} & \multicolumn{2}{c|}{5.6 \iab}   &  \multicolumn{2}{c|}{20 \iab}     \\ \cline{2-5} 
                  & \multicolumn{1}{c|}{ $\frac{\Delta_{tot}}{(\sigma\times\Br)_{SM}}$ }  &  $\frac{\Delta_{stat}}{(\sigma\times\Br)_{SM}}$      & \multicolumn{1}{c|}{ $\frac{\Delta_{tot}}{(\sigma\times\Br)_{SM}}$ }      & $\frac{\Delta_{stat}}{(\sigma\times\Br)_{SM}}$   \\ \hline
\qqyy             & \multicolumn{1}{c|}{0.101}    &  0.098    & \multicolumn{1}{c|}{0.056}    &  0.052      \\ \hline
\mmyy             & \multicolumn{1}{c|}{0.373}    &  0.371    & \multicolumn{1}{c|}{0.202}    &  0.200      \\ \hline
\nnyy             & \multicolumn{1}{c|}{0.130}    &  0.127    & \multicolumn{1}{c|}{0.071}    &  0.067      \\ \hline
Combined          & \multicolumn{1}{c|}{0.079}    &  0.077    & \multicolumn{1}{c|}{0.046}    &  0.040      \\ \hline
\end{tabular}
\caption{Expected precisions on $\sigma(\ZH)\times\Br(\Higgs\to\gamgam)$ from Asimov data fit in 3 channels and the combination. Results in 20 \iab are obtained by re-fitting the workspace with the scaled signal and background yields. The statistical precision includes the contribution from background modeling. }
\label{tab:result}
\end{table}

\section{ Dependence of $Br(H\to \gamma\gamma)$ measurement precision on ECAL energy resolution}
\label{sec:EcalRes}

While fitting the \myy shape, the width of the signal peak is a direct connection between the measurement precision in $H\to \gamma\gamma$ channel and the ECAL resolution. 
Currently a new detector design for CEPC is under development~\cite{CEPCPhysicsStudyGroup:2022uwl, instruments6030040, Liu_2020_CrystalCalo}, in which the present Si-W sampling ECAL will be replaced by a homogeneous crystal ECAL. 
This new ECAL is expected to have the energy resolution of $\sigma_{E}/E = 3\%/\sqrt{E}$, which is almost five times higher than the sampling Si-W ECAL $\sigma_{E}/E = 16\%/\sqrt{E} \oplus 1\% $~\cite{CEPC_CDR_V2}.
This can benefit the photon detection and the neutral meson ($\pi^{0}$) reconstruction, and further contribute to the Higgs study in $H\to \gamma\gamma$ channel and the flavor physics in $\pi^{0}\to \gamma\gamma$ final state, e.g. $B^0_{(s)} \to \pi^0 \pi^0 $~\cite{Wang:2022nrm}.
The jet energy resolution may not be improved much from this ECAL, since the detector granularity is the dominant factor in the PFA-based jet reconstruction. 

\begin{figure}[hbtp]
\centering
  \includegraphics[width=0.9\linewidth]{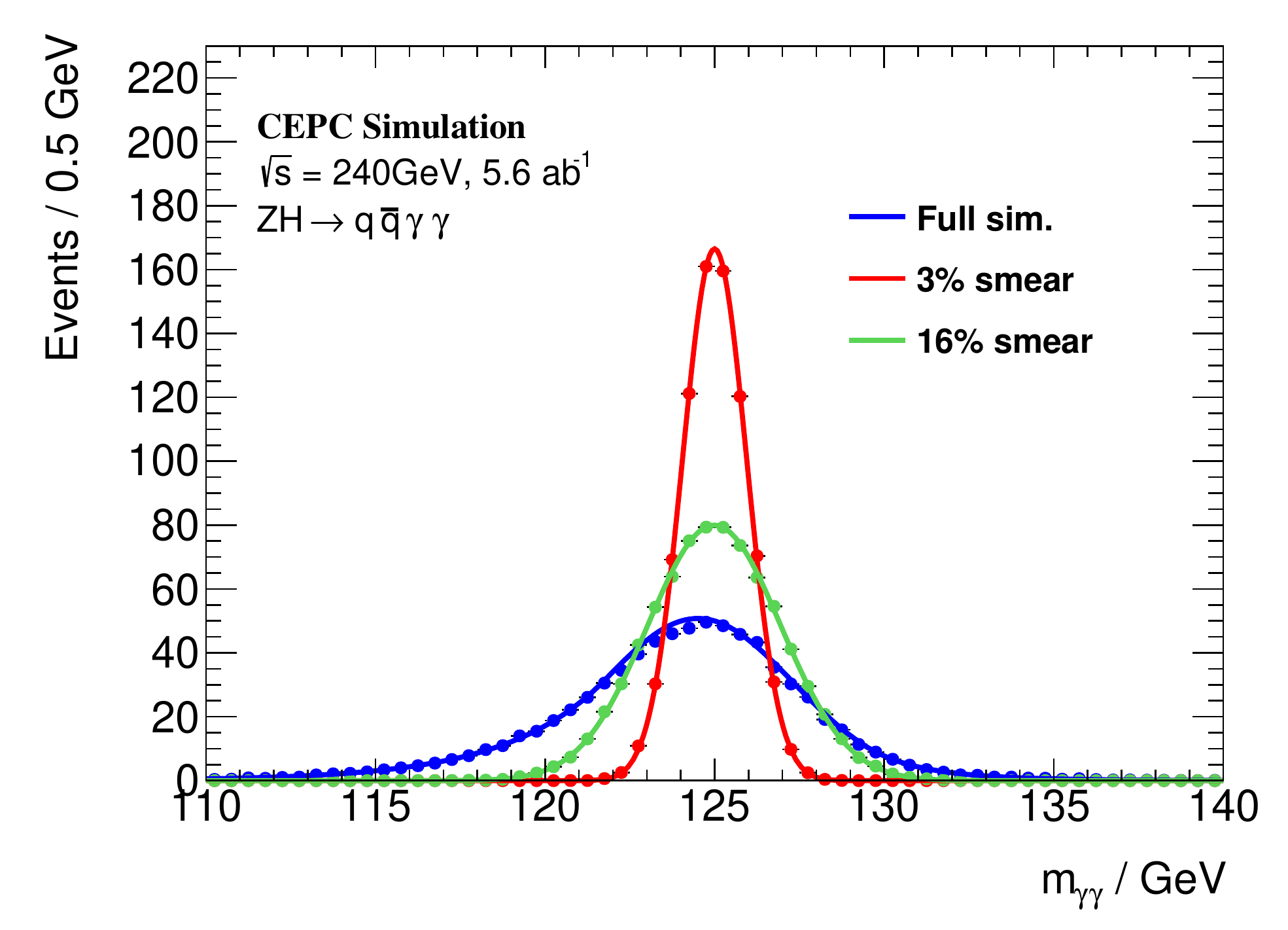}
  \caption{The signal shape for the full simulated $\Higgs\to \gamma\gamma$ sample (blue) and for two samples with smeared photon energy (3\% in red and 16\% in green). The fitted signal width are 2.81 GeV, 0.94 GeV and 1.96 GeV respectively. }
  \label{fig:SigCompare}
\end{figure}

\begin{figure}[hbtp]
\centering
  \includegraphics[width=0.9\linewidth]{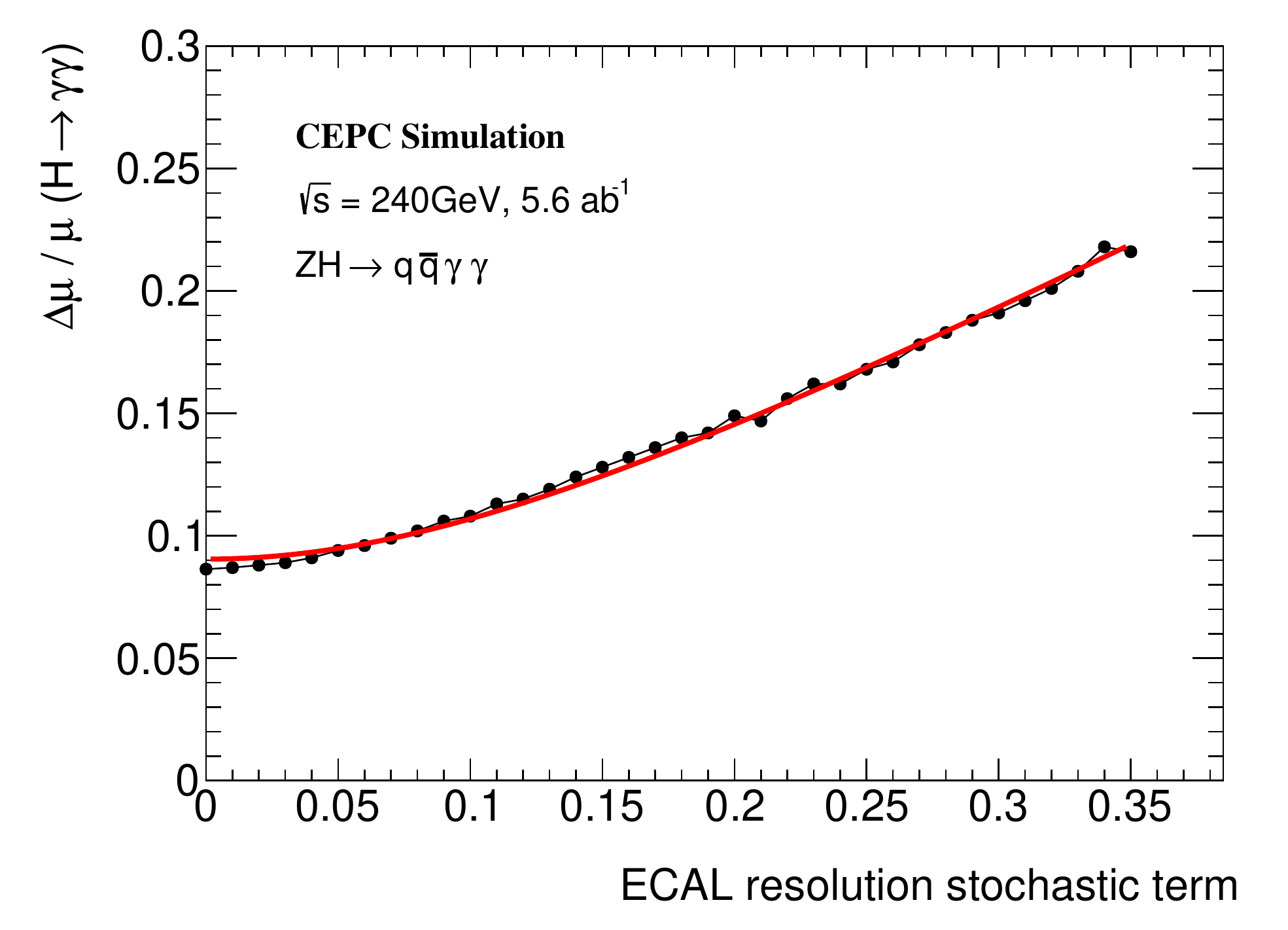}
  \caption{The signal strength measurement precision in $\ZH \to \qqyy$ channel as a function of the stochastic term in ECAL resolution from a fast analysis. The points are fitted with Eq.~\ref{eq:Bvsdeltamu}. }
  \label{fig:Bvsdeltamu}
\end{figure}

\begin{figure}[hbtp]
\centering
  \includegraphics[width=0.9\linewidth]{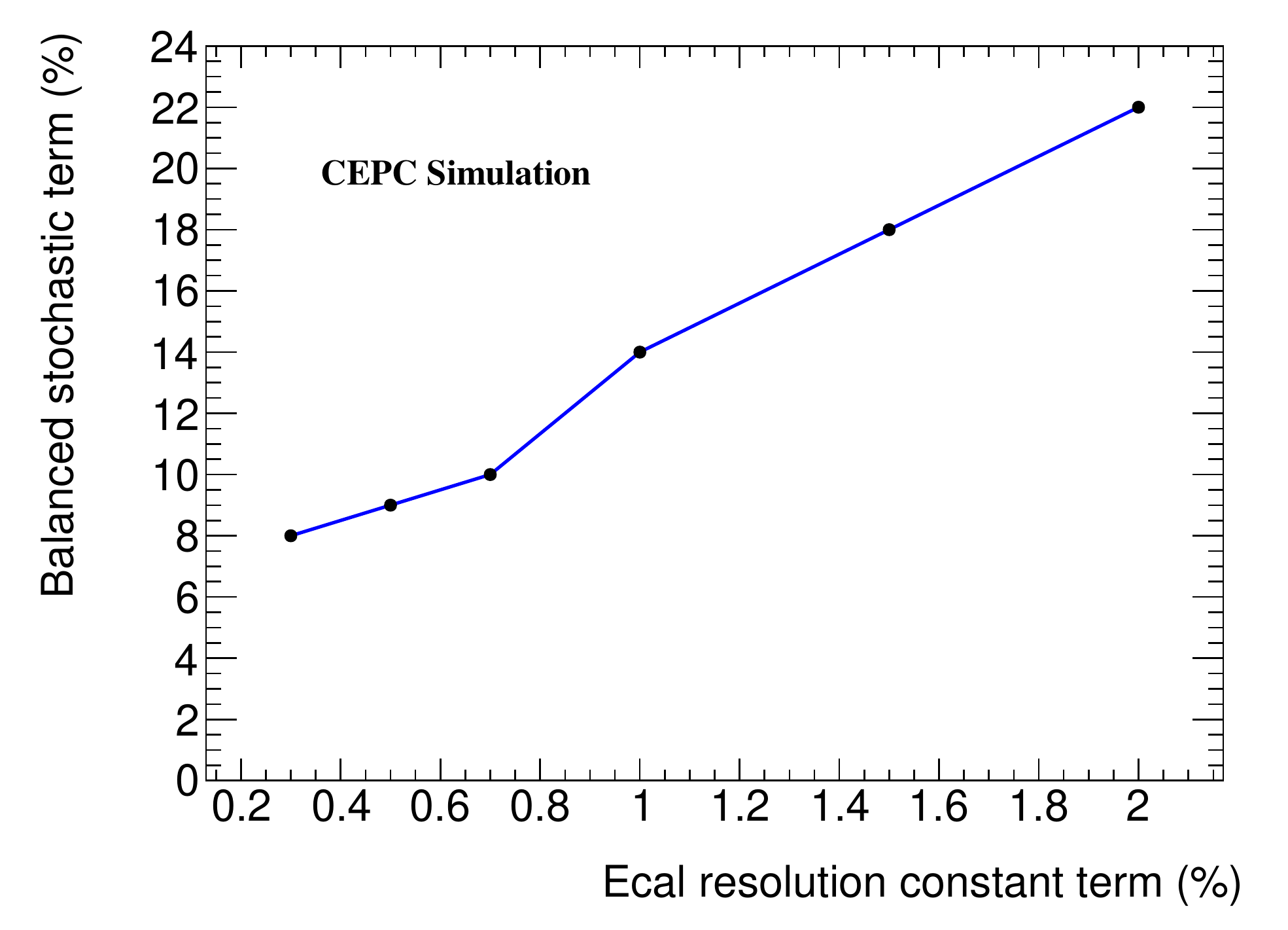}
  \caption{The balanced ECAL stochastic resolution points with different configurations of the constant term. }
  \label{fig:ABrelation}
\end{figure}

We performed a rough estimation in the \qqyy channel within the strategy of this work to study ECAL resolution impact on the $H\to \gamma\gamma$ measurement. 
In the estimation the selected photon is replaced by the truth photon with a smearing in its energy. Normally the ECAL energy is approximated as: 
\begin{equation}
\frac{\sigma_{E}}{E} = A \oplus \frac{B}{\sqrt{E}} \oplus \frac{C}{E},
\end{equation}
where $A$ stands for the constant term like energy leakage, readout threshold, etc. 
$B$ represents the stochastic term from photoelectron statistics and depends on the sensitive material. 
$C$ comes from the electronic noises. 
Presently the noise term $C$ is expected to be 0, and the constant term $A$ is expected to be at the level of 1\%. 
The photon energy is smeared with the stochastic term $B$ varying from 1\% to 35\%. 
Figure~\ref{fig:SigCompare} shows a comparison between the \myy shape from the full simulation and 2 smearing points 3\% and 16\%. 
The jet performance is kept consistent with the baseline Si-W sampling ECAL, assuming there is no impact from the new detector.  
The same selection criteria are applied as in Sec.~\ref{sec:selection}, while the BDT is not employed in this simplified study to focus on the photon detection only, which is expected to have a roughly 30\% decrease comparing with the result in Sec.~\ref{sec:results}. 
A Gaussian function is used to describe the signal model from the energy smearing. 
The 2-dimension model is replaced with a 1-dimension \myy model, 
and a similar unbinned maximum likelihood fit is performed to extract the signal strength precision $\delta\mu/\mu$, without systematic uncertainties.
Considering the \myy and the BDT are independent, this simplification is expected to have very little impact on the relative improvement. 
Figure~\ref{fig:Bvsdeltamu} shows the relationship between energy resolution $B$ and the fitted precision $\delta\mu/\mu$. 
These points can be fitted with the following function: 

\begin{equation}
\frac{\delta \mu}{\mu} = p_{0} \oplus (p_{1}\times B), 
\label{eq:Bvsdeltamu}
\end{equation}
where $p_{0}$ and $p_{1}\times B$ represent the contributions from constant term and stochastic term respectively. 
From this relation the homogeneous ECAL is able to bring a 28\% improvement in the statistic precision of signal strength measurement.
Also a "critical point" can be defined: the two components in resolution have the same contribution to $\delta\mu/\mu$, 
i.e. $p_{0}=p_{1}\times B$.
When the constant term $A$ is fixed to 1\%, the critical point for $B$, within this definition, is 14\%. 
This indicates the constant term in resolution would become the dominant contribution at new ECAL design point with $B$ = 3\%. 
A scanning for a series of constant terms and the corresponding balanced stochastic terms is shown in Figure~\ref{fig:ABrelation}.

\section{Conclusion}
\label{sec:conclusion}

This paper presents the expected precision for the measurement of the cross section times branching ratio in the CEPC via $\ZH\to\qqyy$, $\ZH\to\mmyy$, $\ZH\to\nnyy$ channels. 
The physics events are reconstructed with the CEPC-v4 detector simulation, and selected by a set of criteria. 
A BDTG is developed for further signal/background separation, and is used along with \myy as discriminating variables in the maximum likelihood fit when extracting the signal strength. 
We build a preliminary framework for the systematic uncertainty analysis in CEPC with the nuisance parameters, 
and take several leading terms into account.
With the scheduled integrated luminosity of 5.6 \iab a precision of 7.9\% (7.7\% stat.) is expected to be achieved at the CEPC,
With 20 \iab data this precision can be 4.6\% (4.0\% stat.). 
More mature results await further development of this framework and better knowledge of systematic terms in the CEPC. 
Meanwhile the ECAL performance is studied by smearing photon energy resolution in \qqyy channel. 
A direct relationship between the ECAL resolution and the $\sigma\times\Br$ precision is foreseen.

\section{Acknowledgments}
The authors would like to thank the CEPC software group for the technical supports of simulation, reconstruction packages, 
as well as the CEPC physics group for the valuable discussions. 
This study is supported by the IHEP innovative project on sciences and technologies under Project No. E2545AU210.



\bibliographystyle{elsarticle-num-names}

\bibliography{CEPC_HGam_manuscript_v5}

\clearpage
\appendix
\section{Appendix}
\label{sec:App}

\begin{figure}[!h]
  \centering
  \subfigure[$pT_{\gamma 1}$]           { \includegraphics[width= 0.40\linewidth]{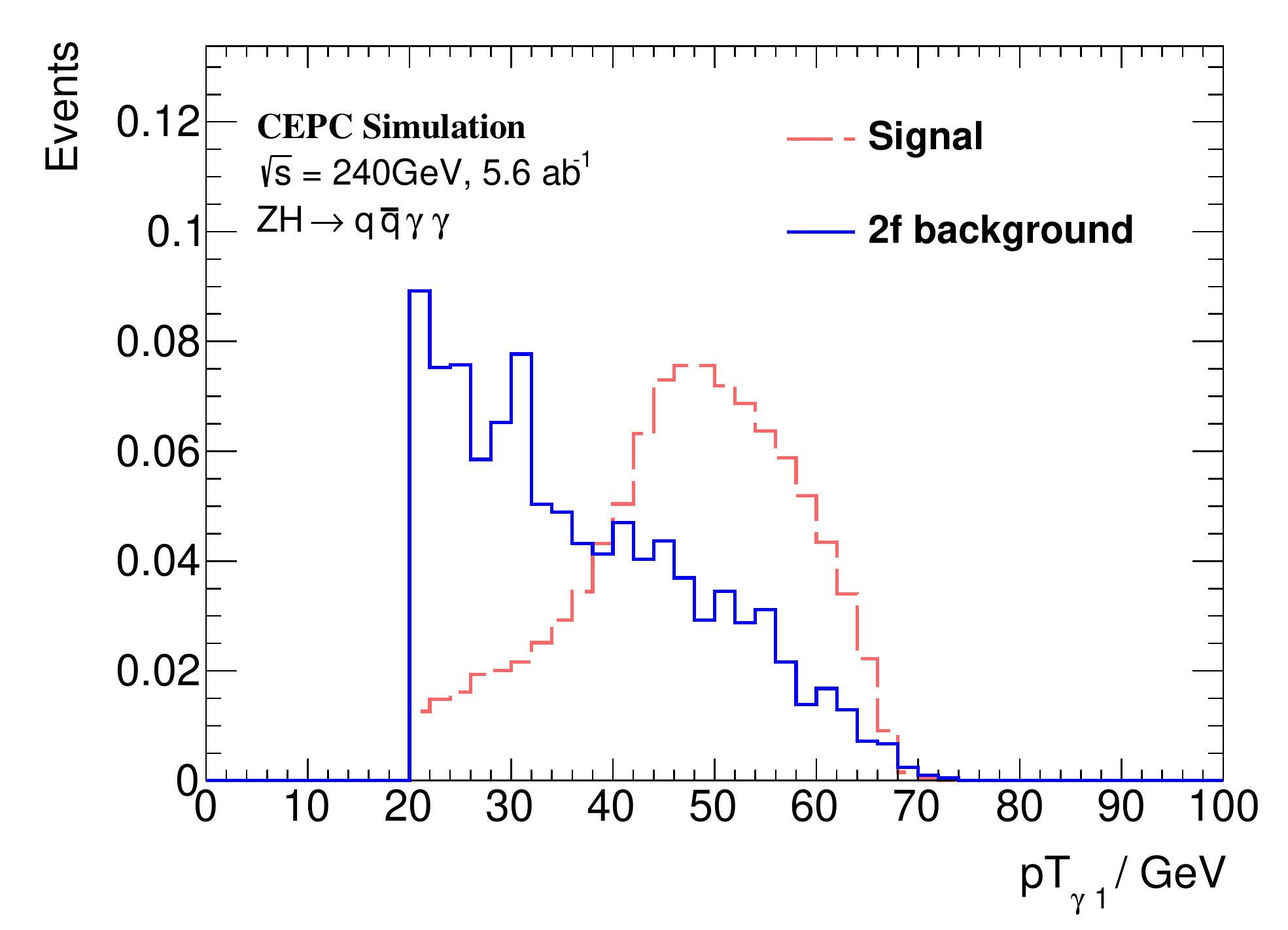} }
  \subfigure[$\cos\theta _{\gamma 2}$]   { \includegraphics[width= 0.40\linewidth]{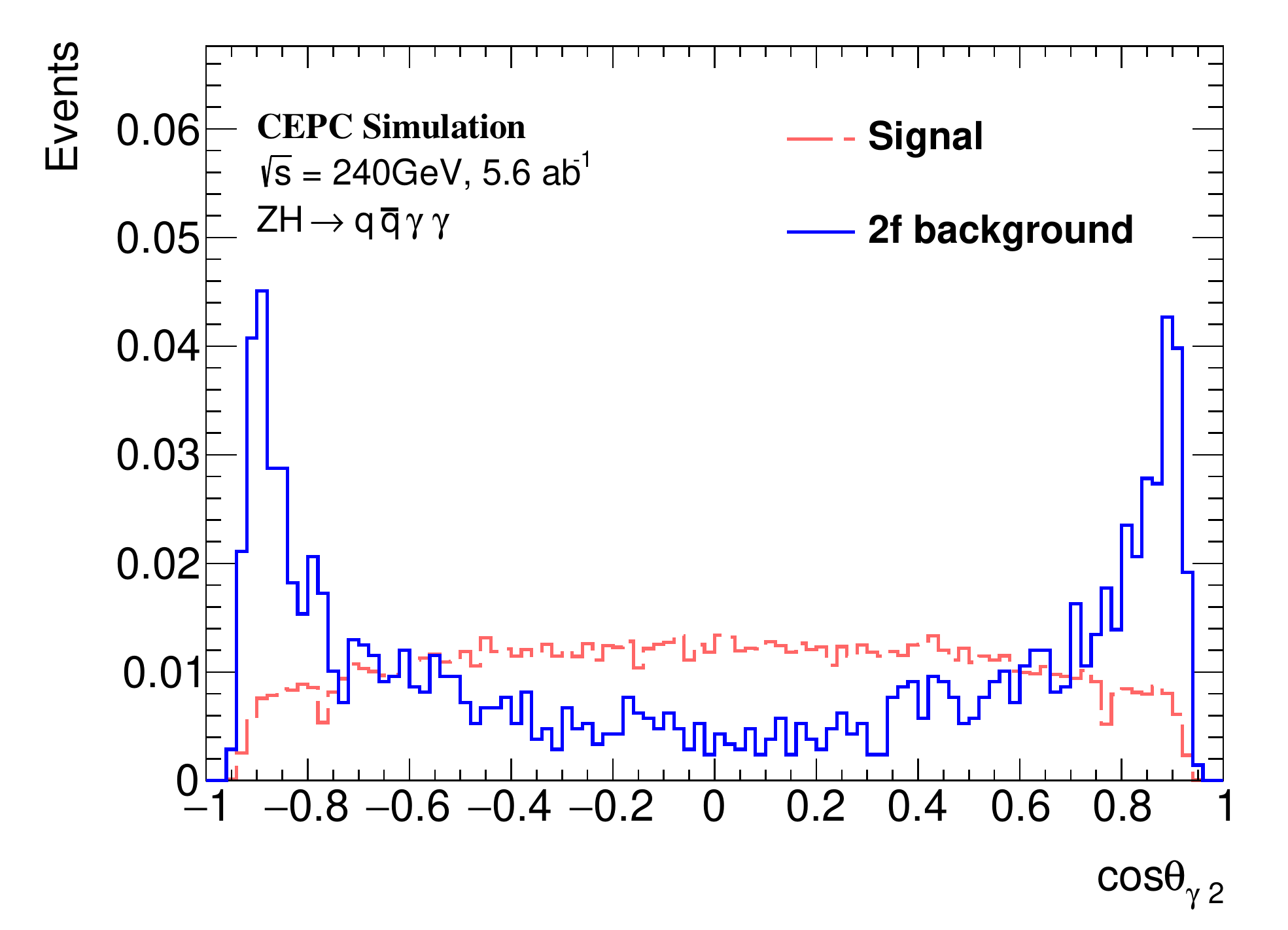} } \\
  \subfigure[$\Delta\Phi_{\gamgam}$]    { \includegraphics[width= 0.40\linewidth]{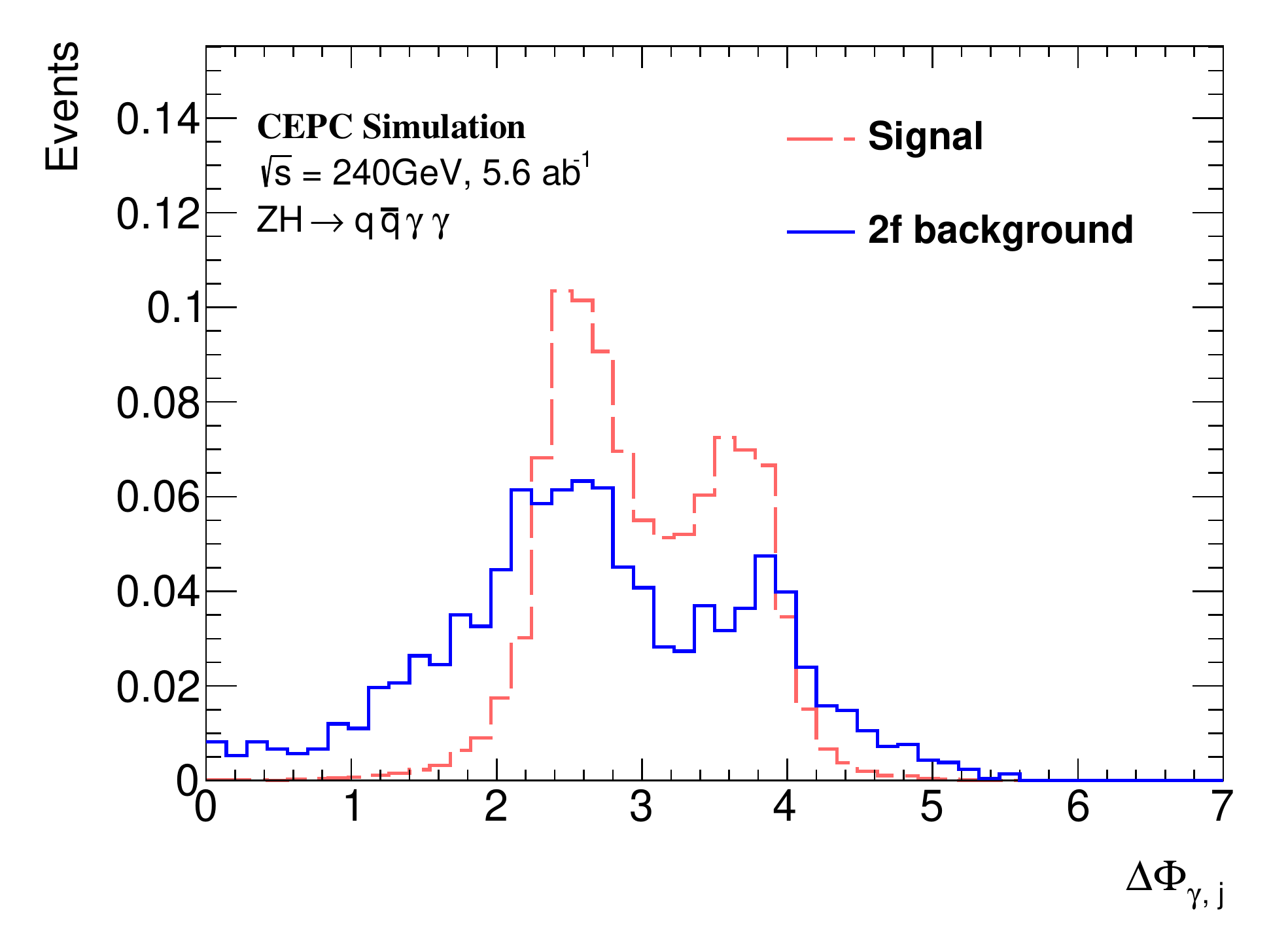} }
  \subfigure[$min\Delta R_{\gamma, j}$] { \includegraphics[width= 0.40\linewidth]{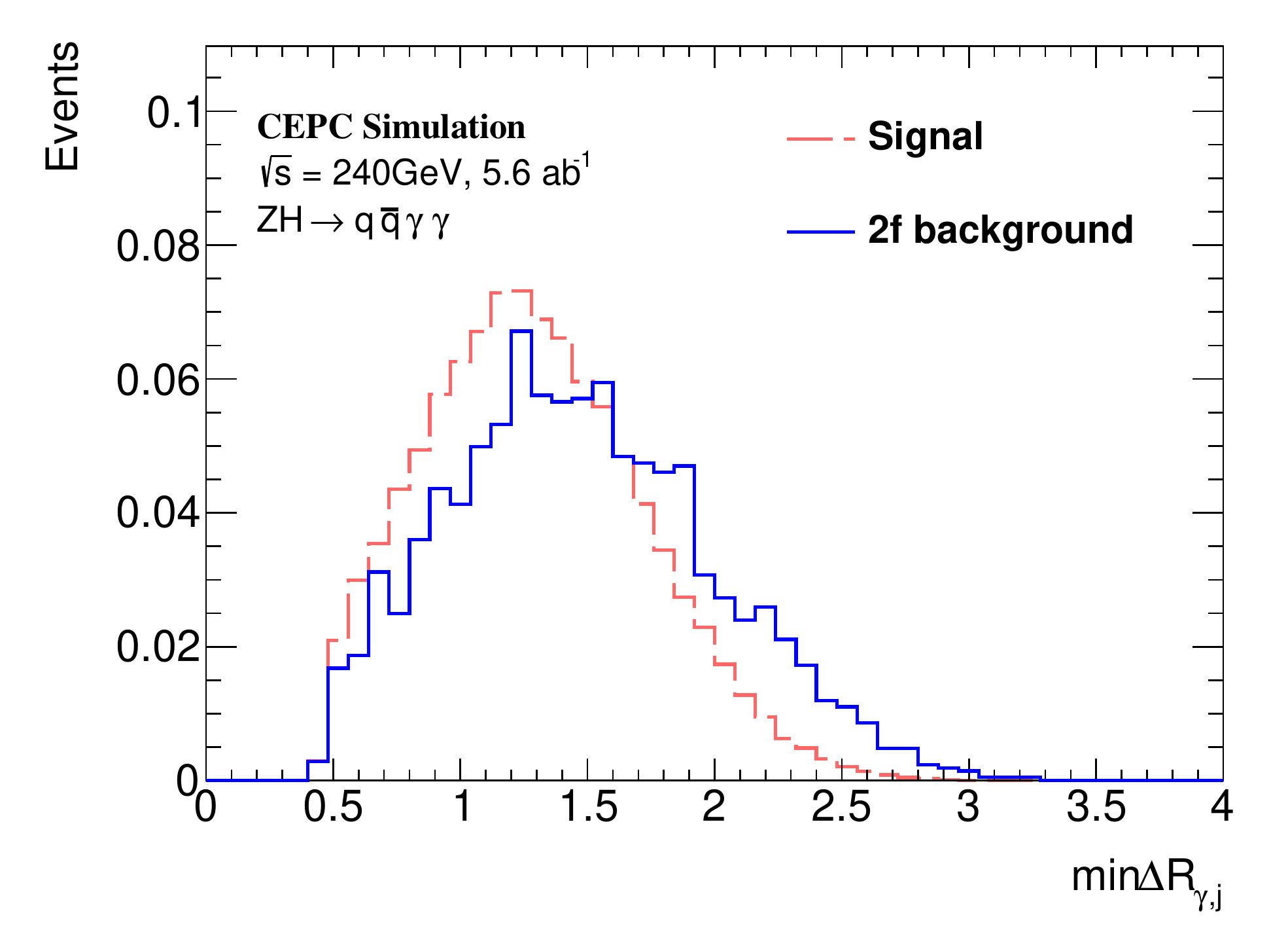} } \\
  \subfigure[$E_{j1}$]                  { \includegraphics[width= 0.40\linewidth]{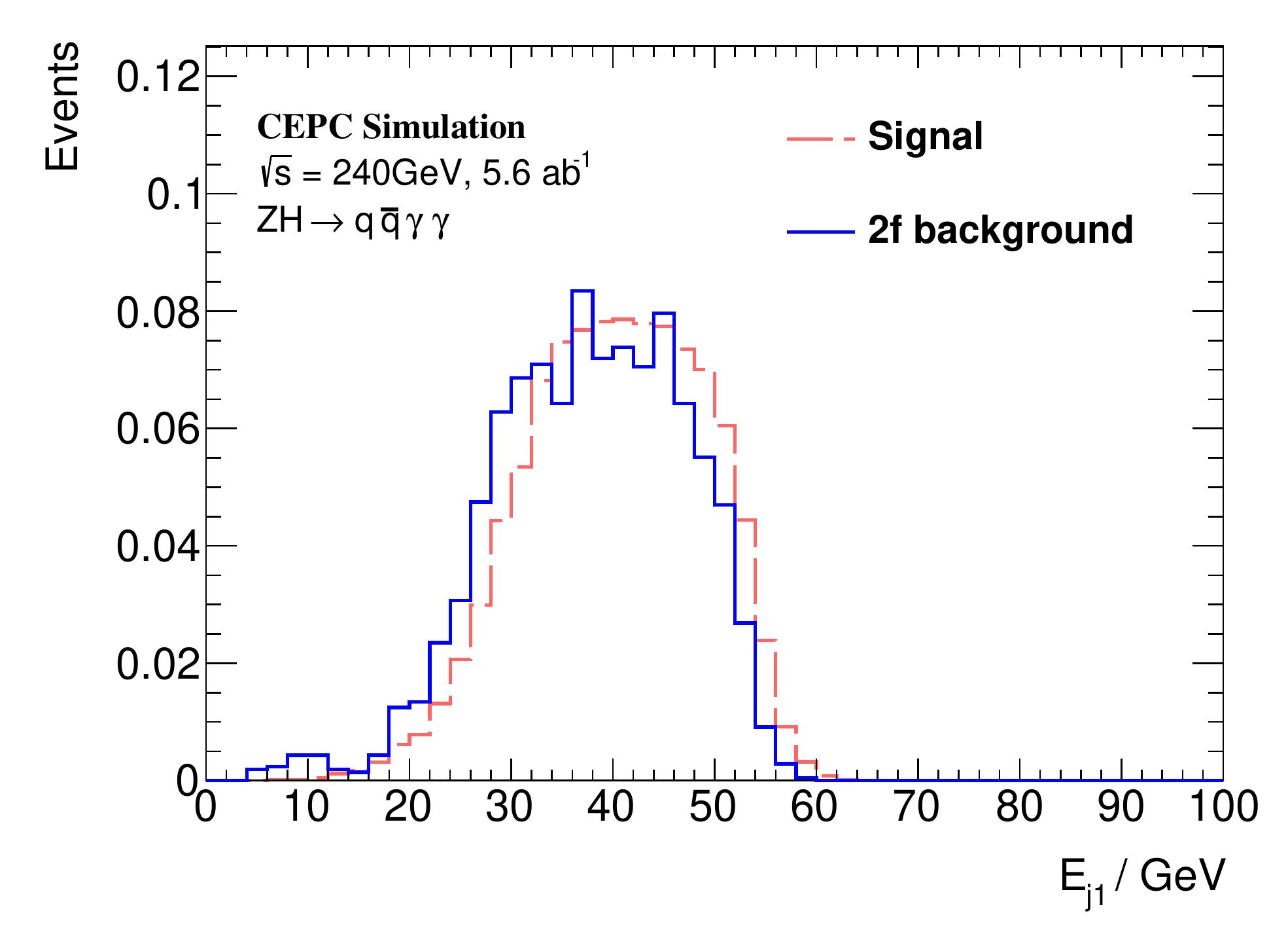} } 
  \subfigure[$\Delta\Phi_{\gamgam, jj}$]{ \includegraphics[width= 0.40\linewidth]{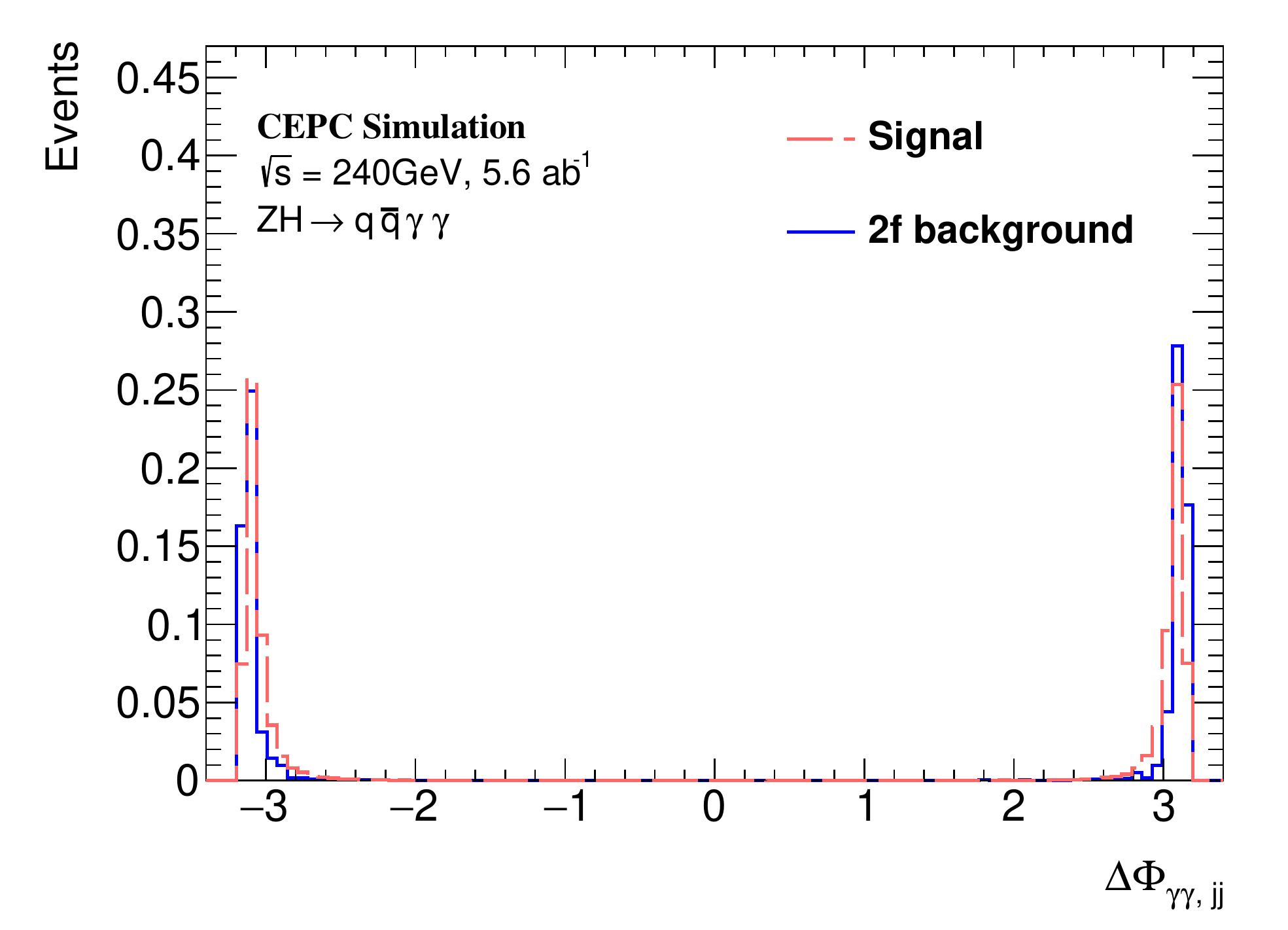} } \\
  \subfigure[$pT_{j2}$]                 { \includegraphics[width= 0.40\linewidth]{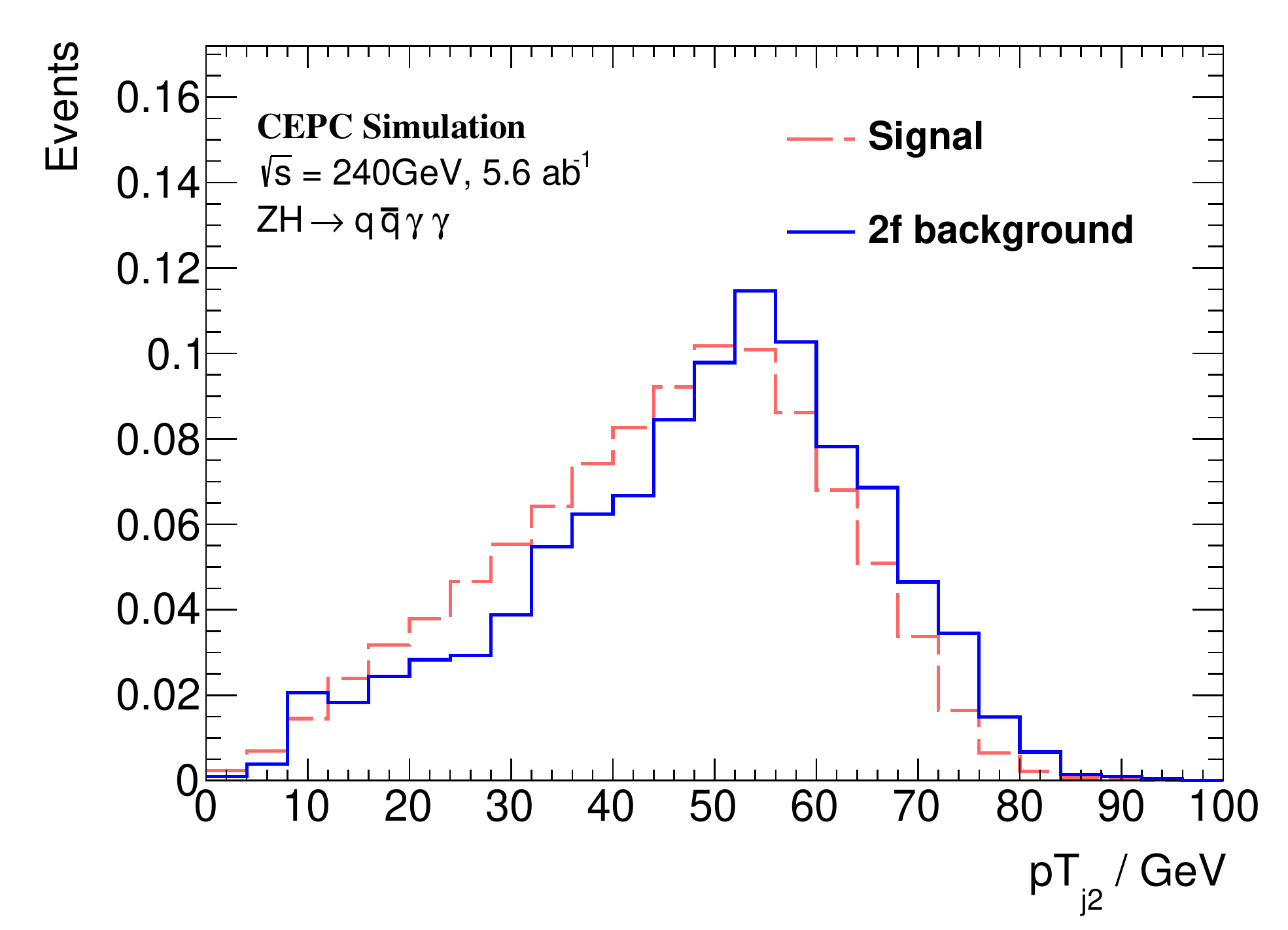} }
  \subfigure[$\cos\theta_{j1}$]          { \includegraphics[width= 0.40\linewidth]{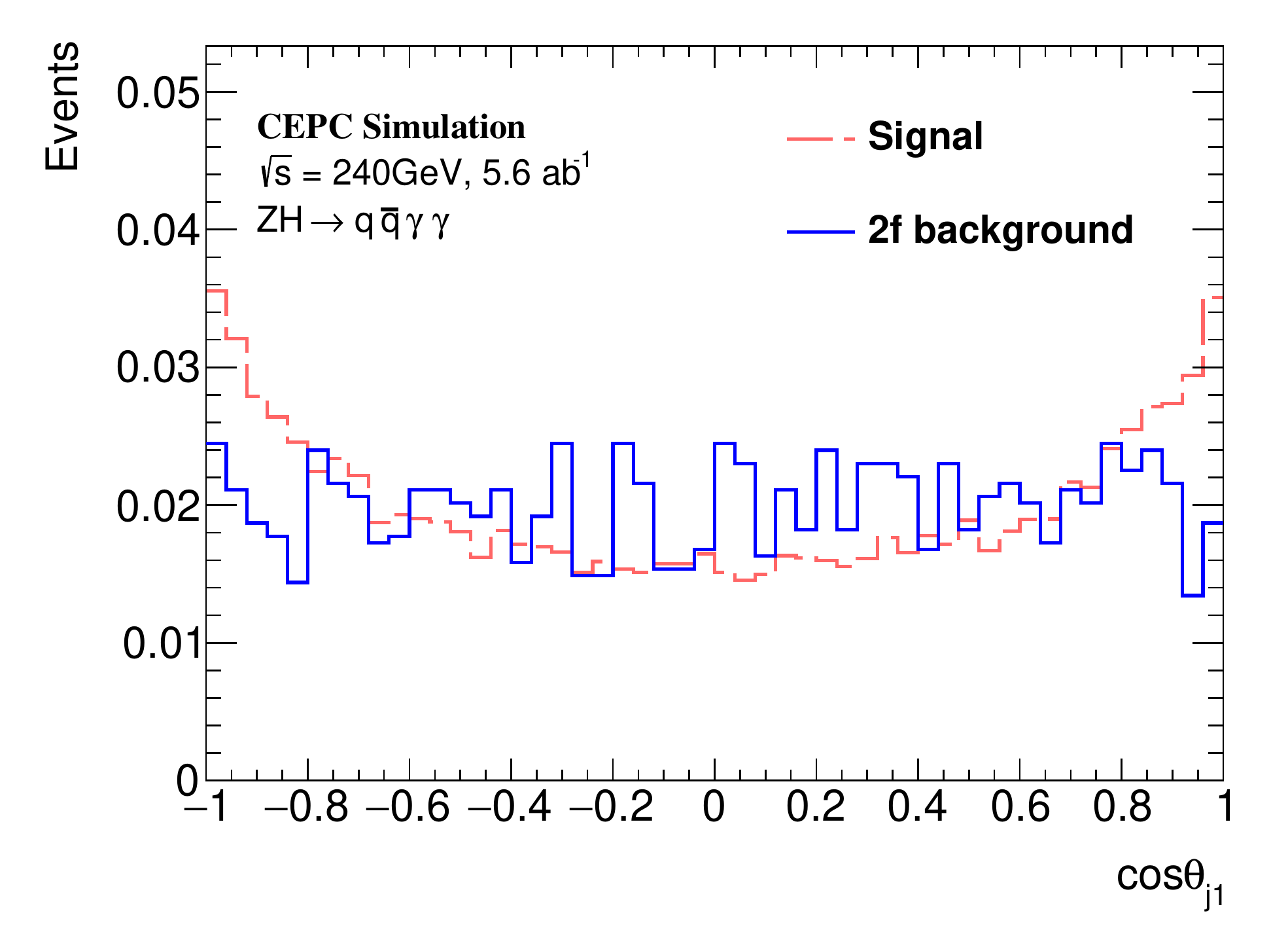} } \\
  \subfigure[$\cos\theta_{\gamgam, jj}$] { \includegraphics[width= 0.40\linewidth]{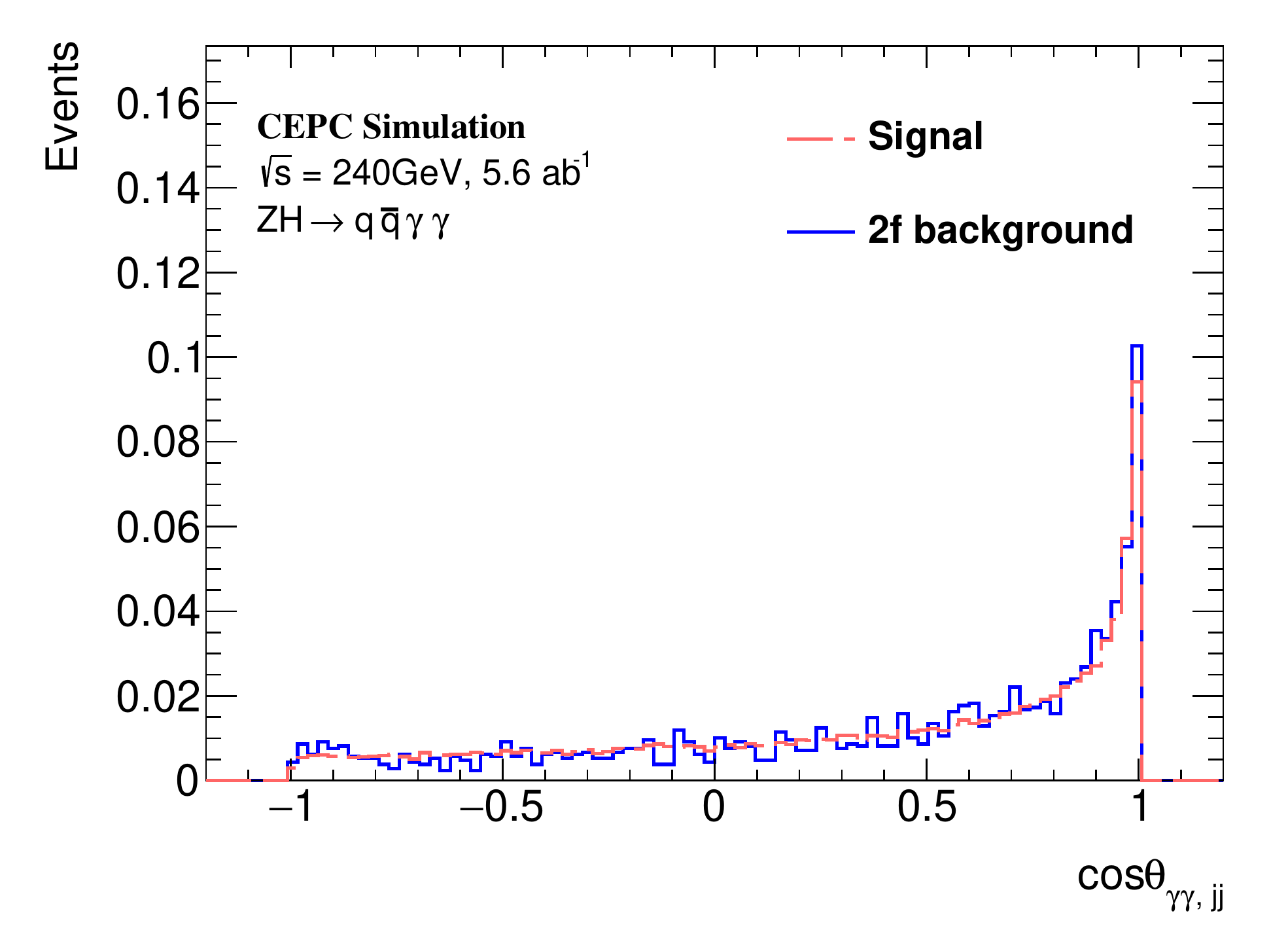} }
  \subfigure[$\cos\theta_{\gamma 1, j1}$]{ \includegraphics[width= 0.40\linewidth]{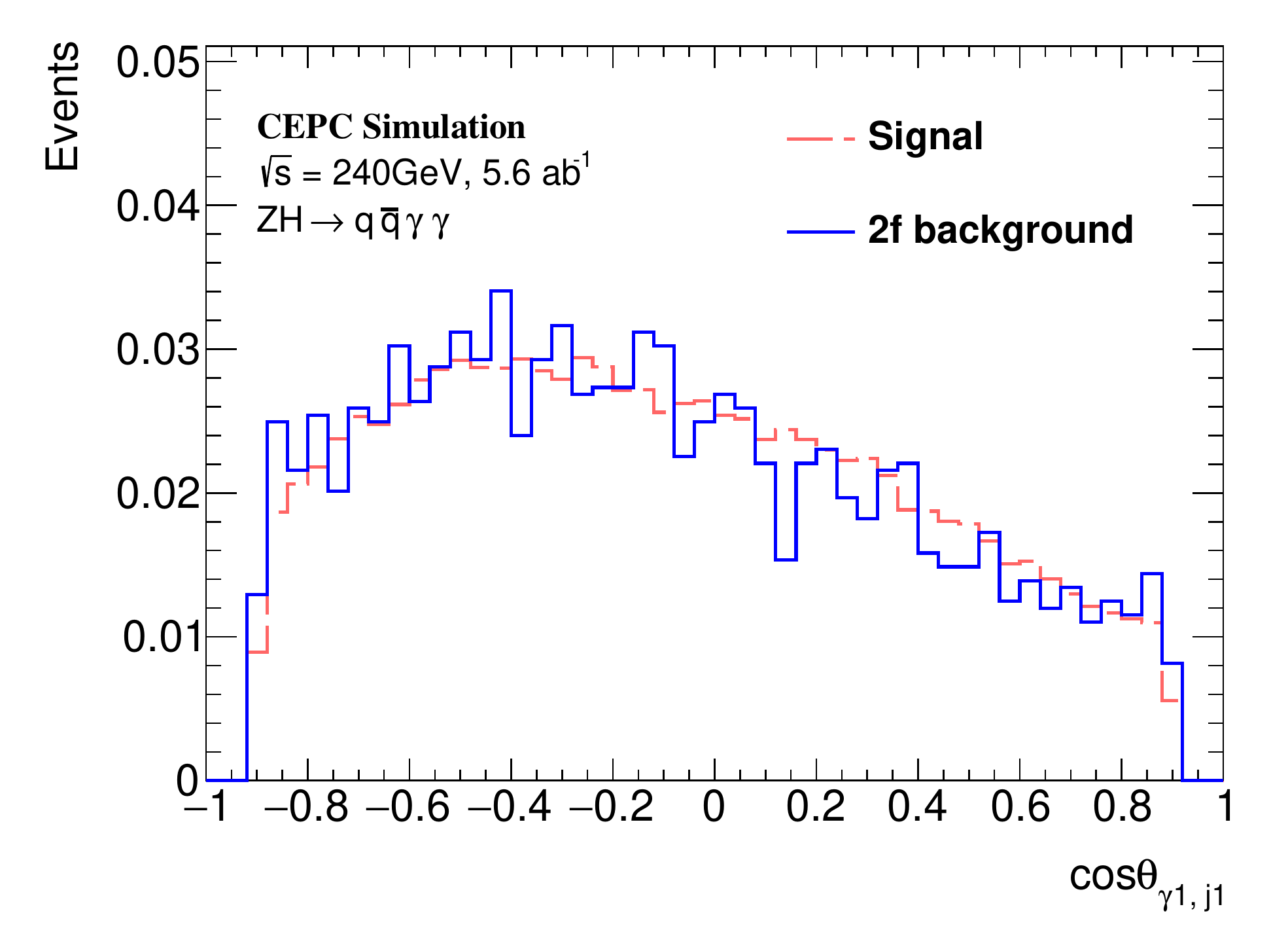} } \\
\caption{Training variables in \qqyy channel. The signal and background yields are normalized. }
\label{fig:InputVar_qq}
\end{figure}

\begin{figure}[h!]
  \centering
  \subfigure[]{ \includegraphics[width= 0.45\linewidth]{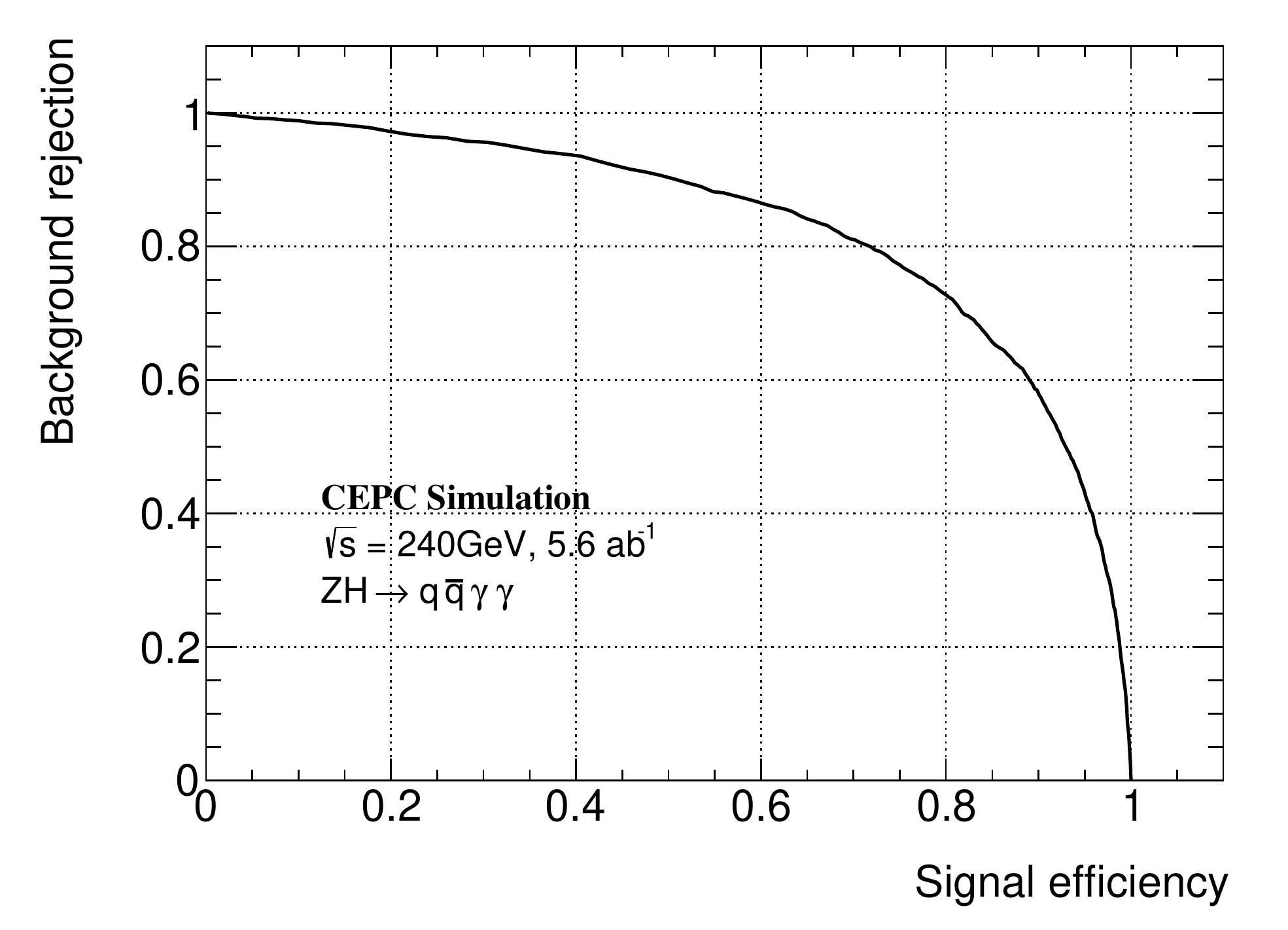} }
  \subfigure[]{ \includegraphics[width= 0.45\linewidth]{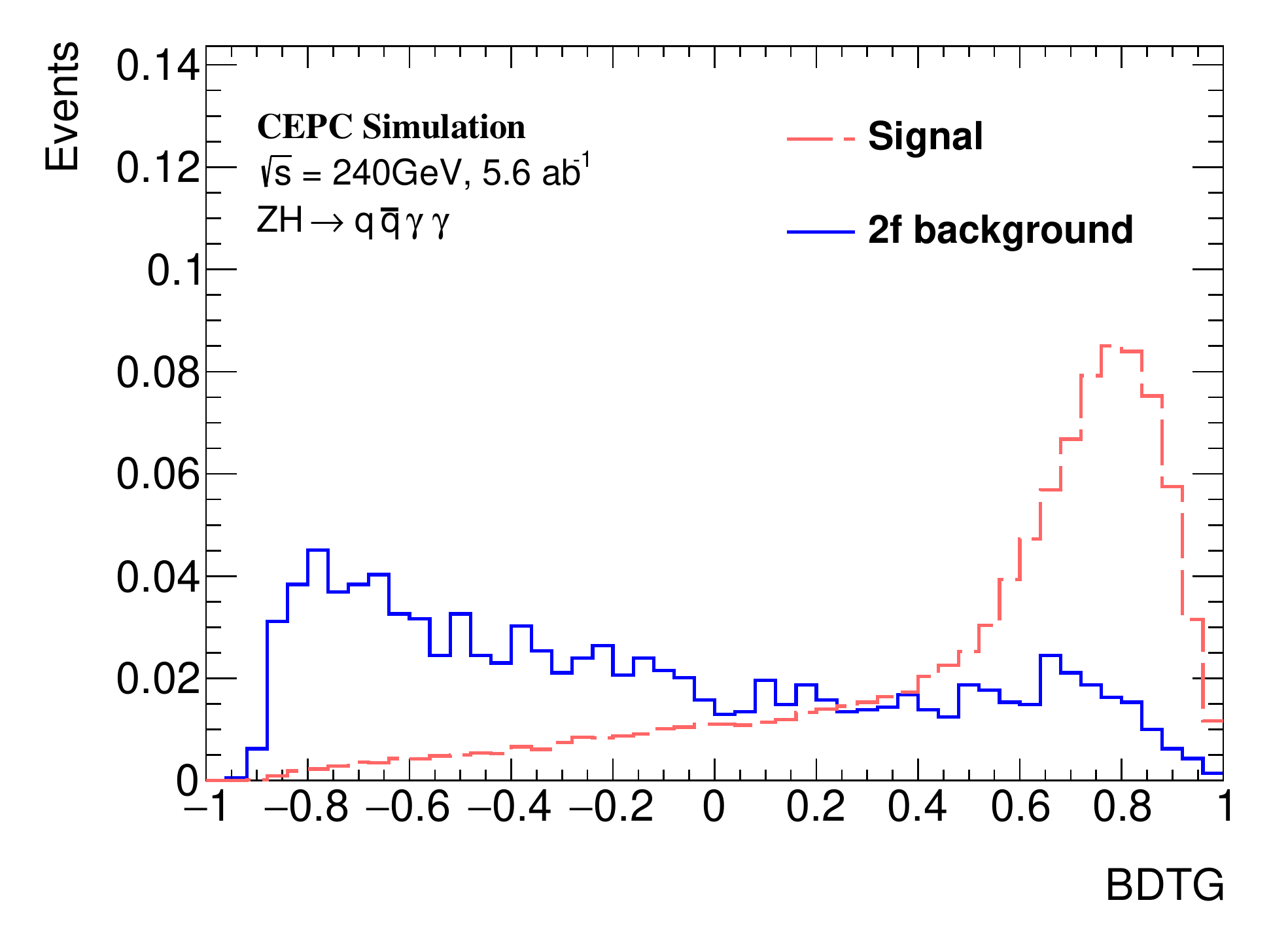} }
\caption{The ROC curve (left) and output BDTG distribution (right) in \qqyy channel. }
\label{fig:ROC_qq}
\end{figure}

\begin{figure}[]
  \centering
  \subfigure[$min\Delta R_{\gamma, \mu}$]    { \includegraphics[width= 0.40\linewidth]{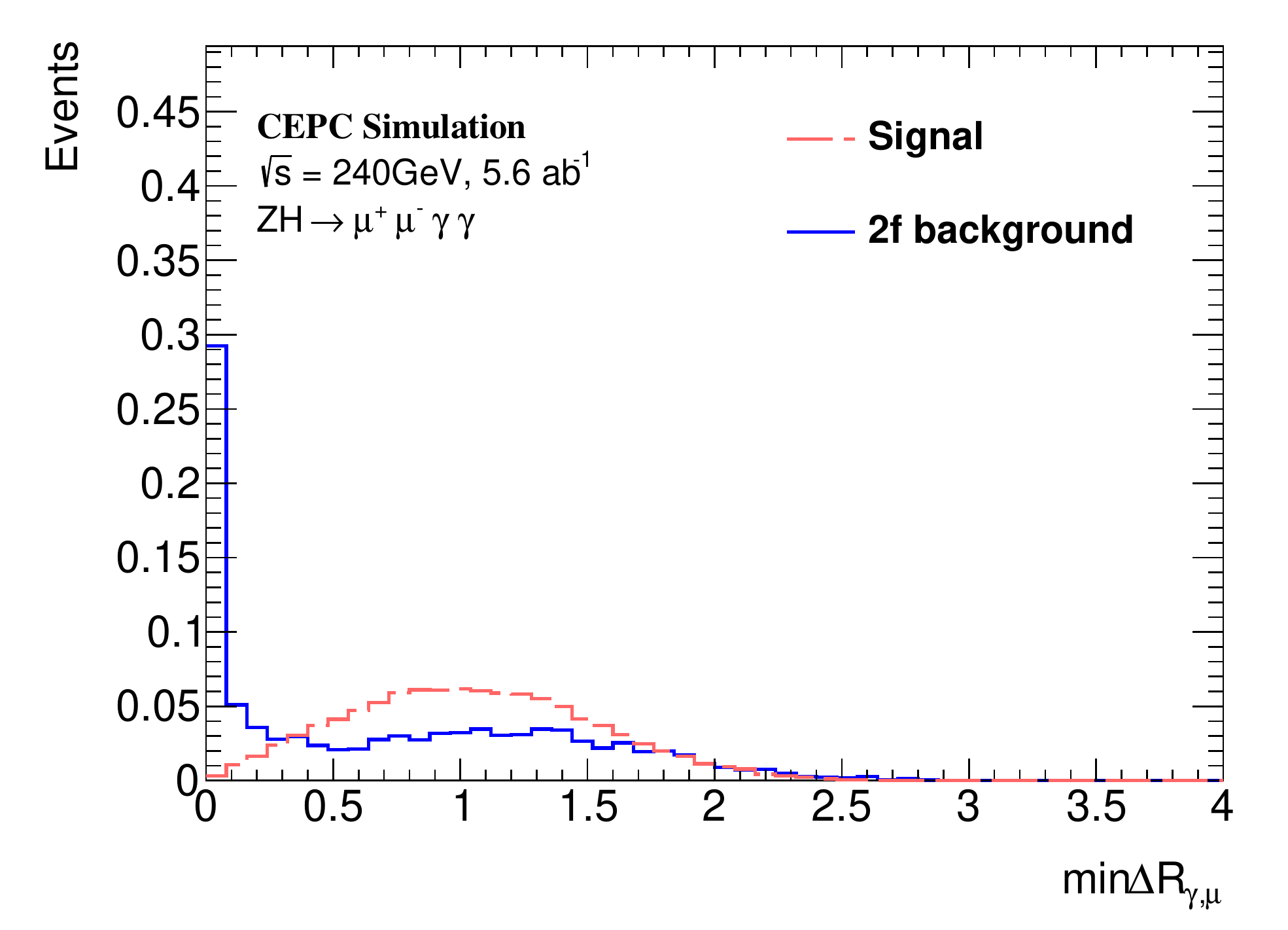} }
  \subfigure[$E_{\mu\mu}$]                   { \includegraphics[width= 0.40\linewidth]{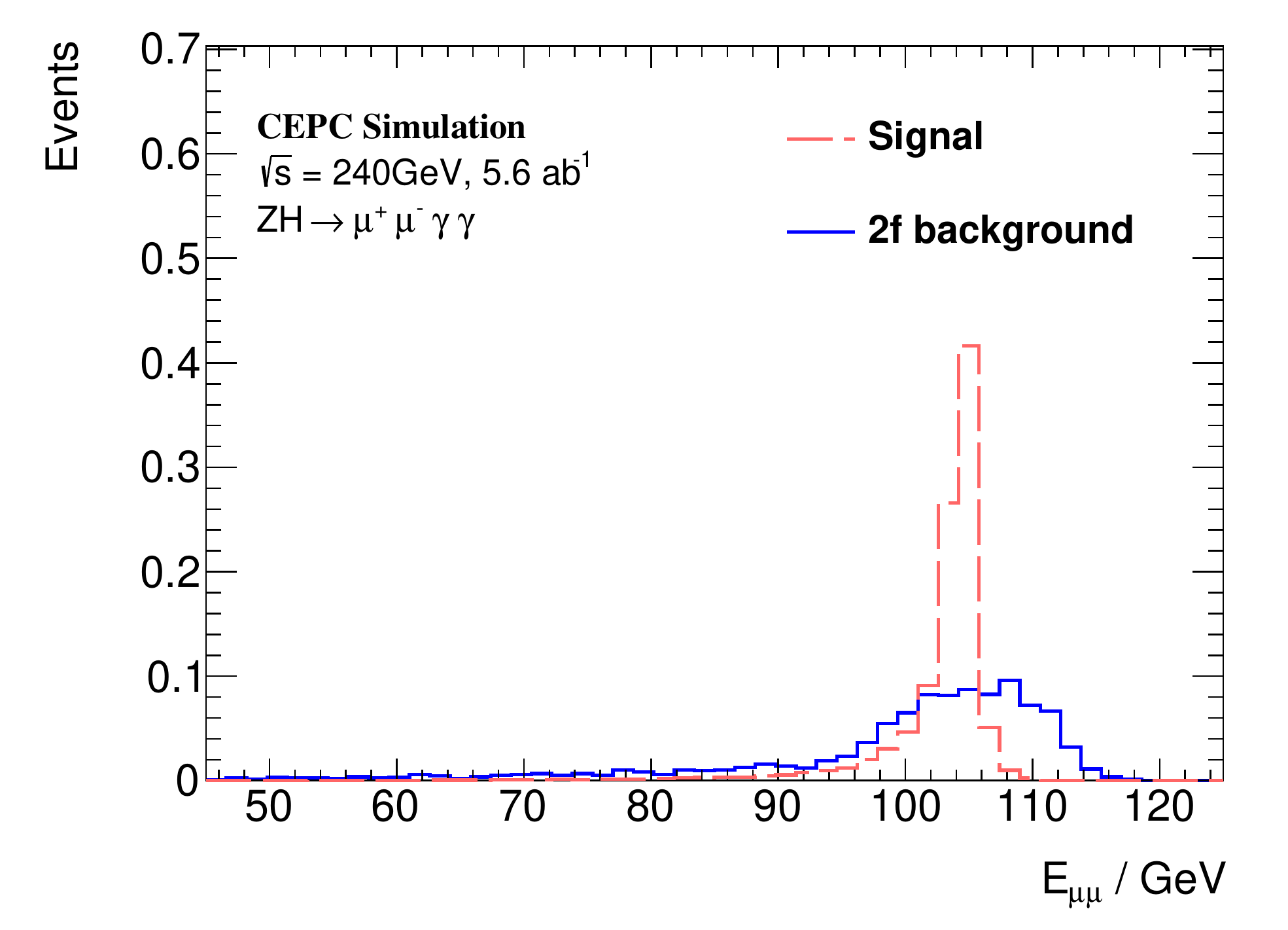} } \\
  \subfigure[$\cos\theta_{\gamma 1, \mu1}$]   { \includegraphics[width= 0.40\linewidth]{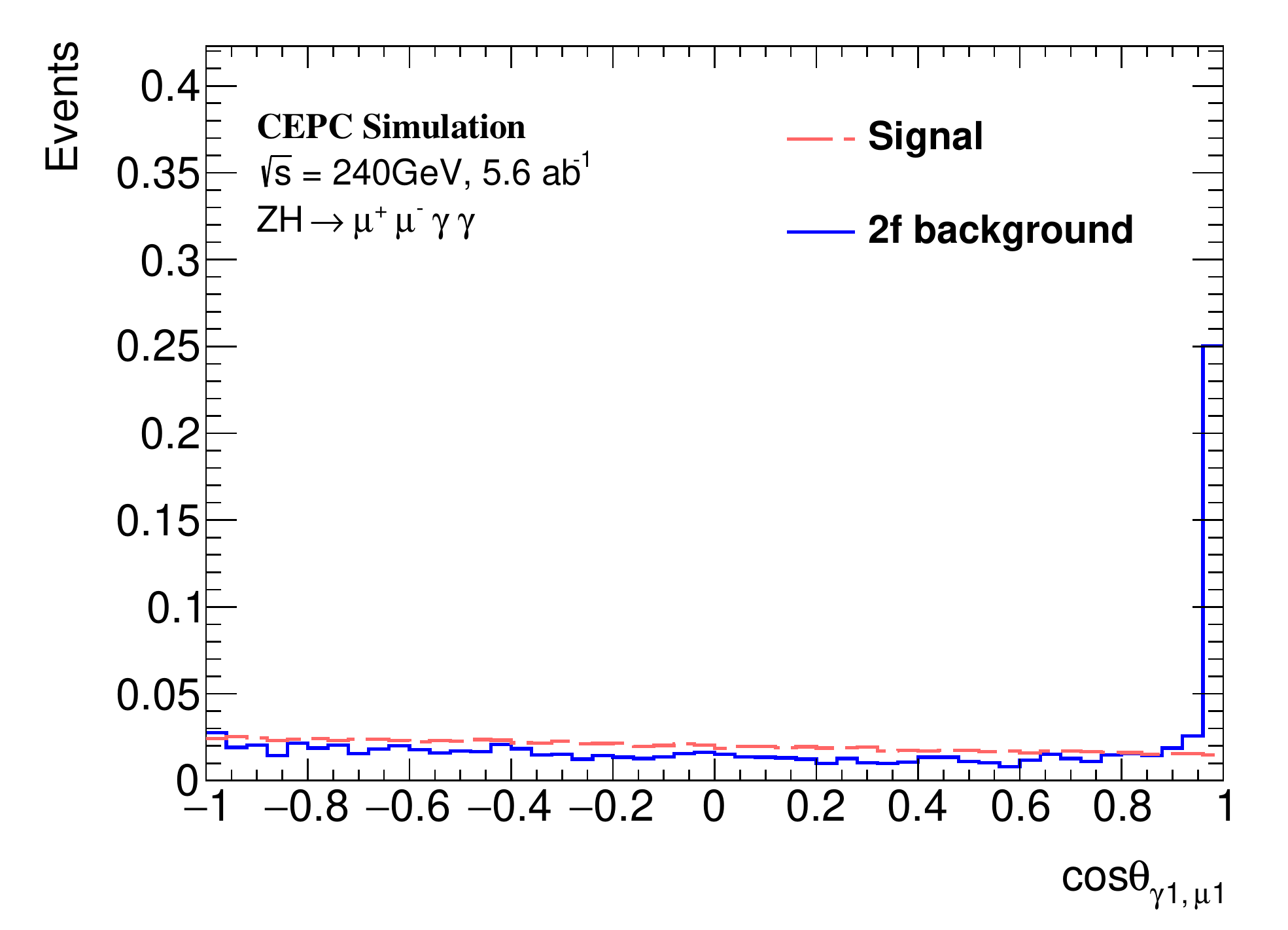} } 
  \subfigure[$E_{\gamma 2}$]                 { \includegraphics[width= 0.40\linewidth]{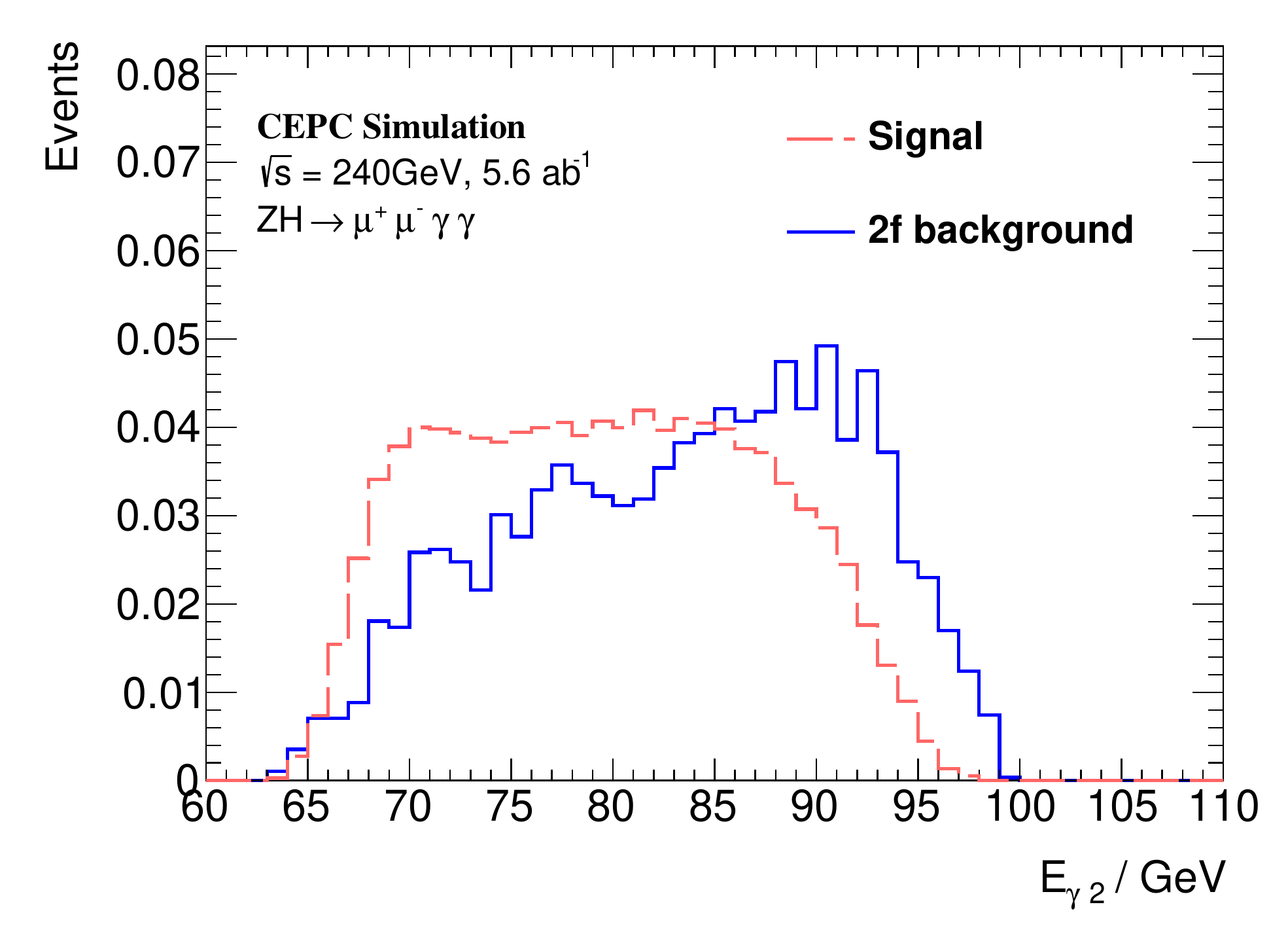} } \\
  \subfigure[$\Delta\Phi_{\gamgam}$]         { \includegraphics[width= 0.40\linewidth]{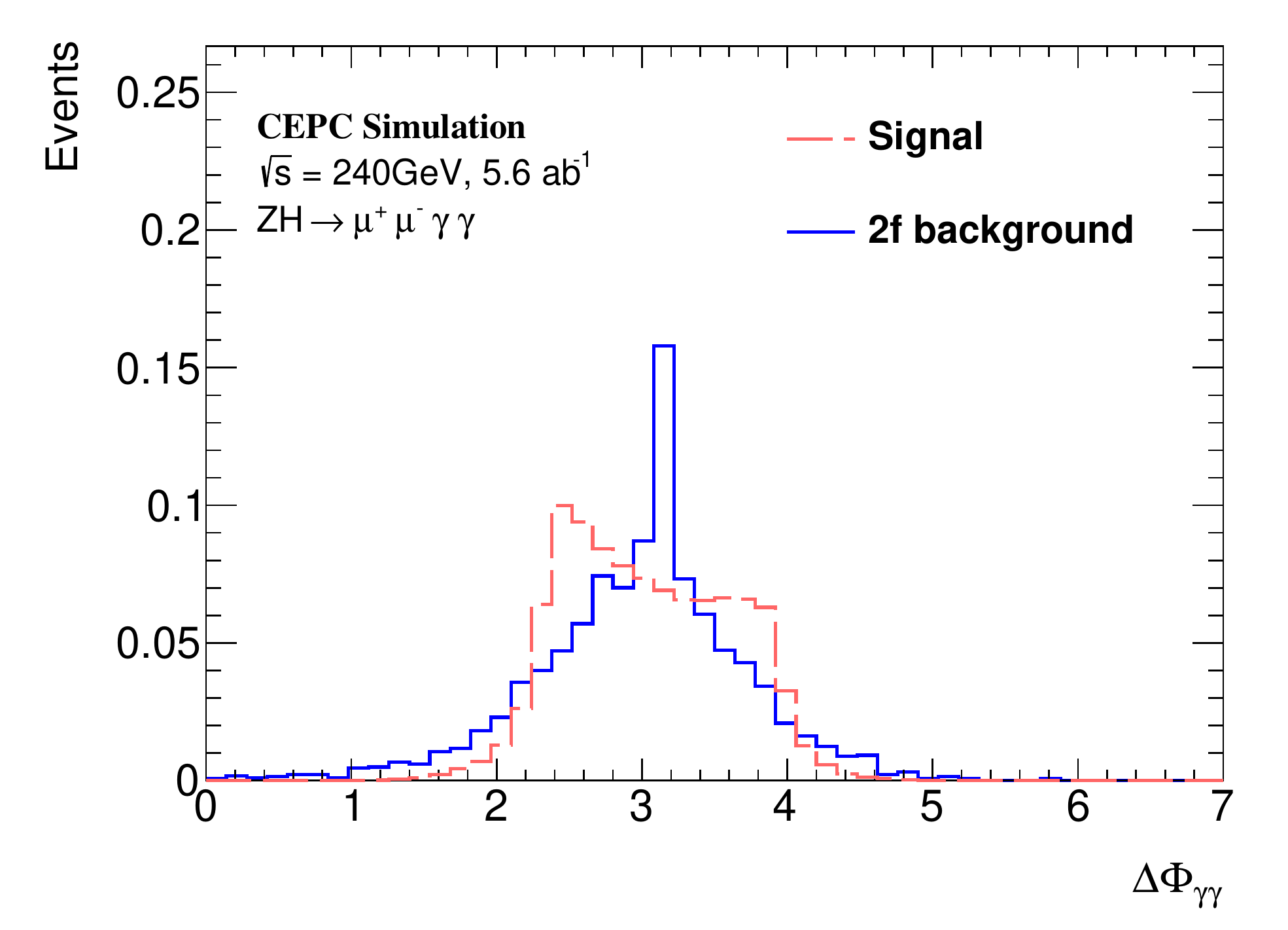} } 
  \subfigure[$\cos\theta_{\gamma 2}$]         { \includegraphics[width= 0.40\linewidth]{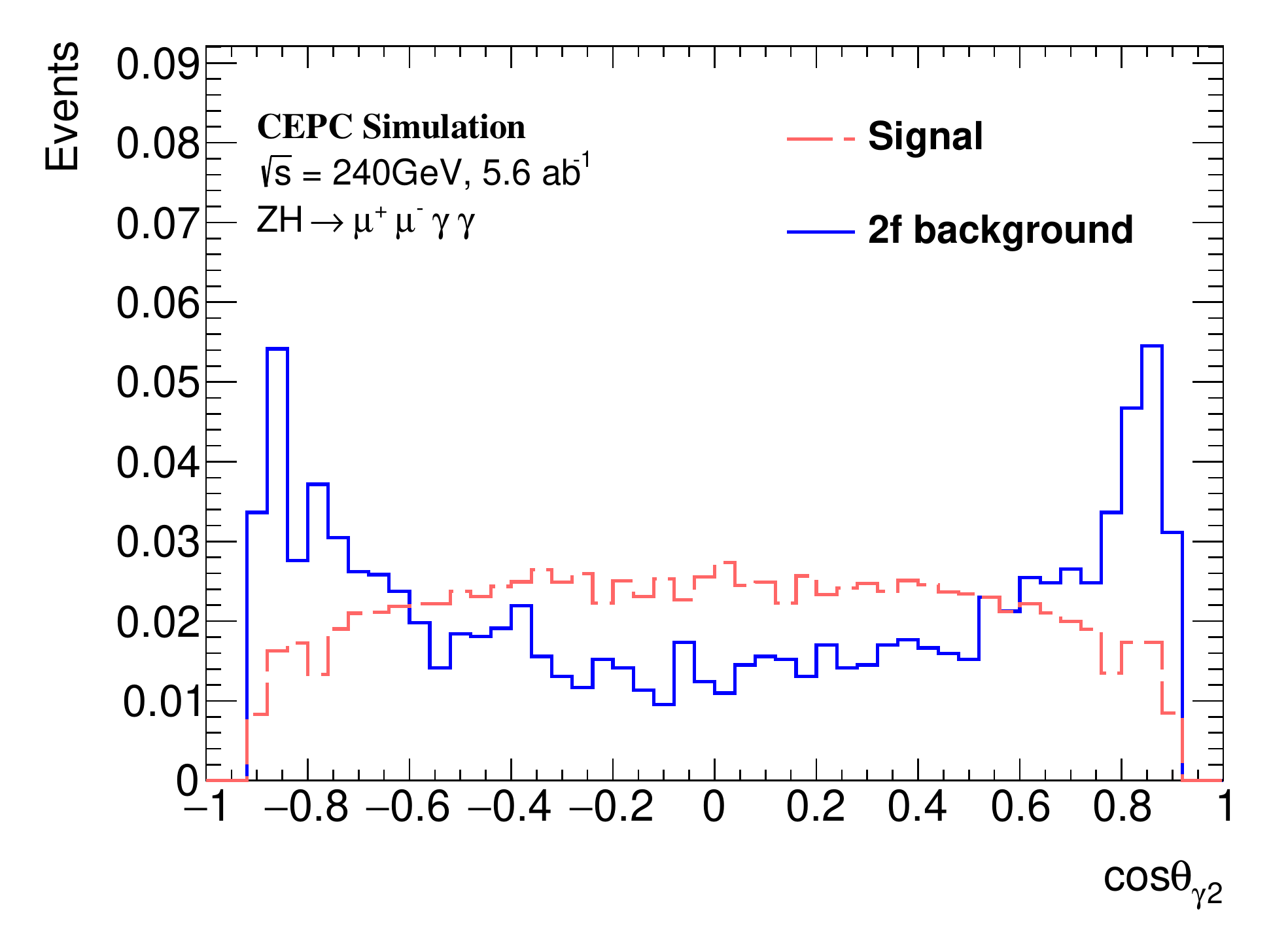} } \\
  \subfigure[$\Delta\Phi_{\gamgam, \mu\mu}$] { \includegraphics[width= 0.40\linewidth]{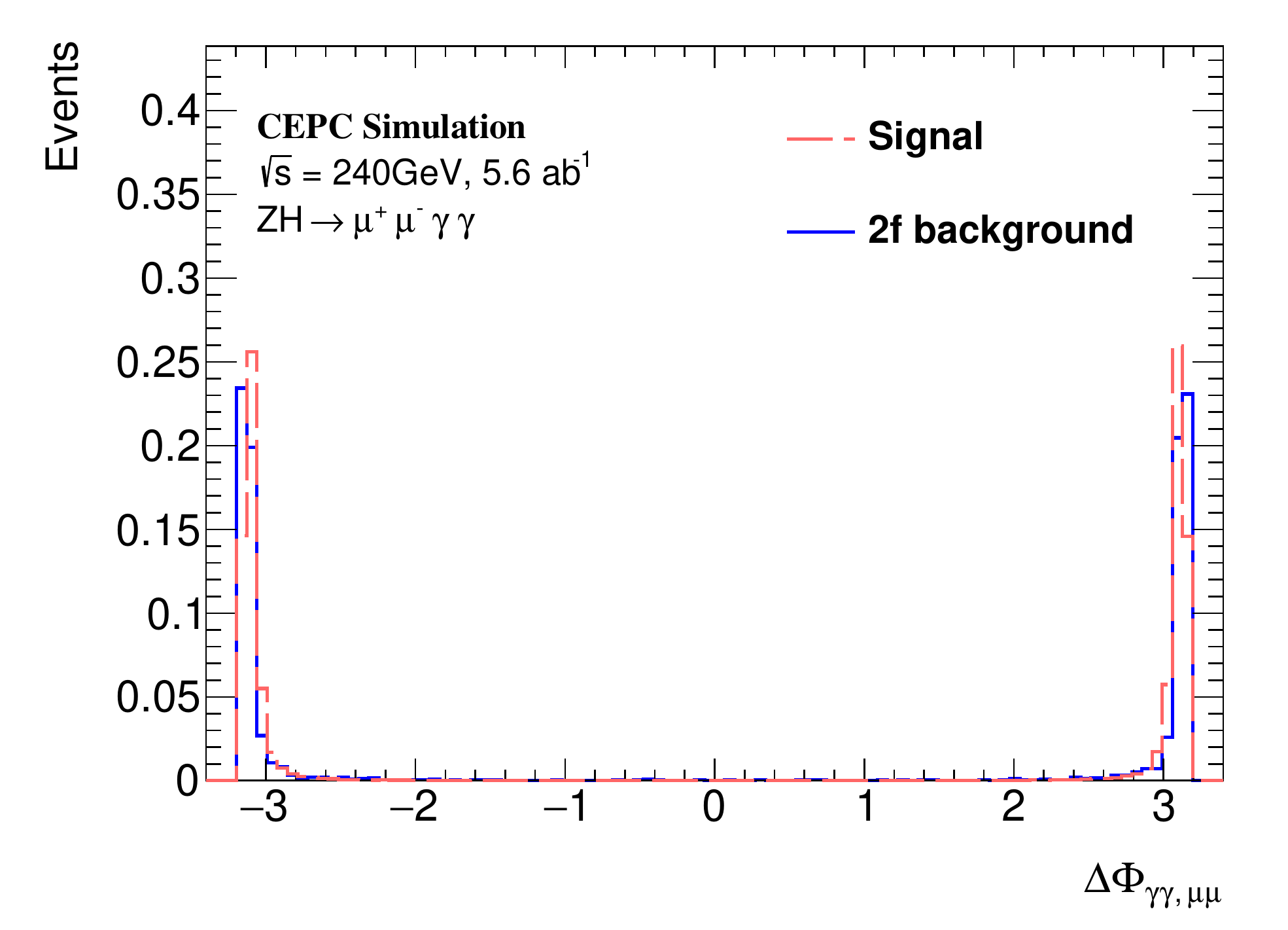} }
  \subfigure[$\cos\theta_{\mu 1}$]            { \includegraphics[width= 0.40\linewidth]{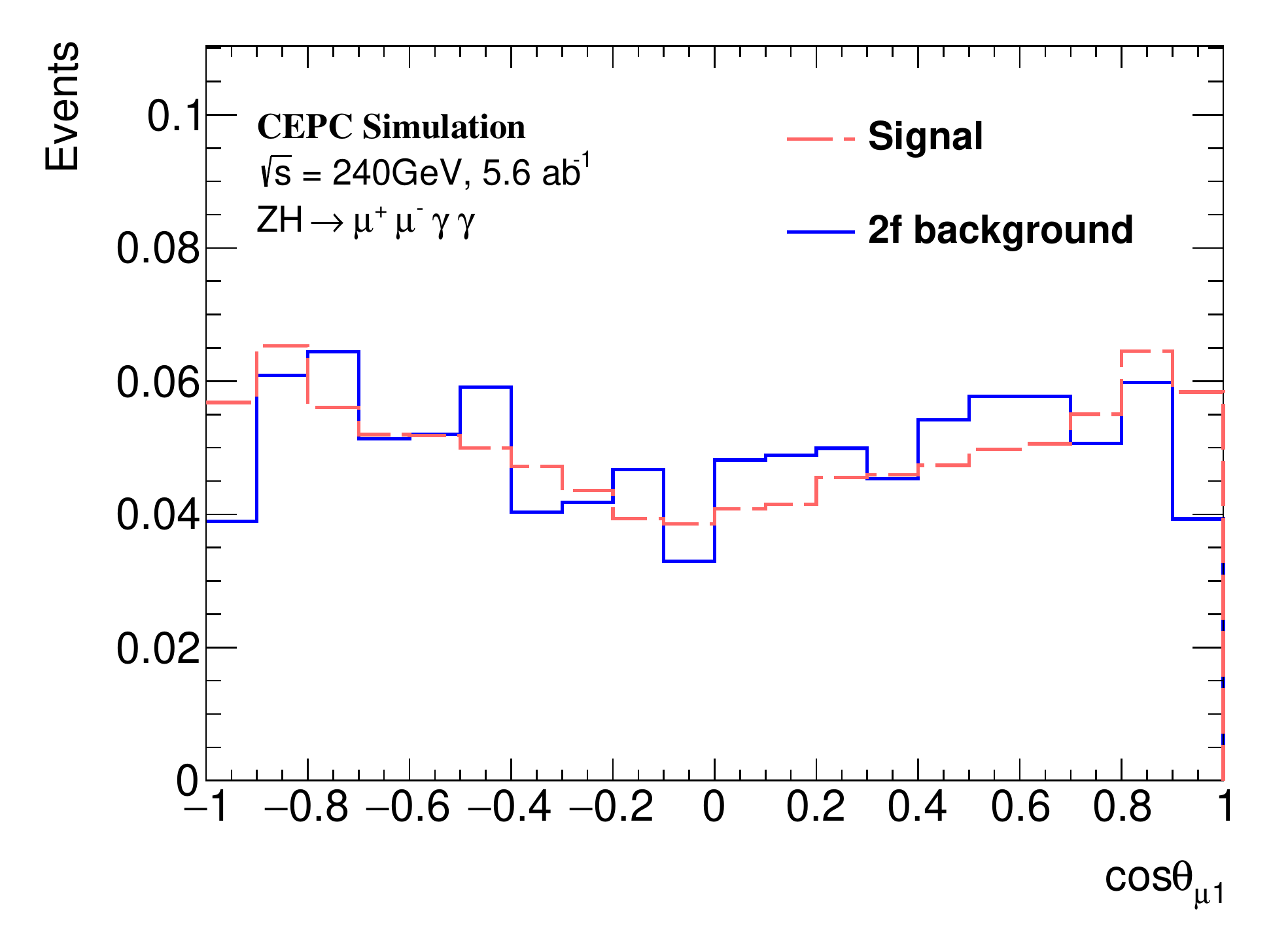} } \\
\caption{Training variables in \mmyy channel. The signal and background yields are normalized. }
\label{fig:InputVar_mm}
\end{figure}

\begin{figure}[]
  \centering
  \subfigure[]{ \includegraphics[width= 0.45\linewidth]{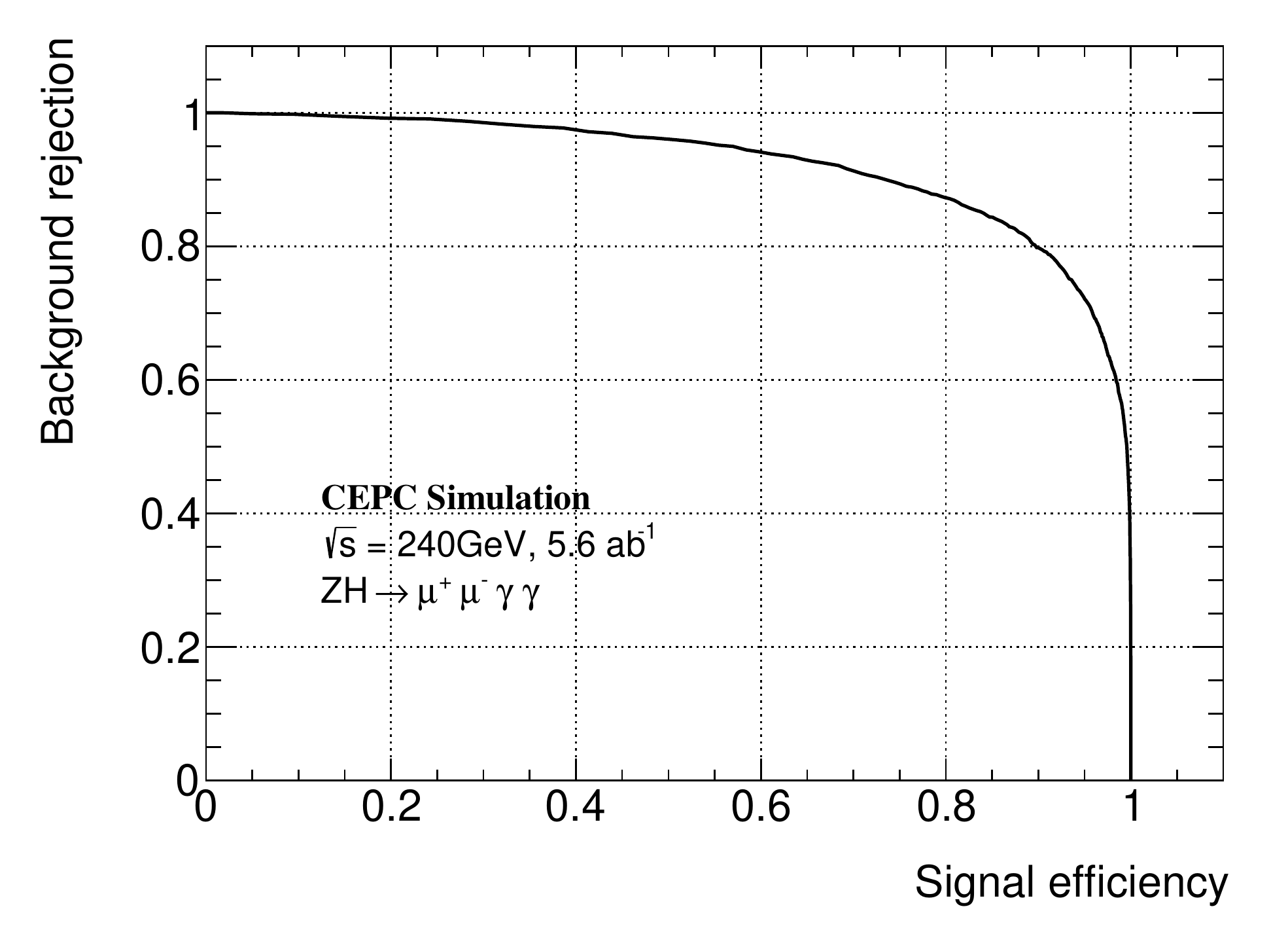} }
  \subfigure[]{ \includegraphics[width= 0.45\linewidth]{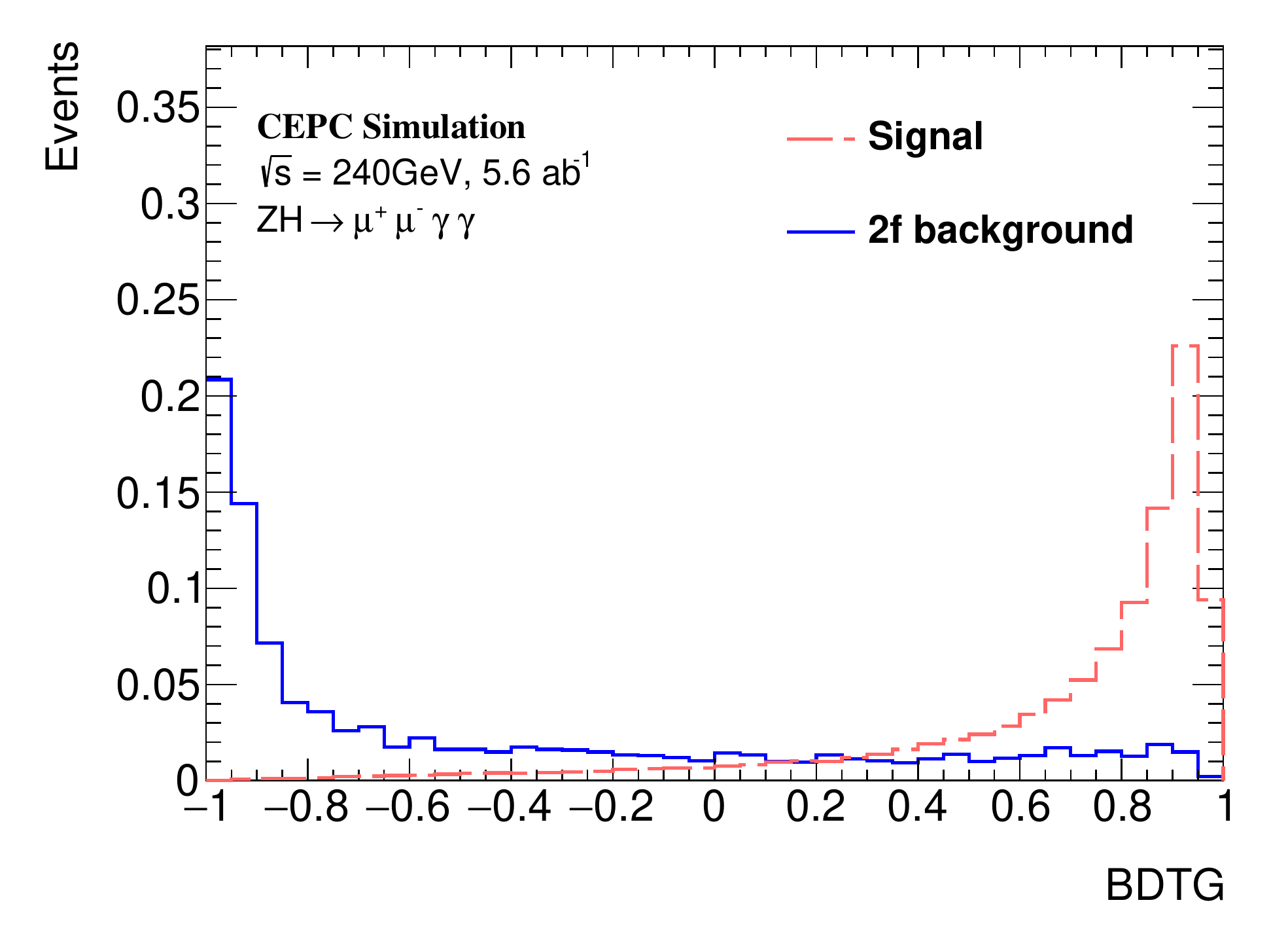} }
\caption{The ROC curve (left) and output BDTG distribution (right) in \mmyy channel. }
\label{fig:ROC_mm}
\end{figure}

\begin{figure}[]
  \centering
  \subfigure[$pT_{\gamma 1}$]         { \includegraphics[width= 0.40\linewidth]{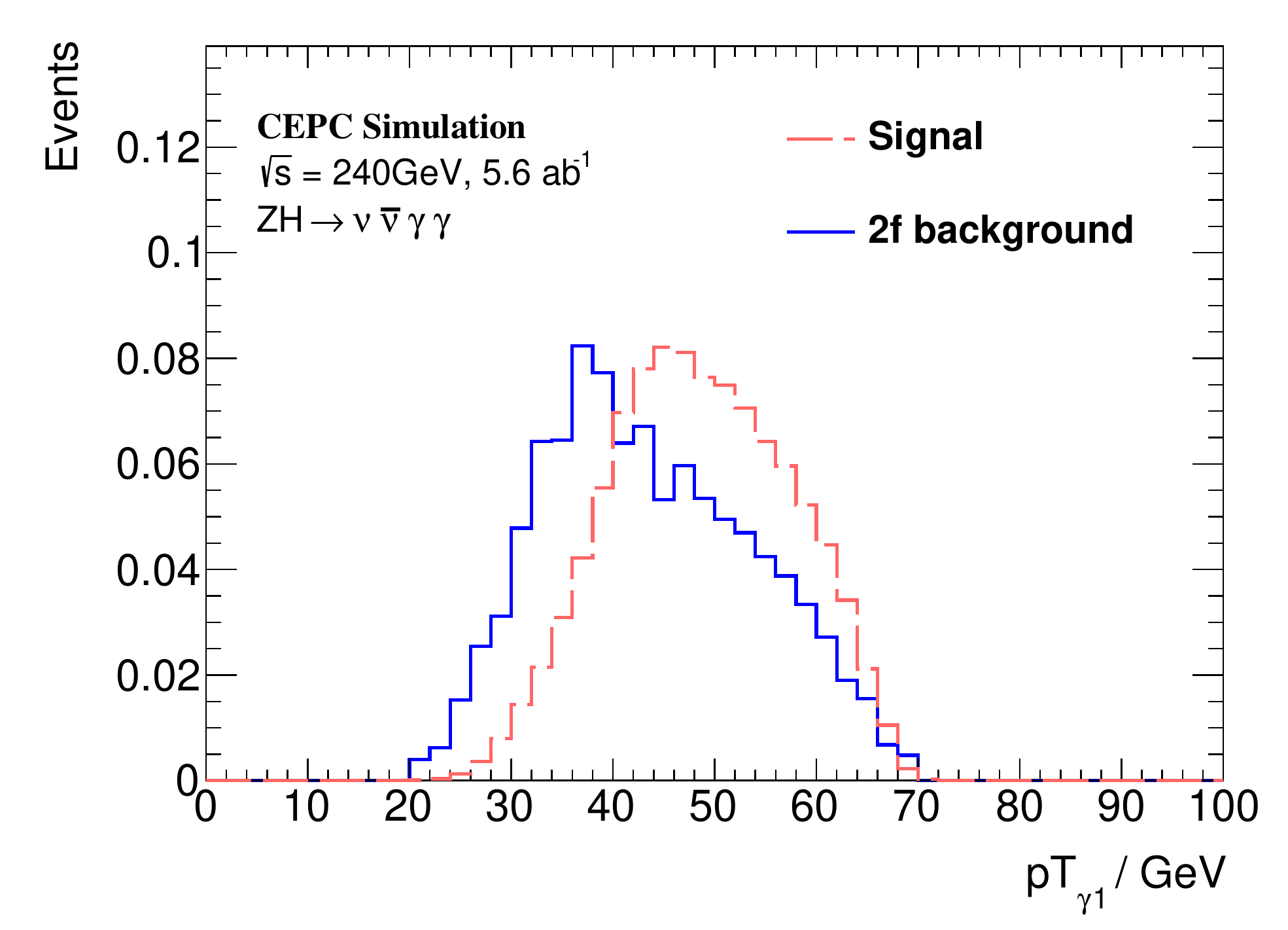} }
  \subfigure[$\cos\theta _{\gamma 2}$] { \includegraphics[width= 0.40\linewidth]{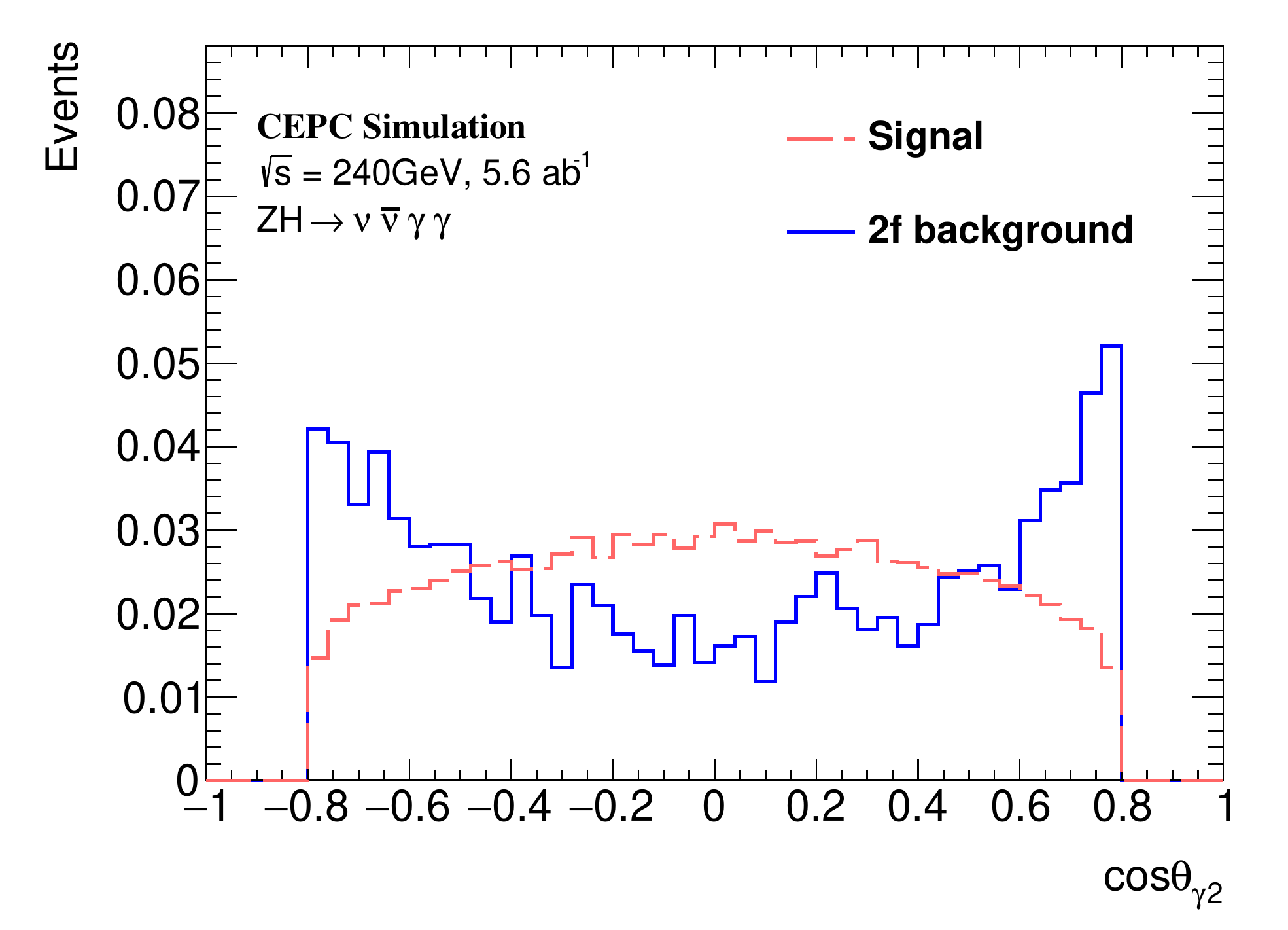} } \\
  \subfigure[$\Delta\Phi_{\gamgam}$]  { \includegraphics[width= 0.40\linewidth]{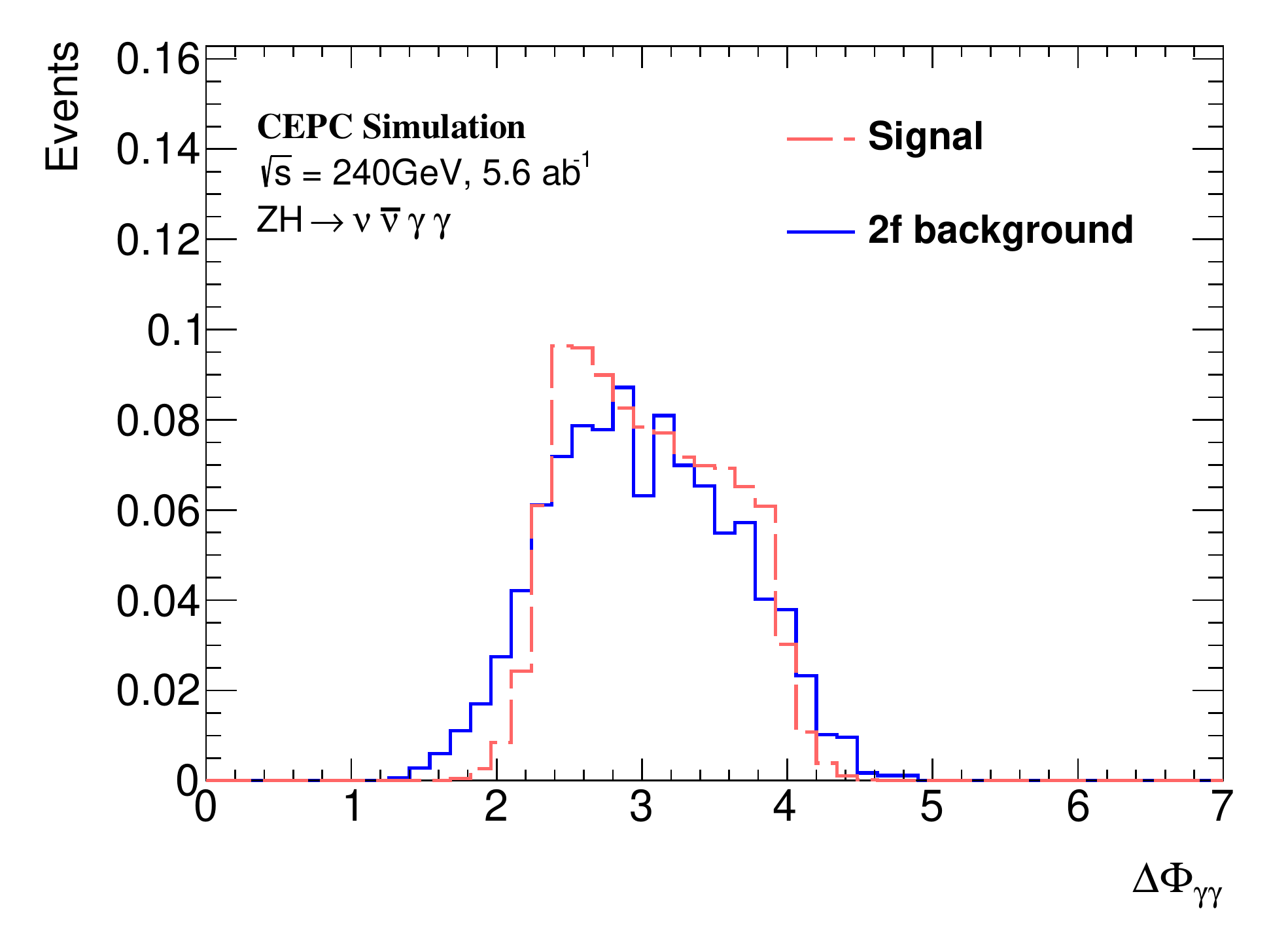} } 
  \subfigure[$pTt_{\gamgam}$]         { \includegraphics[width= 0.40\linewidth]{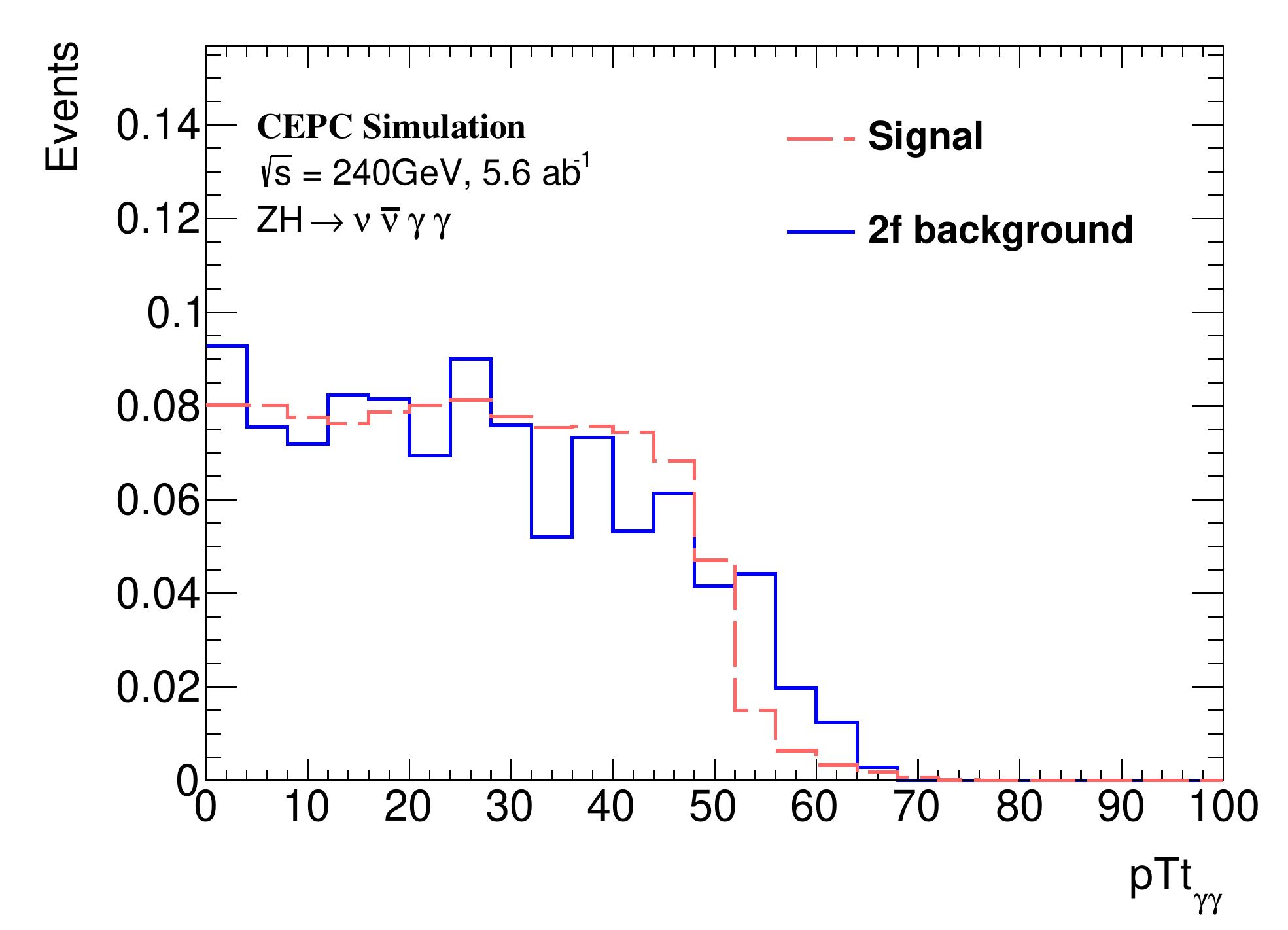} } \\
  \subfigure[$pT_{\gamma 2}$]         { \includegraphics[width= 0.40\linewidth]{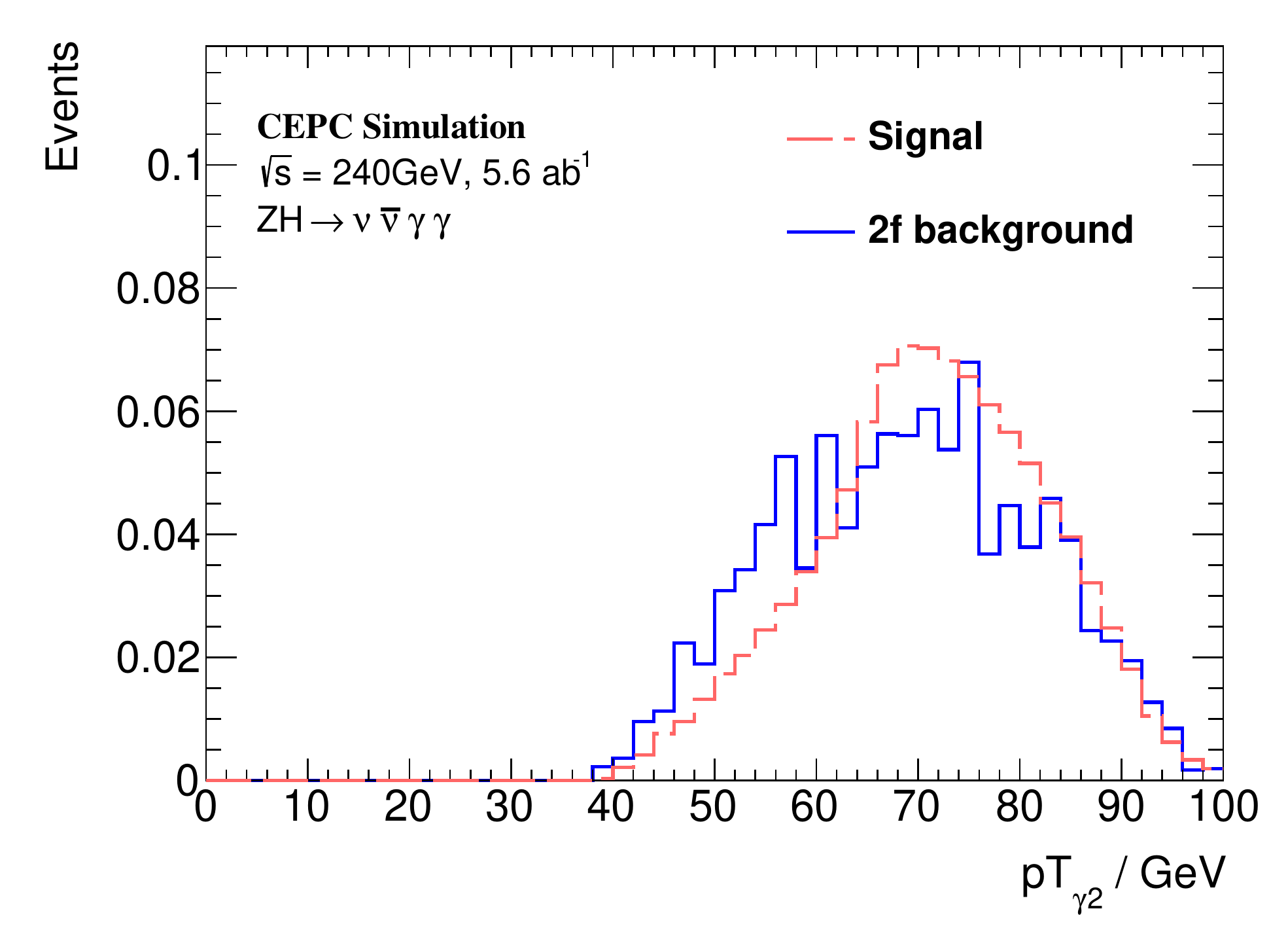} }
\caption{Training variables in \nnyy channel. The signal and background yields are normalized. }
\label{fig:InputVar_nn}
\end{figure}

\begin{figure}[]
  \centering
  \subfigure[]{ \includegraphics[width= 0.45\linewidth]{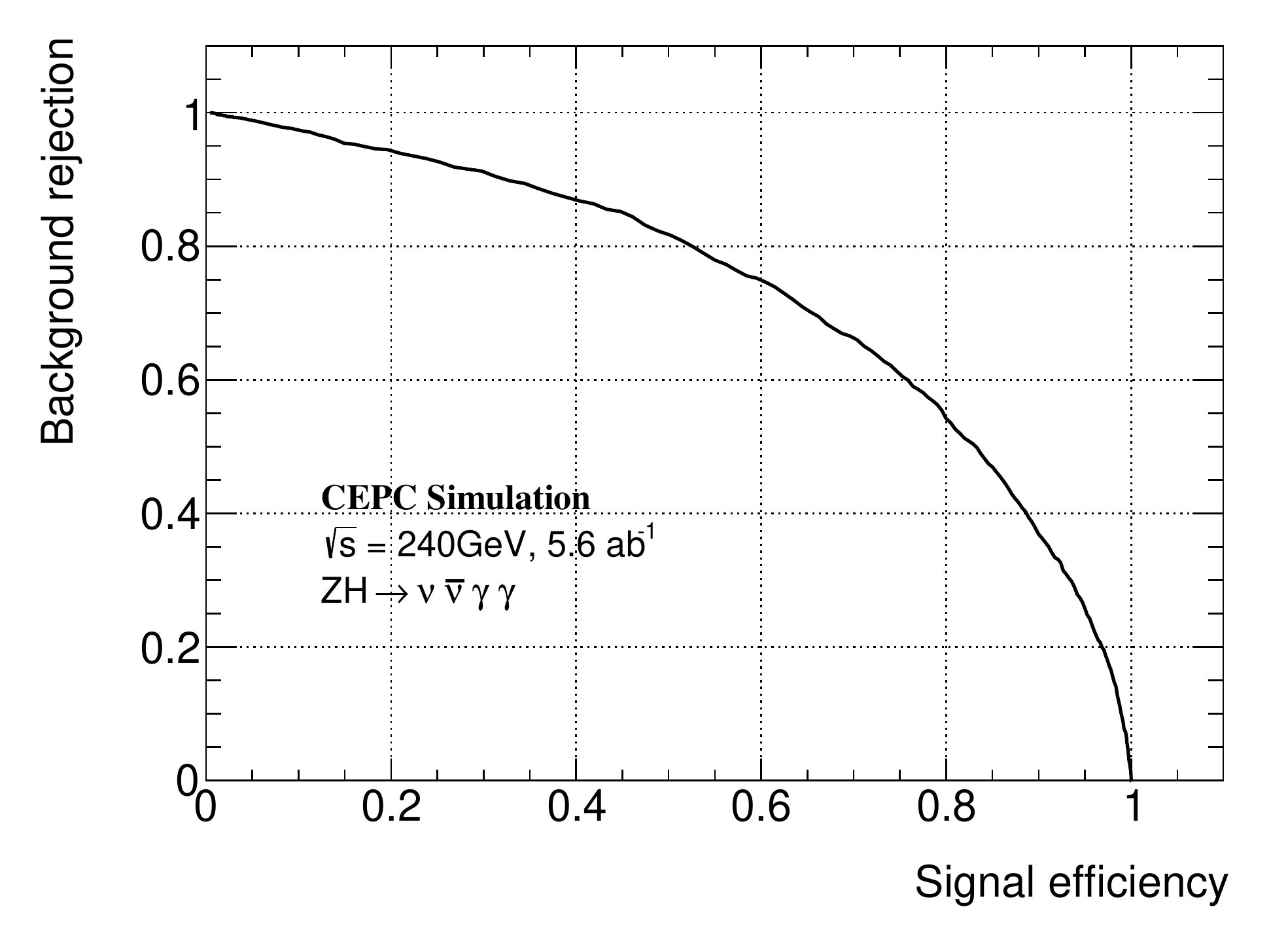} }
  \subfigure[]{ \includegraphics[width= 0.45\linewidth]{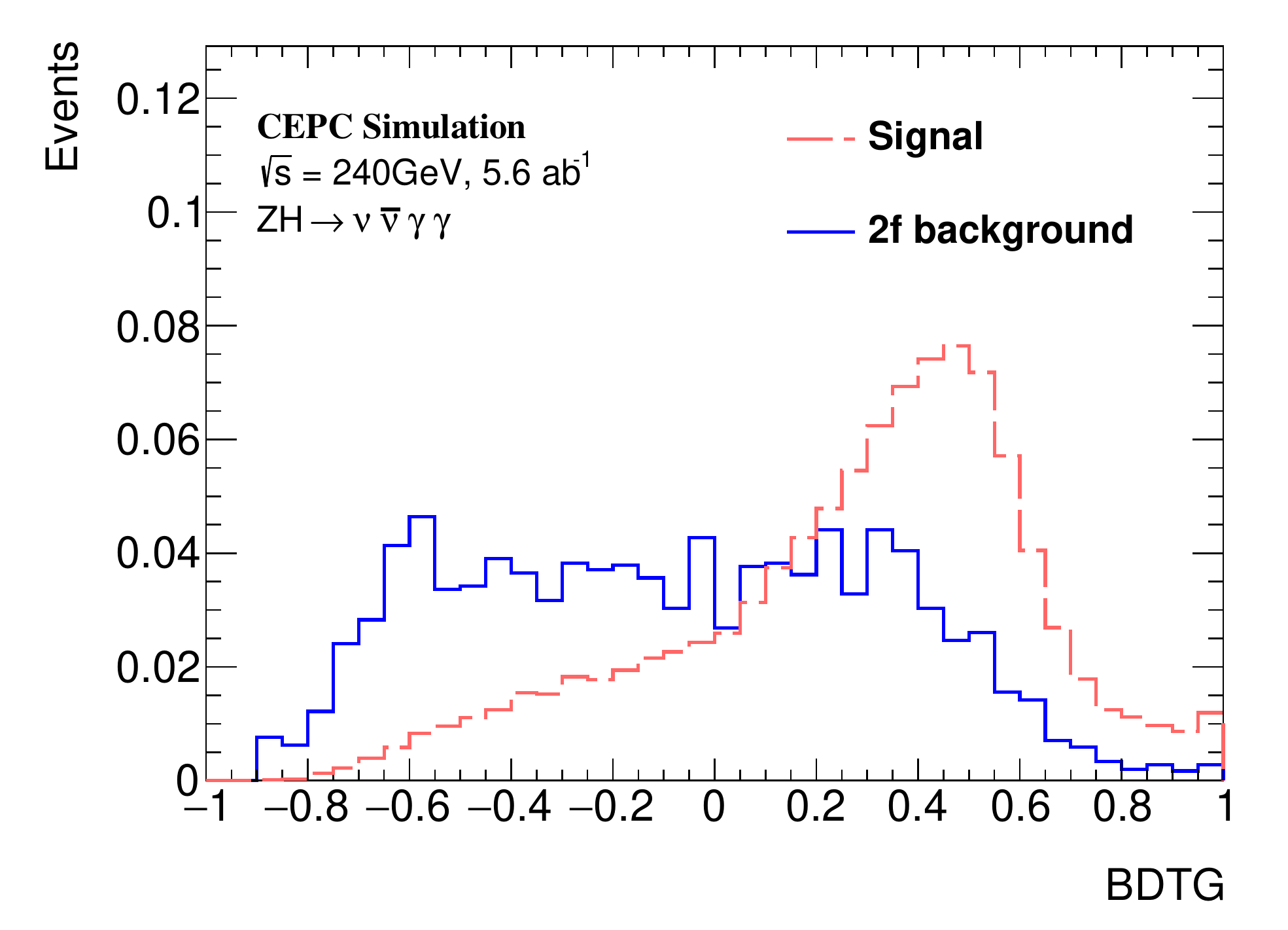} }
\caption{The ROC curve (left) and output BDTG distribution (right) in \nnyy channel. }
\label{fig:ROC_nn}
\end{figure}


\begin{figure}[ht]
  \centering
  \subfigure[\qqyy background]{ \includegraphics[width= 0.9\linewidth]{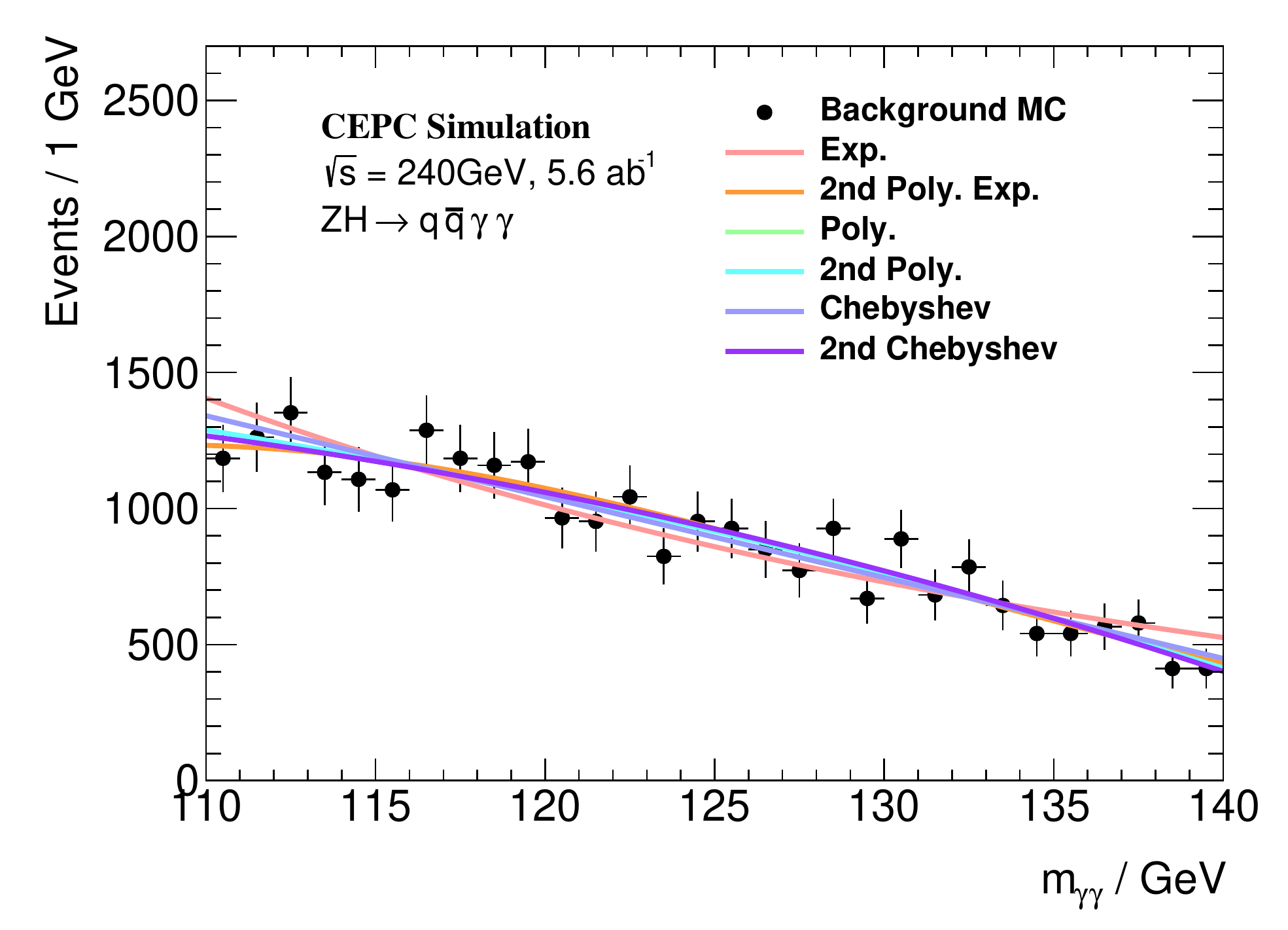} }
  \subfigure[\mmyy background]{ \includegraphics[width= 0.9\linewidth]{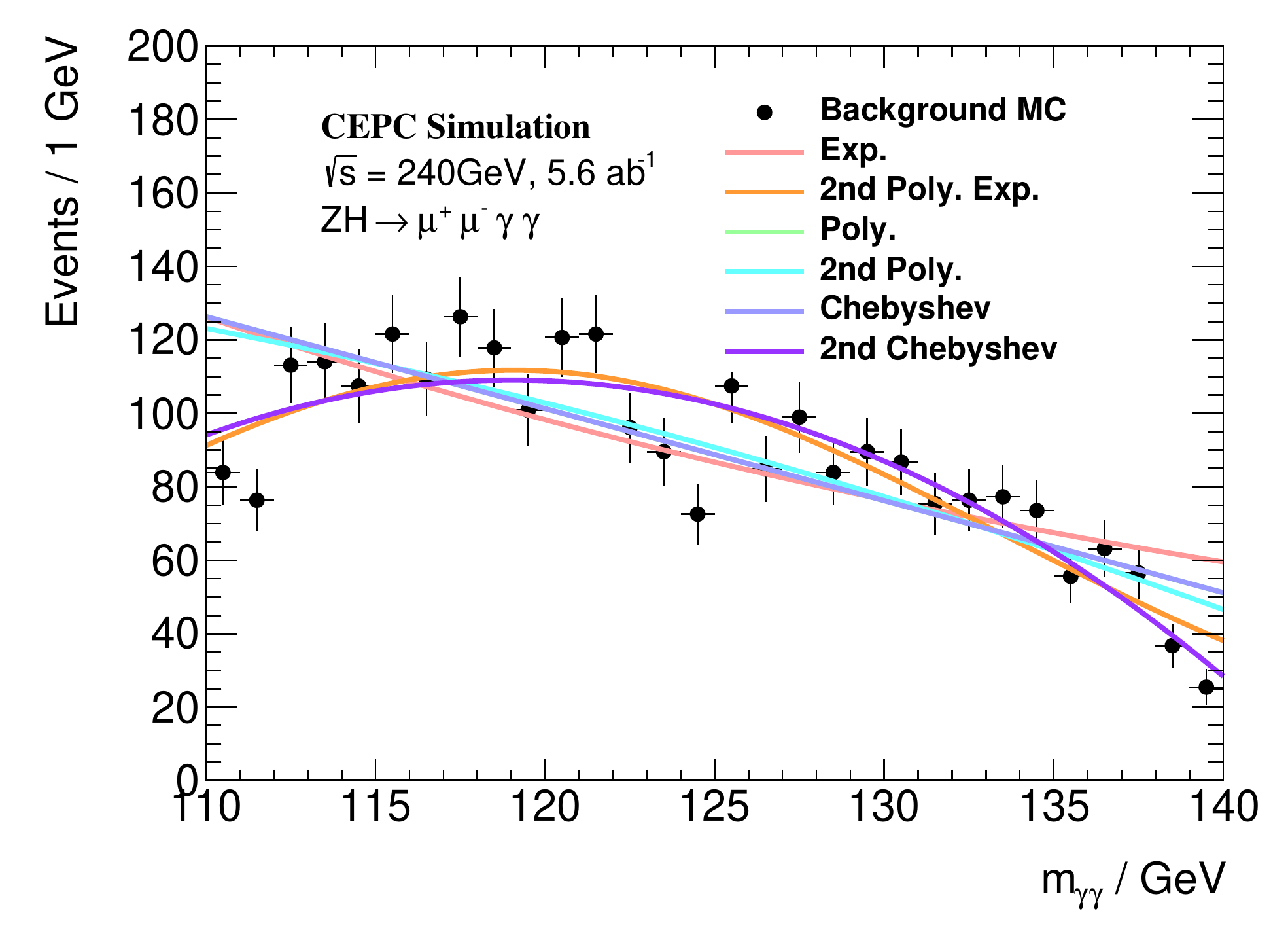} }
  \subfigure[\nnyy background]{ \includegraphics[width= 0.9\linewidth]{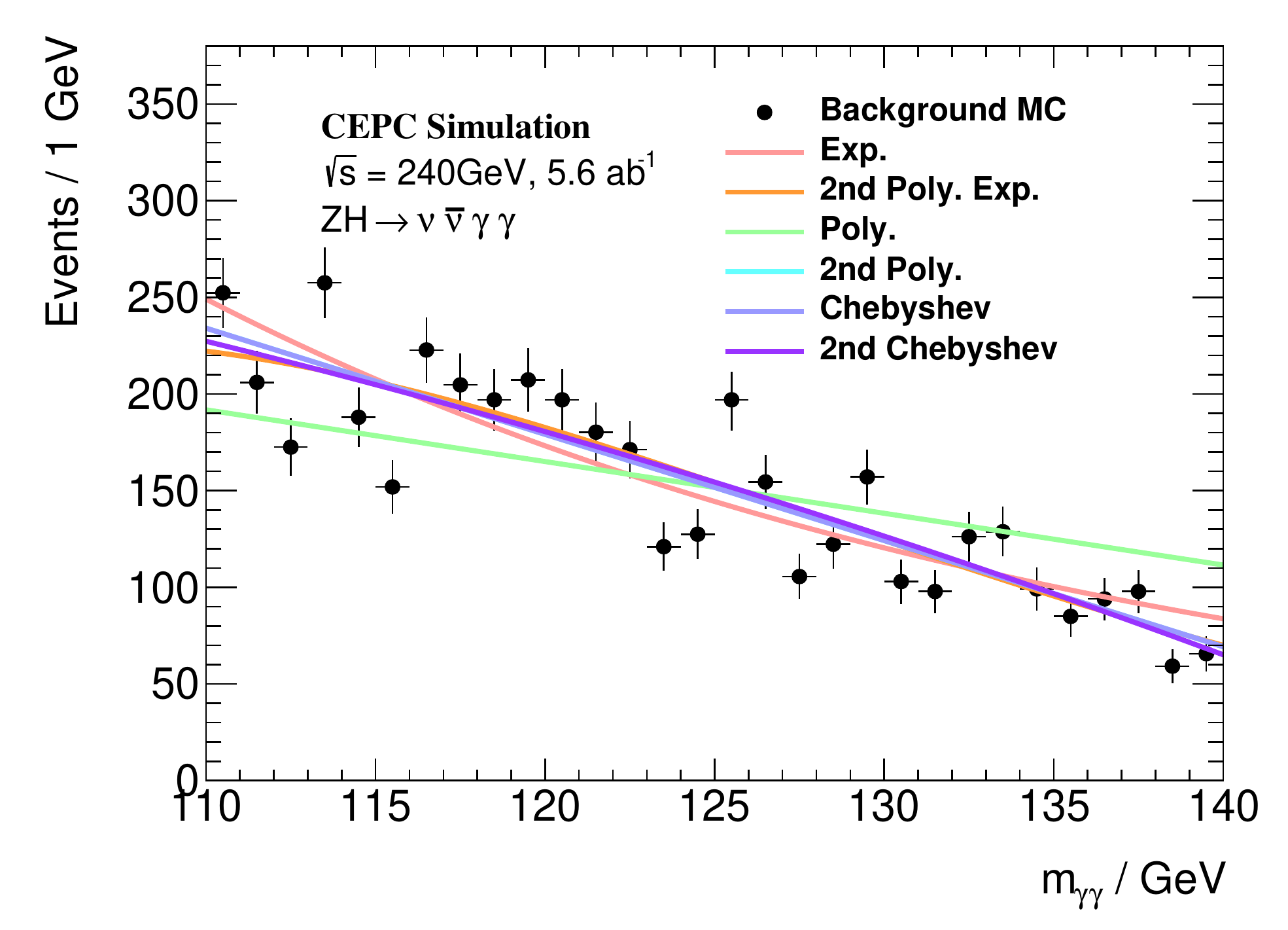} }
\caption{Tested functions for the background modeling. In All 3 channels the second order Chebyshev function gives the smallest $\chi^{2}/Ndof$ value. Detailed numbers are listed in Table~\ref{tab:BkgModel}. }
\label{fig:BkgModel}
\end{figure}

\begin{table}[ht]
\centering
\resizebox{0.9\linewidth}{!}{
\begin{tabular}{|l|c|c|c|}
\hline
                      & \qqyy & \mmyy & \nnyy  \\ \hline
1st order Exp.        & 0.941 & 5.423 & 3.786 \\ \hline
2nd order Exp.        & 0.610 & 2.035 & 3.435 \\ \hline
1st order Poly.       & 0.644 & 4.321 & 7.399 \\ \hline
2nd order Poly.       & 0.600 & 3.758 & 3.439 \\ \hline
1st order Chebyshev   & 0.644 & 4.321 & 3.320 \\ \hline
2nd order Chebyshev   & 0.596 & 1.789 & 3.411 \\ \hline
\end{tabular}}
\caption{The $\chi^{2}$/Ndof values for 6 considered models in the background modeling in 3 channels, 
      including the first and second order exponential, polynomial and Chebyshev functions.} 
\label{tab:BkgModel}
\end{table}



\end{document}